\documentclass[aps,superscriptaddress,showpacs,nofootinbib,11pt]{revtex4-1}
\usepackage{mathrsfs}
\usepackage{bigints}
\usepackage{latexsym,bm}
\usepackage{graphicx}
\usepackage{indentfirst}
\usepackage{slashed}
\usepackage{amssymb}
\usepackage{amsmath}
\usepackage{bbm}
\usepackage{color}
\usepackage[dvipsnames]{xcolor}
\usepackage{epsfig}
\usepackage[titletoc]{appendix}
\usepackage{rotating}
\usepackage{epstopdf}
\usepackage{extarrows}
\usepackage{tabularx}
\usepackage{soul} 
\usepackage{xcolor}
\usepackage{lipsum}
\usepackage[colorlinks=true,allcolors=BlueViolet,hyperfootnotes=false]{hyperref}

\usepackage[utf8]{inputenc}

\DeclareMathOperator\artanh{artanh}

\newcommand{\be}{\begin{equation}} \newcommand{\ee}{\end{equation}}
\newcommand{\ba}{\begin{array}{c}} \newcommand{\ea}{\end{array}}
\newcommand{\bea}{\begin{eqnarray}} \newcommand{\eea}{\end{eqnarray}}

\newcommand\tstrut{\rule{0pt}{2.9ex}}       

\begin{document}

\title{\Large Visible energy and angular distributions of the charged particle from the $\tau-$decay  in 
 $b\to c \tau\, (\mu \bar \nu_\mu \nu_\tau,\pi\nu_\tau,\rho\nu_\tau) 
 \bar\nu_\tau$ reactions}

\author{Neus Penalva}
\affiliation{Instituto de F\'{\i}sica Corpuscular (centro mixto CSIC-UV), 
Institutos de Investigaci\'on de Paterna,
C/Catedr\'atico Jos\'e Beltr\'an 2, E-46980 Paterna, Valencia, Spain}

\author{Eliecer Hern\'andez}
\affiliation{Departamento de F\'\i sica Fundamental 
  e IUFFyM,\\ Universidad de Salamanca, Plaza de la Merced s/n, E-37008 Salamanca, Spain}
\author{ Juan Nieves}
\affiliation{Instituto de F\'{\i}sica Corpuscular (centro mixto CSIC-UV), 
Institutos de Investigaci\'on de Paterna,
C/Catedr\'atico Jos\'e Beltr\'an 2, E-46980 Paterna, Valencia, Spain}


\date{\today}
\begin{abstract}
We study the  $d^2 \Gamma_d /(d\omega d\cos\theta_d) $, $d\Gamma_d /d\cos
\theta_d$ and $ d\Gamma_d /dE_d $ distributions, which are defined  in 
terms of the visible energy and polar angle of the charged particle from 
the $\tau-$decay  in  $b\to c \tau\, (\mu \bar \nu_\mu \nu_\tau,\pi
\nu_\tau,\rho\nu_\tau) \bar\nu_\tau$ reactions. 
These differential decay widths could be measured in the near future with 
certain precision. The first two  contain information on the  transverse tau-spin, 
tau-angular and  tau-angular-spin asymmetries of the $H_b\to H_c\tau\bar
\nu_\tau$ parent decay and, from a  dynamical point of view, they
are richer than the commonly used one, $d^2 \Gamma_d /(d\omega dE_d) $, since the latter 
 only depends on the   tau longitudinal polarization. 
We pay attention 
to the deviations with respect to the predictions of the standard model (SM) 
for these new 
observables, considering new physics (NP) operators constructed using both right- and 
left-handed neutrino fields, within an effective field-theory approach. 
We present 
results for   $\Lambda_b\to\Lambda_c\tau\, (\mu \bar \nu_\mu \nu_\tau,\pi
\nu_\tau,\rho\nu_\tau)  \bar\nu_\tau$  and $\bar B \to D^{(*)}\tau\, (\mu \bar \nu_\mu \nu_\tau,\pi\nu_\tau,\rho\nu_\tau) 
 \bar\nu_\tau$ sequential decays   
 and discuss their use to disentangle between different NP models. In this 
 respect, we show that $d\Gamma_d /d\cos\theta_d$, which should be measured 
 with sufficiently good statistics, becomes quite useful, especially in the 
 $\tau\to \pi \nu_\tau$  mode. The study carried out in this work could be of special relevance due to the  recent  
LHCb measurement of the lepton flavor universality ratio ${\cal R}_{\Lambda_c}$ 
in agreement with the SM. The experiment identified the $\tau$  using its 
hadron decay into $\pi^-\pi^+\pi^-\nu_\tau$, and this result for 
${\cal R}_{\Lambda_c}$, which is in conflict with the phenomenology 
from the $b$-meson sector, needs confirmation from other tau reconstruction channels. 

\end{abstract}

%


\maketitle
\section{Introduction}
%

In the quest to discover new physics (NP) beyond the Standard Model (SM), 
the experimental signals of possible violations of lepton flavor universality 
(LFU) in charged-current (CC) semileptonic $B\to D^{(*)}$ decays reported by 
BaBar~\cite{BaBar:2012obs, BaBar:2013mob}, Belle~\cite{Belle:2015qfa, 
Belle:2016ure,Belle:2016dyj,Belle:2019rba} and LHCb~\cite{LHCb:2015gmp, 
LHCb:2017smo,LHCb:2017rln} have triggered a large activity in  recent years. These 
experiments measured  the ${\cal R}_{ D}=
  \Gamma(\bar B\to D\tau\bar\nu_\tau)/\Gamma(\bar B\to D\ell\bar\nu_\ell)$ 
  and  ${\cal R}_{ D^{*}}=
  \Gamma(\bar B\to D^{*}\tau\bar\nu_\tau)/\Gamma(\bar B\to D^{*}\ell\bar
  \nu_\ell)$ ratios ($\ell=e,\mu
  $), which combined analysis give rise to a $3.1\sigma$ 
  tension with SM results~\cite{HFLAV:2019otj}. The 
  similar  ${\cal R}_{J/\psi}=
  \Gamma(\bar B_c\to J/\psi\tau\bar\nu_\tau)/\Gamma(\bar B_c\to
   J/\psi\mu\bar\nu_\mu)$ observable, measured by the LHCb
Collaboration~\cite{LHCb:2017vlu}, provides also a 1.8 $\sigma$ discrepancy
 with different SM predictions~\cite{Anisimov:1998uk,Ivanov:2006ni,
Hernandez:2006gt,Huang:2007kb,Wang:2008xt,Wen-Fei:2013uea, Watanabe:2017mip, Issadykov:2018myx,Tran:2018kuv,
Hu:2019qcn,Leljak:2019eyw,Azizi:2019aaf,Wang:2018duy}.  Belle has also 
provided results for the averaged  tau-polarization  asymmetry  and the 
longitudinal  $D^*$ polarization ~\cite{Belle:2016dyj, Belle:2019ewo}, 
which together with an upper bound of the leptonic decay rate 
$\bar B_c\to \tau\bar\nu_\tau$ \cite{Alonso:2016oyd}, are commonly used  
to constrain NP contributions in the theoretical global fits to these 
LFU anomalies.

Another reaction that could shed light on the ${\cal R}_{ D^{(*)}}$ puzzle is 
the $\Lambda_b \to \Lambda_c \tau\bar\nu_\tau$ decay, and in particular the universality 
ratio ${\cal R}_{\Lambda_c}$ can be analogously constructed. A result of 
${\cal R}_{\Lambda_c}=0.242 \pm 0.026 \pm 0.040 \pm 0.059$  has just been announced by the
LHCb collaboration~\cite{LHCb:2022piu}, which is in agreement within errors  
with the SM prediction (${\cal R}_{\Lambda_c}^{\rm SM}=0.332 \pm 0.007 \pm 
0.007$~\cite{Detmold:2015aaa}). Contrary, to what is found for the ratios 
measured for the  $ b \to c $ transitions in the meson sector, the central 
value reported in \cite{LHCb:2022piu} turns out to be below the SM result. 
The $\tau^-$ lepton in \cite{LHCb:2022piu} is reconstructed using the three-prong
 hadronic $\tau^-   \to\pi^-\pi^+\pi^-(\pi^0)\,\nu_\tau$ decay, with the same technique 
 used by the LHCb experiment to obtain 
the ${\cal R}_{ D^*}=0.291 \pm 0.019 \pm 0.026 \pm 0.013$ 
measurement~\cite{LHCb:2017rln}, which is only  $1\sigma$ higher than the SM prediction. 
We notice that LHCb reported a significant higher value for ${\cal R}_{ D^*}$ ($0.336 \pm  0.027 \pm  0.030$), $2.1\sigma$ higher 
than that  expected from LFU in the SM, when the  $\tau$ lepton was reconstructed
 using its leptonic decay into a muon~\cite{LHCb:2015gmp}.

One expects  that the  existence of  NP that leads to  LFU violation in semitauonic $b-$meson  decays 
would also affect the  $\Lambda_b \to \Lambda_c \tau\bar\nu_\tau$ reaction,  and thus, a 
confirmation of  the result of Ref.~\cite{LHCb:2022piu} for  ${\cal R}_{\Lambda_c}$, 
using other reconstruction channels will shed light into this puzzling  situation. 
Such research  might provide very stringent constraints on  NP extensions of the SM, 
since scenarios leading to different deviations from SM expectations for 
${\cal R}_{\Lambda_c}$ and ${\cal R}_{ D^{(*)}}$ seem to be required. A new measurement
   of  ${\cal R}_{\Lambda_c}$, through the $\tau\to \mu\nu_\tau\bar\nu_\mu$ decay channel, 
   is in progress at the  LHCb experiment~\cite{Marco}, which in light of the previous 
   discussion will undoubtedly be very relevant.

As we will detail below, we present in this work some energy and angular distributions of a charged particle product from the decay of the $\tau$ produced in the $b\to c \tau \bar\nu_\tau$ transition that, if measured, could  contribute significantly  to clarify the current situation regarding the violation of universality in $b-$hadron decays. 

There is a multitude of theoretical works evaluating NP effects on the LFU ratios and on the outgoing  unpolarized (or longitudinally polarized) tau angular distributions  in $\bar B \to D^{(*)}$~\cite{Nierste:2008qe, Tanaka:2012nw, Fajfer:2012vx, Duraisamy:2013pia,Duraisamy:2014sna,Becirevic:2016hea, Ligeti:2016npd, Ivanov:2017mrj,Bernlochner:2017jka, Blanke:2018yud, Bhattacharya:2018kig, Colangelo:2018cnj,Murgui:2019czp, Shi:2019gxi, Alok:2019uqc, Mandal:2020htr, Kumbhakar:2020jdz,Iguro:2020cpg, Bhattacharya:2020lfm,  Penalva:2021gef,Penalva:2020ftd},  $\bar B_c\to J/\psi,\eta_c$~\cite{Dutta:2017xmj,Tran:2018kuv,Leljak:2019eyw, Harrison:2020nrv, Penalva:2020ftd} or  $\Lambda_b \to \Lambda_c$~\cite{Dutta:2015ueb,Shivashankara:2015cta, Li:2016pdv,Datta:2017aue,Ray:2018hrx,
Blanke:2018yud,Bernlochner:2018bfn,DiSalvo:2018ngq,Blanke:2019qrx,
Boer:2019zmp,Murgui:2019czp,Mu:2019bin,Hu:2020axt, Penalva:2019rgt, Penalva:2020xup, Penalva:2021gef} semileptonic decays. In general, different 
  NP scenarios  usually lead to an equally good reproduction of the LFU ratios, and hence other observables are needed to 
 constrain and determine the
 most plausible NP extension of the SM. Typically, the $\tau$ forward-backward 
 ($A_{FB}$) and longitudinal polarization (${\cal A}_{\lambda_\tau}= \langle P^{\rm CM}_L\rangle$) 
 asymmetries turn out to be more convenient for this purpose\footnote{A greater discriminating power can 
 be also reached by analyzing the four-body $\bar B\to D^*(D\pi, D\gamma)\tau\bar\nu_\tau$
~\cite{Duraisamy:2013pia,Duraisamy:2014sna,Becirevic:2016hea, Ligeti:2016npd, Colangelo:2018cnj,Bhattacharya:2020lfm, Mandal:2020htr} or similarly in the baryon reaction by considering the $\Lambda_c\to \Lambda \pi$ decay ~\cite{Boer:2019zmp,Hu:2020axt}.}.  
The final $\tau$  does not travel far enough for a displaced vertex, and it is very difficult to reconstruct from 
its decay products since they involve at least one more neutrino. 
Thus, the maximal accessible information on the $b \to  c \tau\bar\nu_\tau$ transition is encoded 
in the visible~\cite{Alonso:2016gym,Alonso:2017ktd, Asadi:2020fdo} decay products of the $\tau$ lepton, for which the three dominant 
modes $\tau \to \pi \nu_\tau ,\, \rho \nu_\tau$ and 
$\ell\bar\nu_\ell\nu_\tau$ ($\ell=e,\mu$)  account 
for more than 70\% of the total $\tau$ decay width  ($ \Gamma_\tau$). 

\begin{figure}
\includegraphics[scale=0.65]{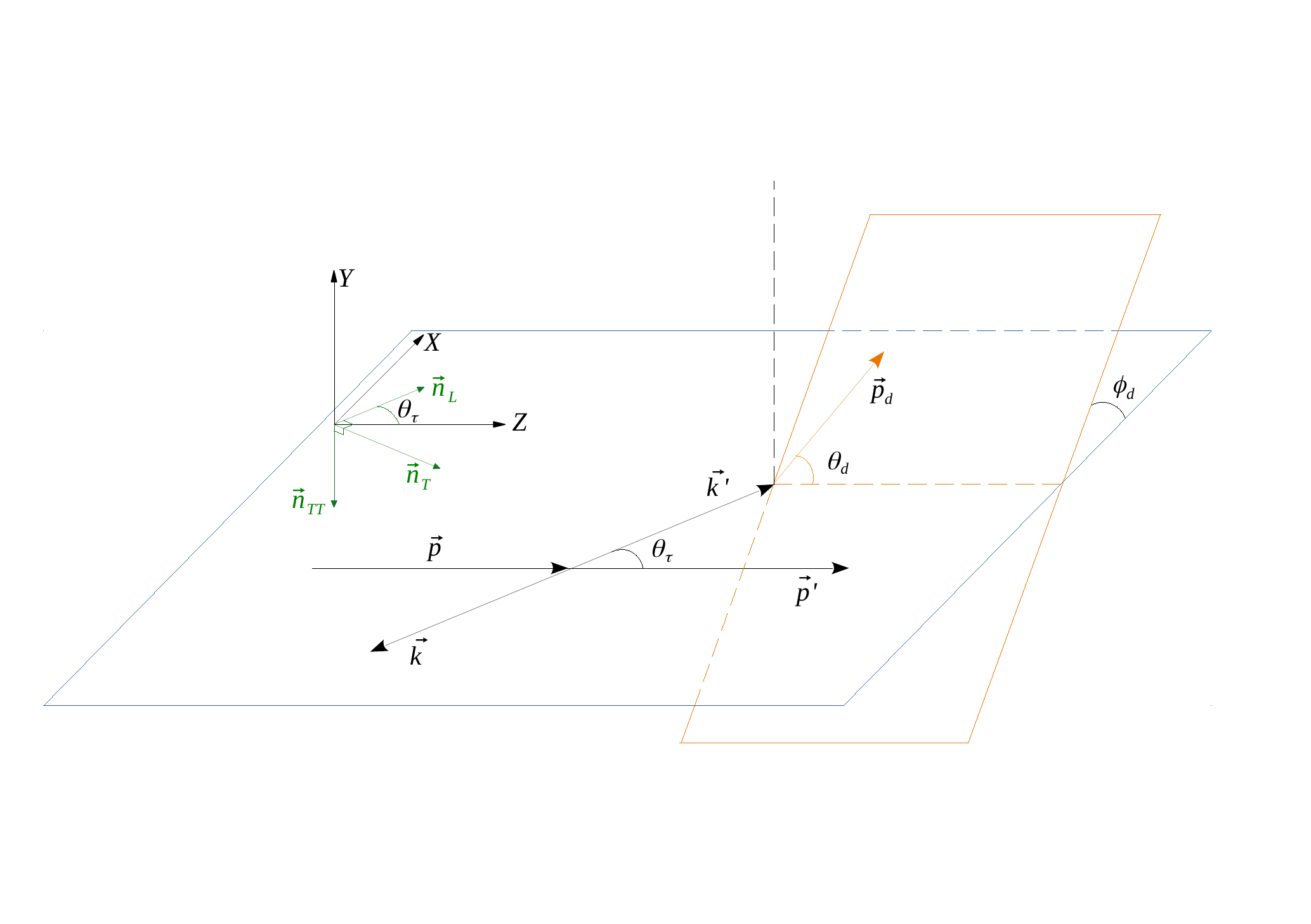}
\vspace{-2cm}
\caption{ Kinematics in the $\tau\bar\nu_\tau$ CM reference system 
associated with Eq.~\eqref{eq:visible-distr}, and used in 
Ref.~\cite{Penalva:2021wye}. The initial and final hadron three-momenta 
are $\vec{p}$ and $\vec{p}^{\,\prime}$, respectively, with 
$\vec{q}= \vec{p}-\vec{p}^{\,\prime}=\vec{0}$, while $\vec{k}^{\,\prime}$ 
and $\vec{k}$ are those of the intermediate $\tau$ and outgoing 
$\bar\nu_\tau$ emerging from the primary CC transition ($\vec{q}= 
\vec{k}+\vec{k}^{\,\prime}=\vec{0}$\,). In addition, $\vec{p}_d$ is the momentum of 
the tau-decay massive product ($\mu, \pi$ or $\rho$). We also show the unit vectors 
($\vec{n}_L, \vec{n}_T$ and $\vec{n}_{TT}$) which define the three independent
 projections of the $\tau-$polarization vector (see Ref.~\cite{Penalva:2021gef}). }
\label{fig:kinematics}
\end{figure}

For
the subsequent decays of the produced $\tau$, after the $b\to c \tau \bar \nu_\tau$
 transition,
\bea
H_b \to H_c &\tau^-& \bar \nu_\tau \nonumber \\
&\,\rotatebox[origin=c]{180}{$\Lsh$} &  \pi^-\nu_\tau ,\, \rho^-\nu_\tau, \, \mu^-\bar\nu_\mu\nu_\tau.
 \, e^-\bar\nu_e\nu_\tau,
\eea
we have~\cite{Penalva:2021wye} (the expression below was 
derived in Refs.~\cite{Alonso:2016gym,Alonso:2017ktd, Asadi:2020fdo} 
for the particular case of $\bar B \to D^{(*)}$ decays) 
\bea
\frac{d^3\Gamma_d}{d\omega  d\xi_d d\cos\theta_d} & = & {\cal B}_{d}
\frac{d\Gamma_{\rm SL}}{d\omega} \Big\{  F^d_0(\omega,\xi_d)+ F^d_1(\omega,\xi_d) 
\cos\theta_d + F^d_2(\omega,\xi_d)P_2(\cos\theta_d)\Big\},\label{eq:visible-distr}
\eea
where all involved kinematical variables are shown in Fig.~\ref{fig:kinematics}. In Eq.~\eqref{eq:visible-distr}, 
$\omega$ is the product of the two hadron four-velocities
which is related to the four-momentum transferred as $q^2=(p-p')^2=M^2+M^{\prime2}-2MM'\omega$, with $M,M'$ the masses of the initial
and final hadrons respectively. In addition,   ${\cal B}_{d}$ is the branching ratio for the
$\tau\to d\nu_{\tau}$ decay, where $d$ stands for  $d=\pi,\rho,\ell\bar\nu_\ell$, $\xi_d=E_d/(\gamma m_\tau)$ is the
ratio of the energies of the tau-decay massive product
 ($\pi,\rho$ or $\ell$) and the tau lepton measured in the 
 $\tau\bar\nu_\tau$ center of mass frame (CM), with  
 $\gamma=(q^2+m_\tau^2)/(2m_\tau\sqrt{q^2})$, and the  related variable
 $\beta=(1-1/\gamma^2)^{1/2}=(q^2-m_\tau^2)/(q^2+m_\tau^2)$, defining the boost from the tau-rest frame to
 the CM one. $\theta_d$ is the angle made by the tree-momenta of the final hadron and the  tau-decay massive 
 product in the  CM reference system and $P_2$  is the Legendre polynomial of order two. Besides, $d\Gamma_{\rm SL}/d\omega$ is
the unpolarized  differential semileptonic $H_b\to H_c\tau\bar\nu_\tau$ decay 
width that can be written as
\bea
\frac{d\Gamma_{\rm SL}}{d\omega}=\frac{G_F^2|V_{cb}|^2M^{\prime3}M^2}{24\pi^3} \label{eq:n0w}
\sqrt{\omega^2-1}\Big(1-\frac{m_\tau^2}{q^2}\Big)^2n_0(\omega),
\eea
where $n_0(\omega)=
3a_0(\omega)+a_2(\omega)$,
with $a_{0,2}(\omega)$ given in Refs.~\cite{Penalva:2020xup,Penalva:2020ftd},  contains all the dynamical effects 
including any possible NP contribution to the $b\to c$ transition.
Finally, the $F^d_{012}(\omega,\xi_d)$ two dimensional functions can be written as\footnote{This angular decomposition was firstly introduced in \cite{Asadi:2020fdo} in the context of the $\tau-$hadronic decay modes in $\bar B \to D^{(*)}$ reactions.   }
 \begin{eqnarray}
 F^d_0(\omega,\xi_d) &=& C_n^d(\omega,\xi_d)+C_{P_L}^d(\omega,\xi_d)\,\langle P^{\rm CM}_L\rangle(\omega), \nonumber \\
 F^d_1(\omega,\xi_d) &=& C_{A_{FB}}^d(\omega,\xi_d)A_{FB}(\omega)+C_{Z_L}^d(\omega,\xi_d)Z_L(\omega)
 + C_{P_T}^d(\omega,\xi_d)\,\langle P^{\rm CM}_T\rangle(\omega), \nonumber \\ 
 F^d_2(\omega,\xi_d) &=& C_{A_Q}^d(\omega,\xi_d)A_{Q}(\omega)+
 C_{Z_Q}^d(\omega,\xi_d)Z_Q(\omega)+ C_{Z_\perp}^d(\omega,\xi_d)Z_\perp(\omega).
 \label{eq:coeff}
\end{eqnarray}
where the $C^d_a(\omega,\xi_d)$ are kinematical coefficients that depend on the 
tau-decay mode. Their analytical expressions can be found,
for the $\pi\nu_\tau,\rho\nu_\tau$ and $\ell\bar\nu_\ell\nu_\tau$ cases, in
  Appendix G of Ref.~\cite{Penalva:2021wye}. In the 
leptonic mode we have kept effects due to the finite  
mass of the outgoing muon/electron, although making $m_\ell=0$ in those expressions
 should be a very good approximation, since both $m_e,/m_\tau$ and $m_\mu/m_\tau$ are much smaller than one. The rest of the
quantities in   Eq.~(\ref{eq:coeff})
 are  the tau-spin ($\langle P^{\rm CM}_{L,T}\rangle(\omega)$), 
 tau-angular  ($A_{FB,Q}(\omega)$) and  tau-angular-spin ($Z_{L,Q,\perp}(\omega)$) asymmetries of the $H_b\to H_c\tau\bar\nu_\tau$
 decay. Actually, these asymmetries and $d\Gamma_{\rm SL}/d\omega$ provide  
 the maximal information that can be extracted from the study of polarized $H_b\to H_c \tau \bar\nu_\tau$  transitions, without considering CP non-conserving  contributions~\cite{Penalva:2021gef, Penalva:2021wye}\footnote{As discussed  in these two references (see also \cite{Asadi:2020fdo}), 
 the azimuthal angular ($\phi_d$) distribution of the tau decay charged product turn out to be  sensitive to possible CP odd effects, which are produced by the existence of relative phases between some of the Wilson coefficients
  in the NP  Hamiltonian of Eq.~\eqref{eq:hnp}. However, the measurement of the angle $\phi_d$ (see Fig.~\ref{fig:kinematics}) would require the full reconstruction of the tau three momentum. For $\bar B \to D^*$ ($\Lambda_b\to \Lambda_c$), some CP-odd observables (triple product  asymmetries), defined using angular distributions involving the kinematics of the  products of the $D^*$ ($\Lambda_c$) decay,  have also been  presented~\cite{Duraisamy:2013pia, 
Duraisamy:2014sna, Ligeti:2016npd, Bhattacharya:2020lfm} (\cite{Boer:2019zmp,Hu:2020axt}).} (see Eq.~(3.46) of the latter of these two references and the related discussion). 
 
In Ref.~\cite{Penalva:2021wye}, we  numerically analyzed  the role that each of the observables,  $d\Gamma_{\rm SL}/d\omega$, $\langle P^{\rm CM}_{L,T}\rangle(\omega)$, $A_{FB,Q}(\omega)$ and $Z_{L,Q,\perp}(\omega)$ could play to establish the existence  of NP beyond the SM in $\Lambda_b\to\Lambda_c\tau\bar\nu_\tau$ semileptonic decays. In fact in that work, we obtained their general expressions, valid for any
$H_b\to H_c\tau\bar\nu_\tau$ decay, when  considering
 an extension of the SM  comprising the full set of dimension-6 semileptonic $b\to c$ operators with 
left- and right-handed  neutrinos. The effective low-energy Hamiltonian for that case is given 
 by~\cite{Mandal:2020htr}
\bea
H_{\rm eff}&=&\frac{4G_F V_{cb}}{\sqrt2}\left[(1+C^V_{LL}){\cal O}^V_{LL}+
C^V_{RL}{\cal O}^V_{RL}+C^S_{LL}{\cal O}^S_{LL}+C^S_{RL}{\cal O}^S_{RL}
+C^T_{LL}{\cal O}^T_{LL}\right.\nonumber \\
&+&\left. C^V_{LR}{\cal O}^V_{LR}+
C^V_{RR}{\cal O}^V_{RR}+C^S_{LR}{\cal O}^S_{LR}+C^S_{RR}{\cal O}^S_{RR}
+C^T_{RR}{\cal O}^T_{RR} \right]+h.c.,
\label{eq:hnp}
\eea
with  left-handed neutrino fermionic operators given by 
\be
{\cal O}^V_{(L,R)L} = (\bar c \gamma^\mu b_{L,R}) 
(\bar \ell \gamma_\mu \nu_{\ell L}), \, {\cal O}^S_{(L,R)L} = 
(\bar c\,  b_{L,R}) (\bar \ell \, \nu_{\ell L}), \, {\cal O}^T_{LL} = 
(\bar c\, \sigma^{\mu\nu} b_{L}) (\bar \ell \sigma_{\mu\nu} \nu_{\ell L})
\label{eq:hnp2}
\ee
and the right-handed neutrino ones
\be
{\cal O}^V_{(L,R)R} = (\bar c \gamma^\mu b_{L,R}) 
(\bar \ell \gamma_\mu \nu_{\ell R}), \, {\cal O}^S_{(L,R)R} = 
(\bar c\,  b_{L,R}) (\bar \ell \, \nu_{\ell R}), \, {\cal O}^T_{RR} = 
(\bar c\, \sigma^{\mu\nu} b_{R}) (\bar \ell \sigma_{\mu\nu} \nu_{\ell R}),
\label{eq:hnp2R}
\ee
and where $\psi_{R,L}= (1 \pm \gamma_5)\psi/2$,  $G_F=1.166\times 10^{-5}$~GeV$^{-2}$  and 
$V_{cb}$ is the corresponding Cabibbo-Kobayashi-Maskawa matrix element. 

The asymmetries introduced in Eq.~\eqref{eq:coeff}
depend on the  pure hadronic structure functions  and  ten (complex) Wilson 
coefficients $C^X_{AB}$ ($X= S, V,T$ and $A,B=L,R$), which parameterize the possible
 deviations from the SM. The former depend on the form factors
 that parameterize the hadronic current and we have obtained them for 
 $1/2^+\to 1/2^+$~\cite{Penalva:2020xup} and $0^-\to 0^-,1^-$~\cite{Penalva:2020ftd} decays.
 
 The  $d^3\Gamma_d/(d\omega  d\xi_d d\cos\theta_d)$  distribution, together with the 
 combined analysis of its $(\xi_d,\cos\theta_d$) dependence, gives access to all the above asymmetries 
 as  functions of $\omega$. The feasibility of such studies can be severely limited, however, by the 
 statistical precision in the measurement of the triple differential decay width. Statistics can be 
 increased by integrating in the  $\cos\theta_d$ or/and $\xi_d$ variables,  although in this case not 
 all observables can be extracted. Thus, it is well known~\cite{Tanaka:2010se}  that the distribution 
 obtained after accumulating in the polar angle, 
\bea
\frac{d^2\Gamma_d}{d\omega  d\xi_d} & = & 2{\cal B}_{d}
\frac{d\Gamma_{\rm SL}}{d\omega}\Big\{  C_n^d(\omega,\xi_d)+C_{P_L}^d(\omega,\xi_d)\,\langle P^{\rm CM}_L\rangle(\omega)\Big\},\label{eq:wE}
\eea
allows to determine $d\Gamma_{\rm SL}/d\omega$ and  the CM $\tau$ longitudinal polarization 
[$\langle P^{\rm CM}_L\rangle(\omega)]$ since the, transition dependent,  $C_n^d(\omega,\xi_d)$
 and $C_{P_L}^d(\omega,\xi_d)$
coefficients are  known kinematical factors~\cite{Penalva:2021wye} (see also 
\cite{Alonso:2017ktd, Tanaka:2010se}). The averaged CM tau  longitudinal polarization asymmetry,
\be
P_\tau = -\frac{1}{\Gamma_{\rm SL}}\int d\omega \frac{d\Gamma_{\rm SL}}{d\omega}
\langle P^{\rm CM}_L\rangle(\omega)
\ee
measured by Belle~\cite{Belle:2016dyj} for the $\bar B \to D^* \tau \bar\nu_\tau$ decay, 
immediately follows.  

In Refs.~\cite{Penalva:2021gef,Penalva:2021wye} we  presented results for  $\langle 
P^{\rm CM}_L\rangle(\omega)$ 
in the $\Lambda_b\to\Lambda_c\tau\bar\nu_\tau$ and $\bar B \to D^{(*)} \tau \bar\nu_\tau$ 
decays 
evaluated within the SM and different NP extensions\footnote{We would like to 
highlight that in Ref.~\cite{Penalva:2021wye} and for the baryon  reaction, 
we showed  also results for the CP-violating observable $P_{TT}$,  calculated  
using the $R_2$ leptoquark model of Ref.~\cite{Shi:2019gxi}. This is the 
$\tau$-polarization component along an axis perpendicular to the  hadron-tau plane (see Fig.~\ref{fig:kinematics}). The contribution of $P_{TT}$ to the differential $H_b\to H_c \tau\, (\mu \bar \nu_\mu \nu_\tau,\pi\nu_\tau,\rho\nu_\tau) 
 \bar\nu_\tau$ distribution disappears when the azimuthal angle $\phi_d$ is integrated out.  }. 
 We also provided similar comparisons for  $d\Gamma_{\rm SL}/d\omega$  in the 
 $\Lambda_b\to\Lambda_c$ and $\bar B \to D^{(*)}$ and $\bar B_c\to J/\psi,\eta_c$ reactions 
 in Refs.~\cite{Penalva:2019rgt,Penalva:2020xup} and \cite{Penalva:2020ftd}, respectively.

In this work, we take advantage of the analytical results derived in  \cite{Penalva:2021wye}, 
and we study, in secs. ~\ref{sec:domdcos}, \ref{sec:dcos} and \ref{sec:de}, respectively, the alternative 
distributions $ d^2 \Gamma_d / (d\omega d\cos\theta_d) $, $ d\Gamma_d /d\cos\theta_d$
and $ d\Gamma_d /dE_d $, which could also be measured in the near future with 
certain precision. We pay attention to the deviations with respect to the predictions of the SM for 
these new observables, considering NP operators constructed using both right- and left-handed 
neutrino fields, within the effective theory approach established by 
Eqs.~\eqref{eq:hnp}--\eqref{eq:hnp2R}. We will present results for 
the $\Lambda_b\to\Lambda_c\tau\, (\mu \bar \nu_\mu \nu_\tau,\pi\nu_\tau,
\rho\nu_\tau)  \bar\nu_\tau$ (main text) and the $\bar B \to D^{(*)}\tau\, (\mu \bar \nu_\mu \nu_\tau,
\pi\nu_\tau,\rho\nu_\tau) 
 \bar\nu_\tau$ sequential decays (Appendix~\ref{app:bddstar}),  obtained 
 within different beyond the SM scenarios, and we discuss their use to extract 
 some of the tau asymmetries introduced in Eq.~\eqref{eq:coeff}. Details on 
 the used form-factors and references to the original works where they were 
 calculated can be found in \cite{Penalva:2021gef,Penalva:2021wye}.

\section{The $d^2\Gamma/(d\omega d\cos\theta_d)$ distribution}
\label{sec:domdcos}
The limits\footnote{In the case of the lepton mode, the lowest one could be either $y/\gamma$  or $\xi_1$ depending on whether $q^2$ is smaller than or greater than $m^4_\tau/m_d^2$, respectively. Obviously, given the range of $q^2$ values which can be accessed in the semileptonic $H_b \to H_c$ parent decays and the masses of the charged leptons, we are always in the first of the two scenarios.} on  the $\xi_d$ variable are tau-decay mode dependent and thus, one has~\cite{Penalva:2021wye}
\bea
\tau\to \mu\bar\nu_\mu\nu_\tau &\Rightarrow&y/\gamma\le\xi_d \le \xi_2,\nonumber\\
\tau\to (\pi,\rho)\nu_\tau &\Rightarrow& \frac{1-\beta}{2}+  
\frac{1+\beta}{2}y^2=\xi_1  \le \xi_d \le \xi_2= \frac{1+\beta}{2}+   
\frac{1-\beta}{2}y^2,
  \label{eq:ed-range}
\eea
with $y=m_d/m_\tau$ and $m_d$ the mass of the tau-decay massive product 
($\pi,\,\rho$ or $\mu$).  After integration
one obtains the double differential decay width
\bea
\frac{d^2\Gamma_d}{d\omega  d\cos\theta_d} & = & {\cal B}_{d}
\frac{d\Gamma_{\rm SL}}{d\omega} \Big[ \widetilde F^d_0(\omega)+ \widetilde F^d_1(\omega) 
\cos\theta_d +\widetilde  F^d_2(\omega)P_2(\cos\theta_d)\Big],\label{eq:visible-distr_theta_d}
\eea
where the new angular expansion coefficients $\widetilde F^d_{{0,1,2}}(\omega)$ 
correspond to
\bea
\widetilde F^{d=\pi, \rho}_{{0,1,2}}(\omega)=\int_{\xi_1}^{\xi_2}F^d_{0,1,2}(\omega,\xi_d)
\,d\xi_d,\qquad \widetilde F^{d=\ell\bar\nu_\ell}_{{0,1,2}}(\omega)=\int_{y/\gamma}^{\xi_2}F^{d=\ell\bar\nu_\ell}_{0,1,2}(\omega,\xi_d)
\,d\xi_d
\eea
and they can be extracted from the angular analysis of the  statistically enhanced $d^2\Gamma/(d\omega d\cos\theta_d)$ distribution. The overall normalization is recovered since $\widetilde F^d_0(\omega)=1/2$ for all tau-decay modes,  and a further integration in
the polar angle $\theta_d$ provides $d\Gamma_d/d\omega={\cal B}_d\ d\Gamma_{\rm SL}/d\omega$, which  in this way can be experimentally obtained from the tau decay-chain reaction.  

In what follows,  we will focus on 
the non-trivial  $\widetilde F^d_1(\omega)$ and $\widetilde F^d_2(\omega)$ functions, 
which read
\bea
\widetilde F^d_1(\omega)&=&C^d_{A_{FB}}(\omega)\,A_{FB}(\omega)+C^d_{Z_L}(\omega)
\,Z_L(\omega)+C^d_{P_T}(\omega)\,\langle P_T^{\rm CM}\rangle(\omega),\label{eq:F1}\\
\widetilde F^d_2(\omega)&=&C^d_{A_Q}(\omega)\,A_Q(\omega)+C^d_{Z_Q}(\omega)
\,Z_Q(\omega)+C^d_{Z_\perp}(\omega)\,Z_\perp(\omega).\label{eq:F2}
\eea
While the $\xi_d$ integration which gives rise to $\widetilde F_0^d(\omega)$  
loses  information on 
$\langle P^{\rm CM}_L\rangle(\omega)$, the statistically enhanced observables $\widetilde
F_{1,2}^{\mu\bar\nu_\mu}(\omega)$  retain 
all the information on the other six asymmetries. 

\subsection{Tau-decay lepton mode }

\begin{figure}
\includegraphics[scale=0.77]{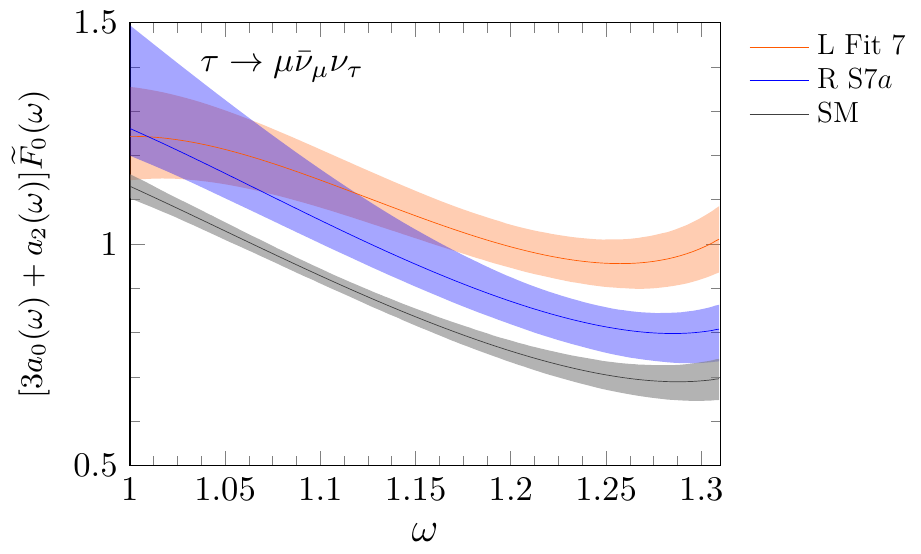}\\
\includegraphics[scale=0.77]{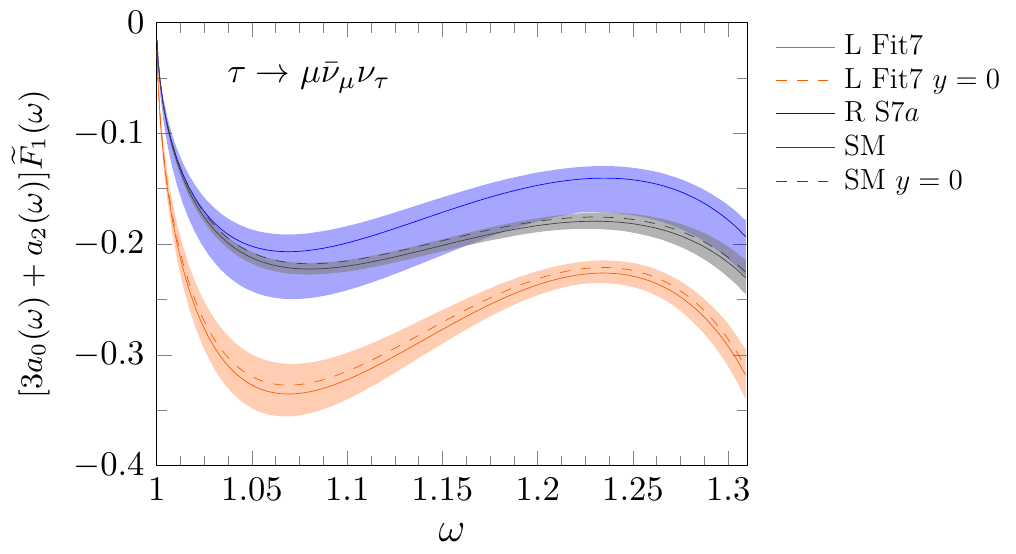}\hspace{.15cm}\includegraphics[scale=0.77]{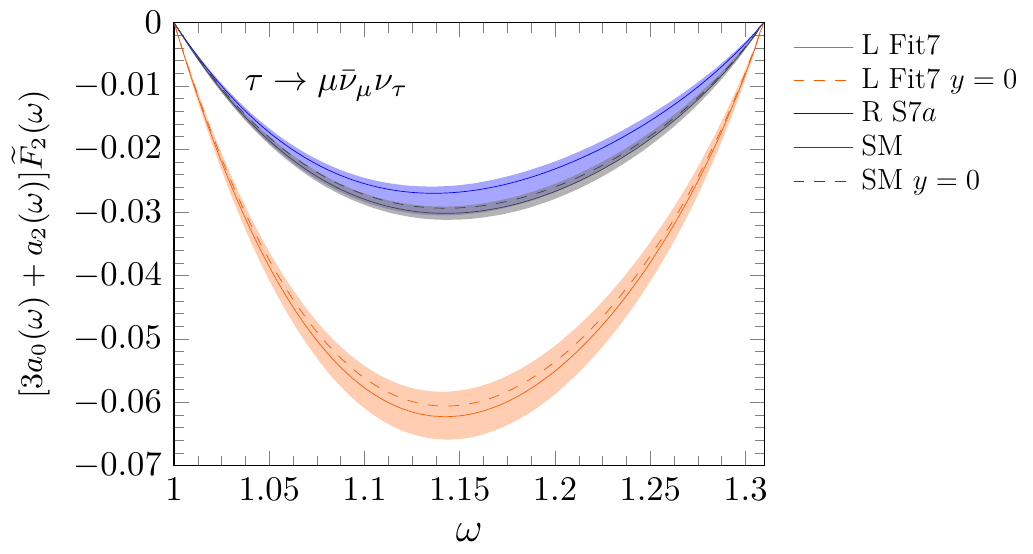}
\caption{ Results for the functions $[3a_0(\omega)+a_2(\omega)]\widetilde F^{\mu\bar \nu_\mu}_{0,1,2}(\omega)$ evaluated for the 
$\Lambda_b\to\Lambda_c\tau(\mu\bar\nu_\mu\nu_\tau)\bar\nu_\tau$ decay, 
keeping the muon mass finite, and obtained from the SM and
the NP models corresponding to Fit 7 of Ref~\cite{Murgui:2019czp} and Fit 7a
 of Ref~\cite{Mandal:2020htr}. Error bands account for uncertainties induced by both form-factors
and fitted Wilson coefficients (added in quadrature). In the SM and Fit 7 cases, we also display the results obtained neglecting the muon mass, as in Eqs.~\eqref{eq:coeffmu-1}-\eqref{eq:coeffmu-3}. }
\label{fig:ftildemu}
\end{figure}

We start with the $\tau \to \mu\bar\nu_\mu\nu_\tau$ channel, since a measurement of the
ratio of branching fractions 
${\cal B}(\Lambda_b\to\Lambda_c\tau(\mu\nu_\tau\bar\nu_\mu)\bar\nu_\tau)/{\cal
 B}(\Lambda_b\to\Lambda_c\mu\bar\nu_\mu)$  is in progress at the  
 LHCb experiment. Moreover, as argued in the introduction, it would be very important 
 to confront the recent LHCb measurement of ${\cal R}_{\Lambda_c}$, reconstructed using 
 the three-prong hadronic $\tau$ decay, with results obtained  
 when the  tau lepton is identified from its leptonic decay into a muon. In the $y=m_\mu/m_\tau=0$ limit, which is a very good 
approximation (${\cal O}(y^2)\sim $ 1\%) in this case, and it is much
 better for the electron tau-decay mode, we  find that the coefficient-functions,  $C^{\mu\bar\nu_\mu}_a(\omega)$, are given by
\bea
C^{\mu\bar\nu_\mu}_{A_{FB}}(\omega)\Big|_{y=0} &=& \frac{3\pi\gamma\beta}{2}C^{\mu\bar\nu_\mu}_{P_T}(\omega)\Big|_{y=0}= 
\frac1{\beta^2}\Big[\beta-\frac{\artanh\beta}{\gamma^2} \Big],\label{eq:coeffmu-1}\\
C^{\mu\bar\nu_\mu}_{A_{Q}}(\omega)\Big|_{y=0} &=& -2\beta\gamma C^{\mu\bar\nu_\mu}_{Z_\perp}(\omega)\Big|_{y=0}= -\frac1{\beta^3}\Big[2\beta^3-3\beta+3\frac{\artanh\beta}{\gamma^2}\Big],  \\
C^{\mu\bar\nu_\mu}_{Z_L}(\omega)\Big|_{y=0}&=& \frac{1}{3\gamma^2\beta^3} \Big[\beta-\artanh\beta
\Big],\quad C^{\mu\bar\nu_\mu}_{Z_Q}(\omega)\Big|_{y=0}= \frac1{2\gamma^2\beta^4}
\Big[3\beta -\big(3-\beta^2\big)\artanh\beta
\Big],\label{eq:coeffmu-3}\
\eea
with $\artanh\beta= \ln\sqrt{\big(1+\beta\big)/\big(1-\beta\big)}$.

Here,  we will present results for $\widetilde F^{\mu\bar\nu_\mu}_{0,1,2}(\omega)$ multiplied by the 
factor $n_0(\omega)$. For  $\widetilde F^{\mu\bar\nu_\mu}_{0}(\omega)$ this amounts to represent 
$n_0(\omega)/2$ and,  since this is the same for all tau-decay modes, it will only 
be shown for the muon tau-decay mode. As mentioned,
the $n_0(\omega)$ function, introduced in Eq.~\eqref {eq:n0w},  
contains all the dynamical effects included in the $d\Gamma_{\rm SL}/d\omega$ 
differential semileptonic decay width, which appears as an overall normalization of the  $d^2\Gamma/(d\omega d\cos\theta_d)$ distribution.
 By showing $\widetilde F^{\mu\bar\nu_\mu}_{0,1,2}
  (\omega)$ times $n_0(\omega)$, we  access to all the
 effects of possible NP beyond the SM on the tau production\footnote{However, we should note that NP contributions to the $\tau$ decay are not considered in this work}. 
 
 The $n_0(\omega)\widetilde F^{\mu\bar\nu_\mu}_{0,1,2}(\omega)$ functions, for the baryon $\Lambda_b\to\Lambda_c\tau(\mu\bar\nu_\mu\nu_\tau)\bar\nu_\tau$ reaction, 
are displayed  in Fig.~\ref{fig:ftildemu}. They  have been 
 evaluated  within the SM and the beyond the SM scenarios of Fit 7 (7a) 
 of Ref~\cite{Murgui:2019czp} (\cite{Mandal:2020htr}), which only includes 
 left- (right-)handed neutrino NP operators.  These two NP scenarios have been 
 adjusted to reproduce the  anomalies observed in the LFU ${\cal R}_{ D}$ and  ${\cal R}_{ D^{*}}$ ratios in $\bar B-$meson decays.  However, in all cases,  we see the results from  
Fit 7 of Ref~\cite{Murgui:2019czp} can be 
distinguished clearly from  SM and Fit 7a  model (R S7$a$ in the plots) ones. 
The results for the Fit 7a  model  are closer to the SM  
and  in the case of the $\widetilde F^{\mu\bar\nu_\mu}_{1,2}(\omega)$ 
functions the uncertainty bands  overlap in the whole $\omega$ interval. 
This is a reflection of what is obtained for the
  tau-asymmetries themselves, as can be seen in Fig.2 of Ref.~\cite{Penalva:2021wye}.

\begin{figure}[h!]
\includegraphics[scale=0.675]{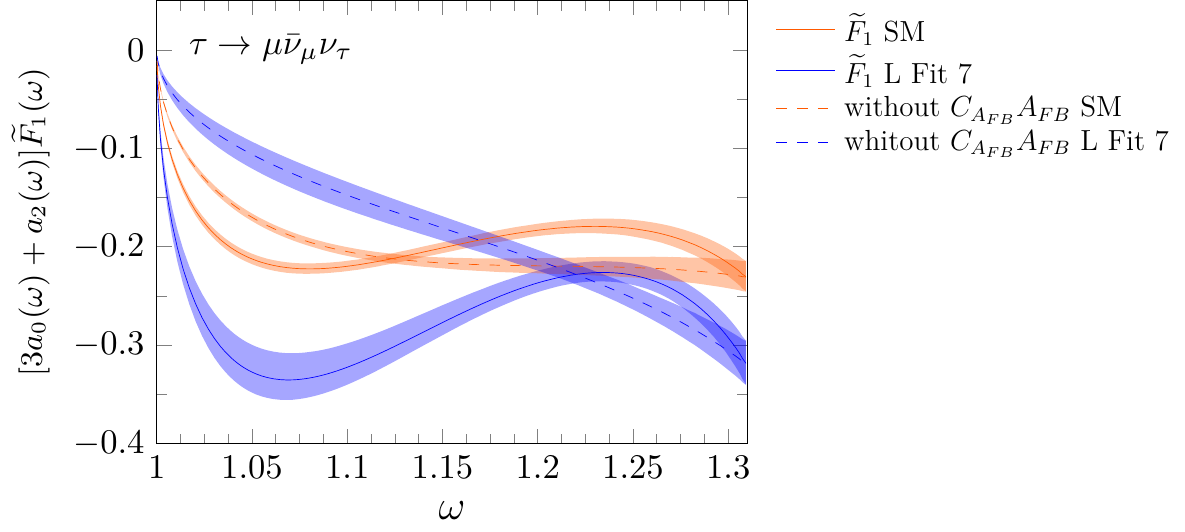}\hspace{.1cm}
\includegraphics[scale=0.675]{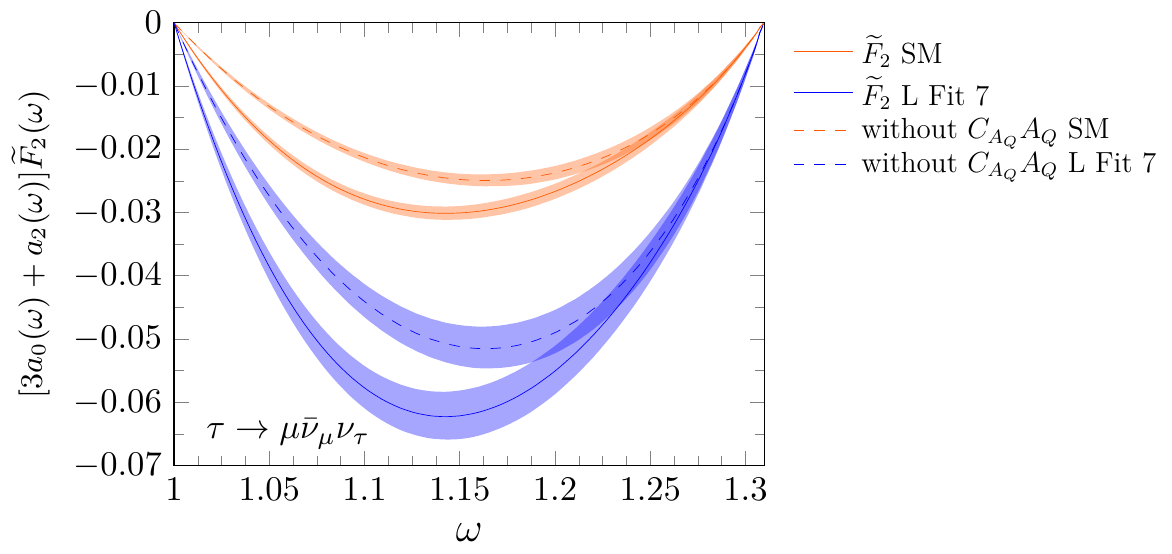}\\
\caption{Comparison of the full results (solid lines) for 
$[3a_0(\omega)+a_2(\omega)]\widetilde
 F^{\mu\bar\nu_\mu}_{1,2}(\omega)$ 
with those obtained setting $A_{FB}(\omega)$ and $A_Q(\omega)$ 
to zero (dashed lines). The muon mass is kept finite. The results
have been obtained for the $\Lambda_b\to\Lambda_c\tau(\mu\bar\nu_\mu\nu_\tau)
\bar\nu_\tau$ 
sequential decay, within the SM and
the NP model corresponding to Fit 7 of Ref~\cite{Murgui:2019czp}.}
\label{fig:comp}
\end{figure}

It is also very instructive to compare the full results for 
$n_0(\omega)\widetilde F^{\mu\bar\nu_\mu}_{1,2}(\omega)$  with those evaluated setting
 $A_{FB}(\omega)$ and  $A_{Q}(\omega)$ to zero. This comparison is presented in  
 Fig.~\ref{fig:comp}. What 
can be inferred from this comparison is that the contribution of the spin ($\langle P^{\rm CM}_T\rangle(\omega)$) and 
angular-spin ($Z_{L,\,Q,\,\perp}(\omega)$) asymmetry terms are sizable and dominant in most of  the 
 $\omega$ interval. This is clearly the case in the vicinity of the end-point of the 
 distributions, 
 $q^2=m_\tau^2$ ($\beta=0$). In fact, using Eqs.~\eqref{eq:coeffmu-1}-\eqref{eq:coeffmu-3}, we find in the $y\to 0$ limit
\bea
\widetilde F^{\mu\bar\nu_\mu}_1(\omega)&\sim& \frac{1}{9\pi}\Big[ 4\langle P_T^{\rm CM}\rangle(\omega_{\rm max})-\pi Z_L(\omega_{\rm max})\Big] + \frac{\beta}{9}\Big[6A_{FB}(\omega_{\rm max})\nonumber\\
&&-\frac{m^2_\tau}{\pi MM'}\big[4\langle P_T^{\rm CM}\rangle^\prime(\omega_{\rm max})-\pi Z_L^\prime(\omega_{\rm max})\big]\Big]+  {\cal O}(\beta^2), \\
\widetilde F^{\mu\bar\nu_\mu}_2(\omega)&\sim&-\frac{\beta}{15}\Big[3Z_\perp(\omega_{\rm max})+2Z_Q(\omega_{\rm max})\Big] + \frac{\beta^2}{15}\Big[   6A_{Q}(\omega_{\rm max})\nonumber\\
&&+\frac{m^2_\tau}{MM'}\big[3Z^\prime_\perp(\omega_{\rm max})+2Z^\prime_Q(\omega_{\rm max})\big]\Big]+  {\cal O}(\beta^3),
\eea
with $\omega_{\rm max}=\omega(q^2=m^2_\tau)=(M^2+M^{\prime2}-m^2_\tau)/(2MM')$, which show that the contributions of the tau-angular asymmetries $A_{FB}$ and $A_Q$ are suppressed by a factor $\beta$ with respect to those proportional to $\langle P_T^{\rm CM}\rangle$ and $Z_{L,Q,\perp}$.

Thus, these two $\widetilde F^{\mu\bar\nu_\mu}_{1,2}(\omega)$ 
 observables, which  have an increased  statistics over
 $F^{\mu\bar\nu_\mu}_{1,2}(\omega,\xi_d)$, could be ideal to measure tau-spin related   asymmetries
 other than the commonly reported $\langle P_L^{\rm CM}\rangle(\omega)$, extracted from the 
 $d^2\Gamma_d/(d\omega dE_d)$ distribution.
\subsection{Tau-decay hadron modes }
The behavior seen in  Fig.~\ref{fig:comp} of the previous section for 
the muon is enhanced in the pion decay mode. After performing the integration 
over the variable $\xi_d$, we have that,  neglecting $y^2$ ($m^2_\mu/m^2_\tau $ and 
$m_\pi^2/m_\tau^2$) corrections,
the coefficients multiplying the two angular asymmetries $A_{FB,\,Q}(\omega)$ are 
the same as in the leptonic mode, while for the rest of the spin and angular-spin asymmetries
 there is an extra  factor of $-3$. This is to say
\be
C^\pi_{A_{FB}, A_Q}(\omega)=C^{\mu\bar\nu_\mu}_{A_{FB},A_Q}(\omega) + {\cal O}(y^2), \quad  C^\pi_{P_T,Z_L,Z_Q,Z_\perp }(\omega)=-3\,C^{\mu\bar\nu_\mu}_{P_T,Z_L,Z_Q,Z_\perp}(\omega) + {\cal O}(y^2)
\ee
This difference in the spin analyzing power makes the pion tau-decay mode a better
candidate for the extraction of information on the spin and angular-spin asymmetries. 
Exact expressions, without the $y=0$ approximation,  for the  $\pi$ and
$\rho$ decay modes are given  in Appendix~\ref{app:pirho}, although neglecting $m_\pi^2/m_\tau^2$ contributions is again  an excellent approximation  
for the pion case. For the $\rho$ decay mode, the spin analyzing power is  suppressed, with 
respect to the pion case, by the factor  $a_\rho=(m_\tau^2-2m_\rho^2)/(m_\tau^2+2m_\rho^2)\approx0.45$ (see Appendix~\ref{app:pirho}), although it is still greater than for the lepton decay mode.

\begin{figure}
\includegraphics[scale=0.675]{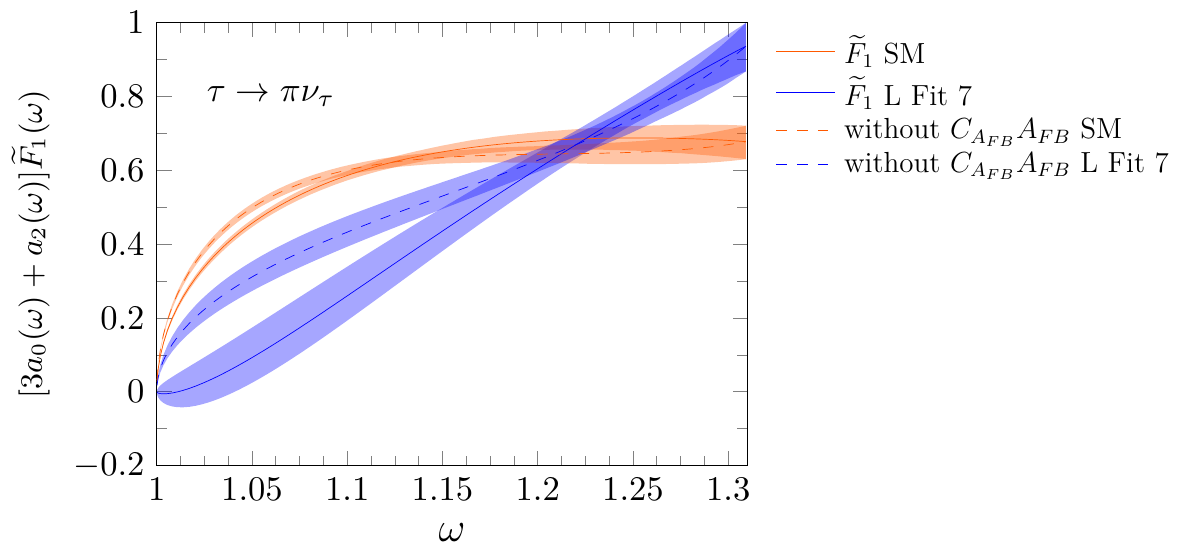}\hspace{.0cm}\includegraphics[scale=0.675]{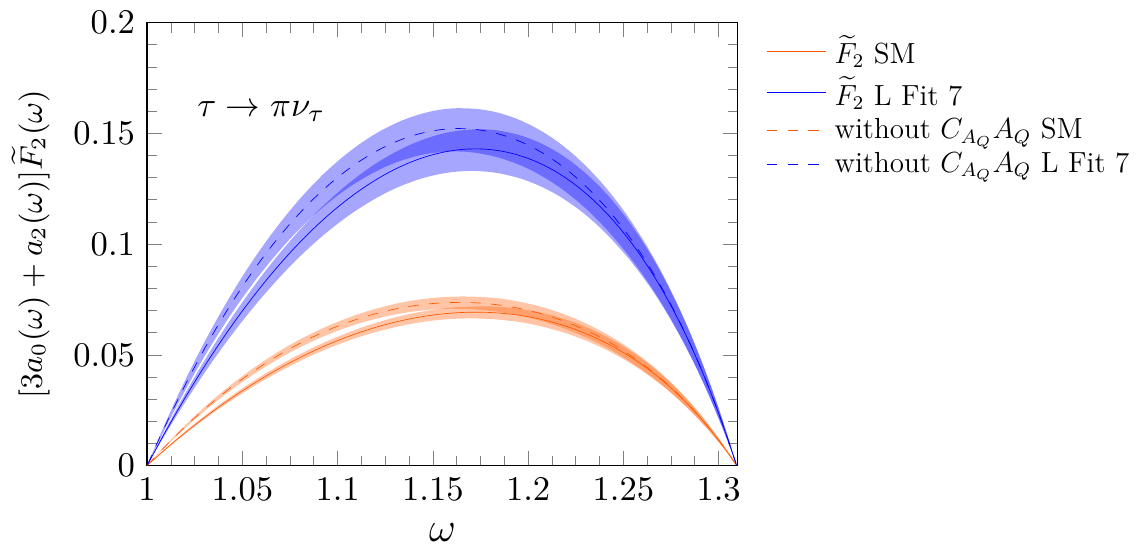}\\
\includegraphics[scale=0.675]{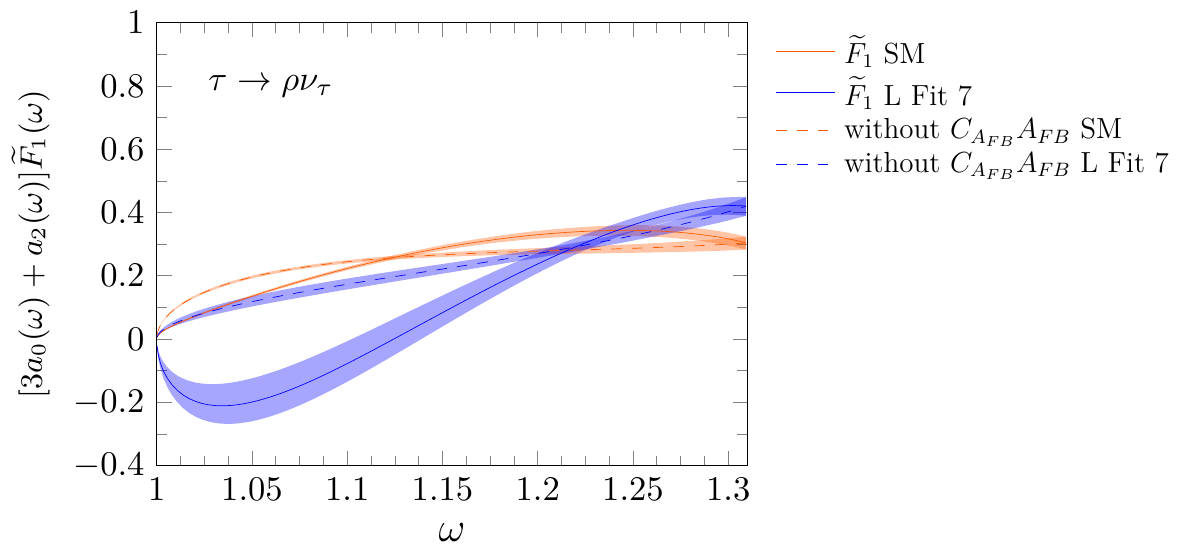}\hspace{.0cm}\includegraphics[scale=0.675]{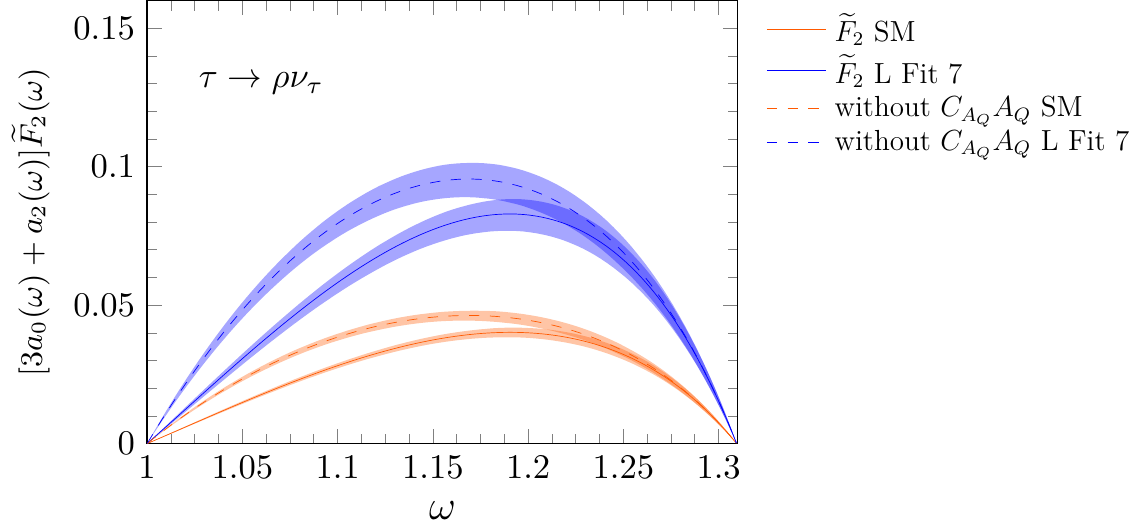}
\caption{ Same as Fig.~\ref{fig:comp}, but for the $[3a_0(\omega)+a_2(\omega)]\widetilde F^{\pi,\,\rho}_{1,2}(\omega)$ hadron-mode distributions. We use the expressions for the coefficients collected in Appendix~\ref{app:pirho}, which were obtained  keeping the pion and rho meson masses finite.}
\label{fig:comppirho}
\end{figure}

Full results, as well as results obtained setting the angular $A_{FB,\,Q}(\omega)$ 
asymmetry terms to zero,
for the hadron-mode $\widetilde F^{\pi,\rho}_{1,2}(\omega)$ functions are shown in 
Fig.~\ref{fig:comppirho} for the $\Lambda_b\to\Lambda_c\tau(\pi\nu_\tau,\,\rho\nu_\tau)
\bar\nu_\tau$ decays, accounting for all mass term corrections ($y=m_{\pi,\,\rho}/m_\tau\ne 0$).
 As expected,  we see that the hadron modes, in particular the pion one, show 
a great sensitivity to the spin-angular asymmetries, which could be  extracted 
from $\widetilde F^{\pi,\rho}_{1}(\omega)$ and $\widetilde F^{\pi,\rho}_{2}(\omega)$. 
These new observables are independent of  the $d\Gamma_{\rm SL}/d\omega$ and  
$\langle P^{\rm CM}_L\rangle(\omega)$ distributions~\cite{Penalva:2021gef, Penalva:2021wye}, 
and they will provide new constraints on the physics governing the $\Lambda_b\to\Lambda_c\tau 
\bar\nu_\tau$ parent decay.

The  $ \Lambda_b \to \Lambda_c \tau(\pi\nu_\tau, \rho\nu_\tau)\bar\nu_\tau$ reaction channels 
have a lower reconstruction efficiency at LHCb than the one driven by the 
$\tau$-decay lepton 
mode~\cite{Marco}. However, they might be accessible in the future, or be easier to reconstruct  
in other machines and/or chains initiated by other parent semileptonic decays. 
 For that 
reason, in Appendix~\ref{app:bddstar} we also present results for distributions obtained from the 
sequential $ \bar B \to D^{(*)} \tau(\pi\nu_\tau, \rho\nu_\tau)\bar\nu_\tau$ 
decays.
\section{The $d\Gamma/d\cos\theta_d$ distribution}
\label{sec:dcos}
\begin{figure}
\begin{center}
\includegraphics[scale=1.0]{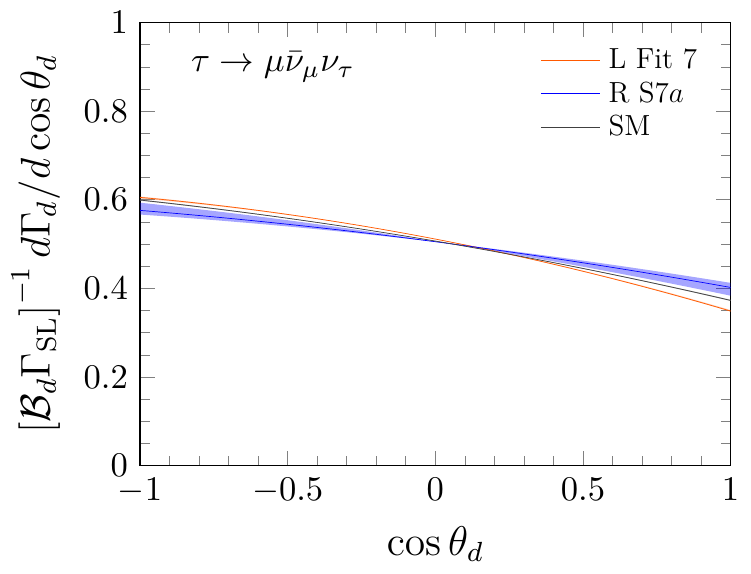}\\
\includegraphics[scale=1.0]{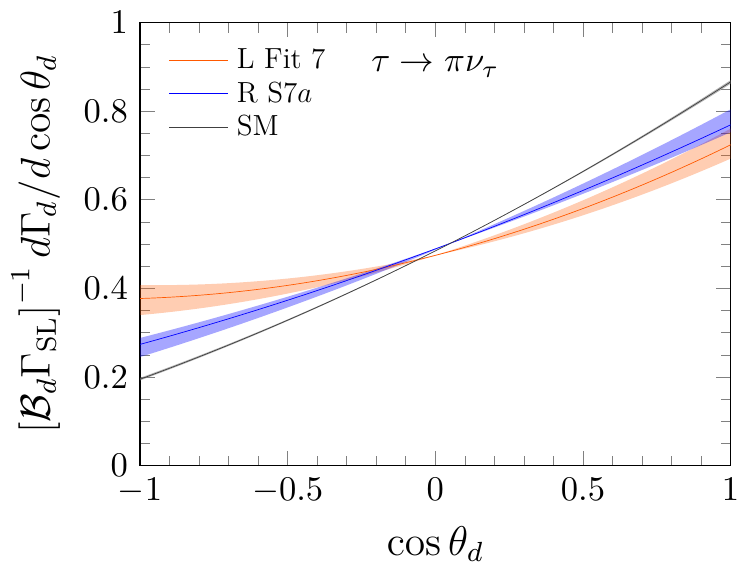}\hspace{.5cm}
\includegraphics[scale=1.0]{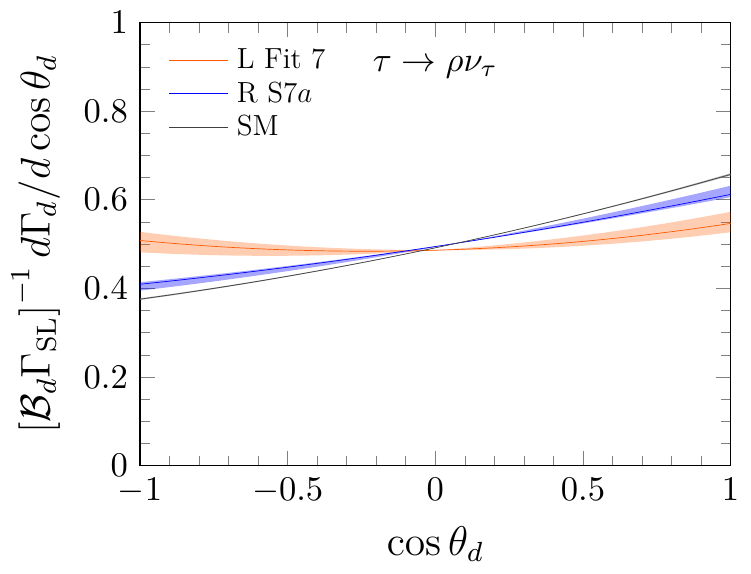}
\caption{Angular $d\Gamma/d\cos\theta_d$ distribution for the 
$\Lambda_b\to\Lambda_c\tau(\mu\bar\nu_\mu\nu_\tau,\,\pi\nu_\tau,\,\rho\nu_\tau)\bar\nu_\tau$ decays, 
keeping $y=m_d/m_\tau$ to its finite value, and obtained within the SM and 
the beyond the SM scenarios of Fit 7 (7a) of Ref~\cite{Murgui:2019czp} 
(\cite{Mandal:2020htr}), which only includes left- (right-)handed neutrino 
NP operators. Error bands account for uncertainties induced by both form-factors
and fitted Wilson coefficients (added in quadrature).}\label{fig:angular}
\end{center}
\end{figure}
A further integration in $\omega$ additionally  enhances the statistics. Although it prevents a separate 
determination of each of the asymmetries, it is still a useful observable in the search 
for NP beyond the SM. This angular distribution reads
\bea
\frac{d\Gamma_d}{d\cos\theta_d}={\cal B}_d\Gamma_{\rm SL}\Big[
\frac12+\widehat F_1^d\cos\theta_d+\widehat F_2^d\, 
P_2(\cos\theta_d)\Big], \quad  \widehat F_{1,2}^d=\frac1\Gamma_{\rm SL}\int_1^{\omega_{\rm max}}
\frac{d\Gamma_{\rm SL}}{d\omega}\widetilde F_{1,2}^d(\omega)\,d\omega.\label{eq:Gcos}
\eea
and an appropriate angular analysis of $d\Gamma/d\cos\theta_d$ should allow to determine the total 
semileptonic width $\Gamma_{\rm SL}$ and the moments $\widehat F_1^d$ and $\widehat F_2^d$. 

The full distributions of Eq.~\eqref{eq:Gcos}, normalized by ${\cal B}_d\Gamma_{\rm SL}$, for the 
$\Lambda_b\to\Lambda_c\tau(\mu\bar\nu_\mu\nu_\tau,\pi\nu_\tau, \rho\nu_\tau)\bar\nu_\tau$ chain-decays, 
evaluated for the SM and different NP models are  presented in Fig.~\ref{fig:angular}. The integrated width 
$\Gamma_{\rm SL}$ and the angular moments $\widehat F_1^d$ and $\widehat F_2^d$ obtained from, each of the physics scenarios 
considered in the figure are collected in Tables~\ref{tab:ratios} and \ref{tab:hatfs}, respectively. As 
already mentioned,  all NP scenarios have been adjusted to reproduce the anomalies observed in the  
${\cal R}_{ D}$ and  ${\cal R}_{ D^{*}}$ ratios in $\bar B-$meson decays,  
and they all  predict values for ${\cal R}_{\Lambda_c}$ that are 
at variance ($2\sigma-3\sigma$) with  both the SM  prediction and the recent 
LHCb measurement, the latter two being within $1\sigma$. In addition, we  also observe differences  in $\widehat F_1^d$ and 
$\widehat F_2^d$, that are hardly accounted for by  errors.  This situation is reflected in 
Fig.~\ref{fig:angular}, where we see that the best discriminating power between the SM and different 
NP extensions is reached for
forward and backward emission in the $\tau$-hadron decay modes, which are more sensitive to 
$\widehat F_{1}^{\pi,\rho}$ and $\widehat F_{2}^{\pi,\rho}$. In fact, these new observables are shown 
as  excellent tools to discern between different inputs for the semileptonic 
$\Lambda_b\to\Lambda_c\tau 
\bar\nu_\tau$ parent reaction.  

In Appendix~\ref{app:bddstar} we collect the
corresponding results for the sequential $ \bar B \to D^{(*)} \tau(\mu\bar\nu_\mu\nu_\tau,\pi\nu_\tau, 
\rho\nu_\tau)\bar\nu_\tau$ decays.

\begin{table}
\begin{center}
\begin{tabular}{c|ccccc}
                        &  SM  & L Fit 7 \cite{Murgui:2019czp}& R S7a 
			\cite{Mandal:2020htr}&LHCb \cite{LHCb:2022piu}
                       \\\hline\tstrut
 $\Gamma_{e(\mu)}$ & ~$2.15\pm 0.08$ \\ \tstrut 
 $\Gamma_\tau$ & ~$0.715\pm 0.015$ 
 &~  $0.89\pm 0.05$&~ $0.81\pm0.06$  \\ \tstrut
 ${\cal R}_{\Lambda_c}$ & ~$0.332 \pm 0.007$   & ~ $0.41\pm 0.02$ 
 & ~ $0.38\pm 0.03$& \ \ $0.242\pm0.026\pm0.040\pm0.059$\\ \hline
\end{tabular}
\end{center}
\caption{Semileptonic decay widths $\Gamma_\tau=\Gamma\left(\Lambda_b\to\Lambda_c\tau\bar\nu_\tau\right)$ 
and  $\Gamma_{e(\mu)}=\Gamma\left(\Lambda_b\to\Lambda_c\, e(\mu)\bar\nu_{e(\mu)}\right)$ [units of  $\left(10\times |V_{cb}|^2 {\rm ps}^{-1}\right)$]
and  ratios ${\cal R}_{\Lambda_c} =\Gamma\left(\Lambda_b\to\Lambda_c\tau\bar\nu_\tau\right)
/\Gamma\left(\Lambda_b\to\Lambda_c\, e(\mu)\bar\nu_{e(\mu)}\right)$ 
obtained in the SM,  the NP model Fit 7 (7a) of Ref~\cite{Murgui:2019czp} (\cite{Mandal:2020htr}), which only includes left- (right-)handed 
neutrino NP operators. Errors induced by the uncertainties in
the form-factors and  Wilson Coefficients are added in quadrature. 
The recent  LHCb~\cite{LHCb:2022piu} measurement of the  ${\cal R}_{\Lambda_c}$ ratio, 
with the tau being reconstructed using 
the $\tau\to\pi^-\pi^+\pi^-(\pi^0)\,\nu_\tau$ decay, is also shown.}
\label{tab:ratios}
\end{table}
\begin{table}[t]
\begin{center}
\begin{tabular}{c|cc|cc|cc}
& $\widehat F_{1}^{\mu\bar\nu_\mu}$& $\widehat F_{2}^{\mu\bar\nu_\mu}$&$\widehat F_{1}^\pi$&$\widehat F_{2}^\pi$
&$\widehat F_{1}^\rho$&$\widehat F_{2}^\rho$\\ \hline \tstrut
SM &$-0.113\pm0.001$&$-0.0137\pm0.0003$&$0.336\pm0.003$&$ 0.0306\pm0.0007$&$0.141\pm0.002$&$0.0166\pm0.0004$\\ \tstrut
L Fit 7&$-0.128\pm0.003$&$-0.0228\pm0.0005$&$0.17\pm0.03$&$0.0507\pm0.0013$&$0.019^{+0.025}_{-0.020}$&$0.0275\pm0.0007$\\ \tstrut
R S7a&$-0.087^{+0.010}_{-0.017}$&$-0.0108^{+0.0006}_{-0.0011} $&$0.25^{+0.03}_{-0.02}$&$0.023^{+0.002}_{-0.003}$
&$0.101^{+0.017}_{-0.005}$&$0.0108^{+0.0027}_{-0.0008}$\\  \hline
\end{tabular}
    \caption{ Predictions for the angular moments $\widehat F^d_{1,\,2}$ for the
    $\Lambda_b\to\Lambda_c\tau(\mu\bar\nu_\mu\nu_\tau,\,\pi\nu_\tau,\,\rho\nu_\tau)
    \bar\nu_\tau$ decays evaluated in the SM and the same NP scenarios 
    considered in Table~\ref{tab:ratios}.}
   \label{tab:hatfs}
   \end{center}
\end{table}

\section{The $d\Gamma/dE_d$ distribution}
\label{sec:de}

Finally in this section we study the energy ($E_d$) distribution of the charged (massive) product 
from the tau-decay.  The idea is to increase the statistics by accumulating events for all 
allowed $\omega$  values 
and provide only the $E_d$ spectrum. Regardless detector efficiencies considerations, the $d\Gamma/dE_d$ 
differential decay width could be determined as  precisely as  $d\Gamma/d\cos\theta_d$ (discussed in 
Sec.~\ref{sec:dcos}) or $d\Gamma_{\rm SL}/d\omega$, with the three distributions giving independent 
information about the dynamics governing the semileptonic $b\to c \tau\bar\nu_\tau$ 
transition~\cite{Penalva:2020xup,Penalva:2021gef}.
From the $d^2\Gamma_d/(d\omega  d\xi_d)$ differential decay width  given in Eq.~\eqref{eq:wE} and 
using $E_d=\gamma m_\tau\xi_d$, we have
\bea
\frac{d\Gamma_d}{dE_d} & = & 2{\cal B}_{d} \int_{\omega_{\rm inf}(E_d)}^{\omega_{\rm sup}(E_d)} d\omega
\frac{1}{\gamma m_\tau}\frac{d\Gamma_{\rm SL}}{d\omega}\Big\{  C_n^d(\omega,E_d)+C_{P_L}^d(\omega,E_d)\,\langle P^{\rm CM}_L\rangle(\omega)\Big\},
\label{eq:EdG}
\eea
The maximum energy,  $E_d^{\rm max}$, of the massive product from the tau-decay is
\be 
E_d^{\rm max} = \frac{(M-M')^2+m_d^2}{2(M-M')}
\ee
while the minimum one, $E_d^{\rm min}$, depends  on the tau-decay mode and the order 
relation between $(M-M')$ and $m_\tau^2/m_d$. For the reactions considered in this work, 
we have $(M-M') \le m_\tau^2/m_d$ and hence
\bea
E_d^{\rm min}&=& m_d, \quad d=\mu\bar\nu_\mu, e\bar\nu_e \\
E_d^{\rm min}&=& \frac{m_d^2(M-M')^2+m_\tau^4}{2m_\tau^2(M-M')}, \quad d=\pi,\rho
\eea
while $E_d^{\rm min}=m_d$ for the hadronic case if  $(M-M')\ge m_\tau^2/m_d$. This latter situation 
occurs for instance in the sequential 
$\bar B \to \pi \tau(\rho\nu_\tau)\bar\nu_\tau$ reaction, involving the CC $b\to u \tau \bar\nu_\tau$ 
transition. 

To perform the $\omega$ integration, first we have  to  obtain the allowed variation of the $\omega$ variable 
for a given $E_d$, i.e., to determine $\omega_{\rm inf}(E_d)$ and $\omega_{\rm sup}(E_d)$ in Eq.~\eqref{eq:EdG}. 
This requires to invert the limits in Eq.~(\ref{eq:ed-range}) and the result depends on the tau-decay  channel
\begin{enumerate}
\item \underline{$\tau\to\mu\bar\nu_\mu\nu_\tau$ and $\tau\to e \bar\nu_e\nu_\tau$}: In this case,  $m_d$ is either the muon or the electron mass, and considering $(M-M') \le m_\tau^2/m_d$, we  find $\omega_{\rm inf}(E_d)=1$, while 
\bea
\omega_{\rm sup}(E_d) &=& \omega_{\rm max} = \frac{M^2+M^{\prime2}-m^2_\tau}{2MM'}, \quad E_d\le\frac{m_\tau^2+m_d^2}{2 m_\tau} \label{eq:invert1}\\
\omega_{\rm sup}(E_d) &=& \frac{M^2+M^{\prime2}-\left(E_d+\sqrt{E_d^2-m_d^2}\,\right)^2}{2MM'}, \quad E_d\ge\frac{m_\tau^2+m_d^2}{2 m_\tau} 
\eea
\item \underline{$\tau\to\pi\nu_\tau$ and $\tau\to\rho\nu_\tau $}: In this case $m_d$ is either the pion 
or rho  mass,  
and considering $(M-M') \le m_\tau^2/m_d$, we also find $\omega_{\rm inf}(E_d)=1$, while 
\bea
\omega_{\rm sup}(E_d) &=&\frac{M^2+M^{\prime2}-m_\tau^4\left(E_d-\sqrt{E_d^2-m_d^2}\,\right)^2/m_d^4}{2MM'} , \quad E_d\le\frac{m_\tau^2+m_d^2}{2 m_\tau} \\
\omega_{\rm sup}(E_d) &=& \frac{M^2+M^{\prime2}-\left(E_d+\sqrt{E_d^2-m_d^2}\,\right)^2}{2MM'}, \quad E_d\ge\frac{m_\tau^2+m_d^2}{2 m_\tau} \label{eq:invert4}
\eea
\end{enumerate}
From the differential distribution of Eq.~\eqref{eq:EdG}, we define a new dimensionless observable 
$\widehat F^d_0(E_d)$,   such that 
$d\Gamma/dE_d=2{\cal B}_d\Gamma_{\rm SL}\widehat F^d_0(E_d)/m_\tau$,  with $\Gamma_{\rm SL}$ the total 
$H_b\to H_c\tau\bar\nu_\tau$ semileptonic decay width, and
\bea
\widehat F^d_0(E_d)=\frac1{\Gamma_{\rm SL}}\int_{1}^{\omega_{\rm sup}(E_d)} \frac1\gamma\frac{d\Gamma_{\rm SL}}{d\omega}\Big\{  C_n^d(\omega,E_d)+C_{P_L}^d(\omega,E_d)\,\langle P^{\rm CM}_L\rangle(\omega)\Big\}\,d\omega, \label{eq:defF0}
\eea
where the corresponding $\omega_{\rm sup}(E_d)$ values can be read out 
from Eqs.~(\ref{eq:invert1})-(\ref{eq:invert4}). This energy function is normalized for all tau-decay
 channels to 
\be
\frac{1}{m_\tau}\int_{E_d^{\rm min}}^{E_d^{\rm min}} dE_d \widehat F^d_0(E_d)= \frac{1}{m_\tau\Gamma_{\rm SL}}\int_{E_d^{\rm min}}^{E_d^{\rm min}} \int_{1}^{\omega_{\rm sup}(E_d)} \frac1\gamma\frac{d\Gamma_{\rm SL}}{d\omega} C_n^d(\omega,E_d)\,d\omega dE_d = \frac12
\ee
Although the CM $\tau$ longitudinal polarization $\langle P^{\rm CM}_L\rangle(\omega)$ does not contribute
 to the normalization 
of $\widehat F^d_0(E_d)$, it  still affects the energy shape of the observable.
 This is in contrast 
to what happens if,  instead, one accumulates on the variable $\xi_d$  in 
the  $d^2\Gamma_d/(d\omega  d\xi_d)$ distribution of Eq.~\eqref{eq:wE} to 
obtain $d\Gamma_d/d\omega$. 
As already mentioned, this $\xi_d$  (or equivalently $E_d$) integration 
removes permanently 
any information about $\langle P^{\rm CM}_L\rangle$.

\begin{figure}
\includegraphics[scale=1]{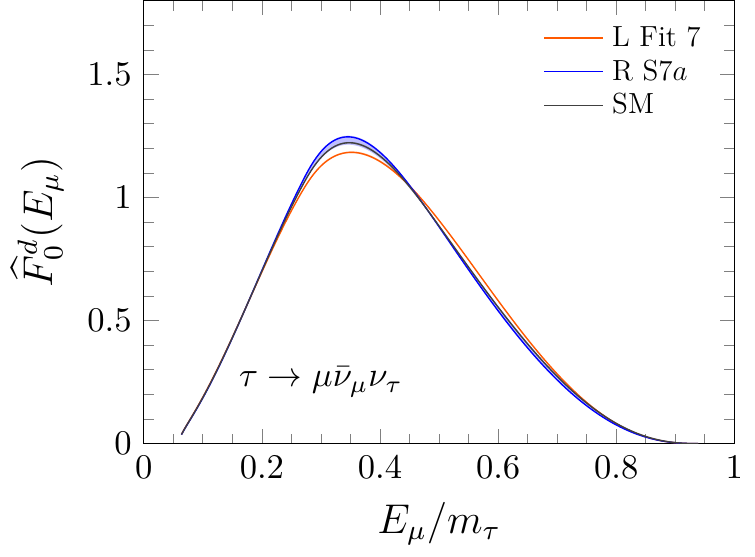}\\
\includegraphics[scale=1]{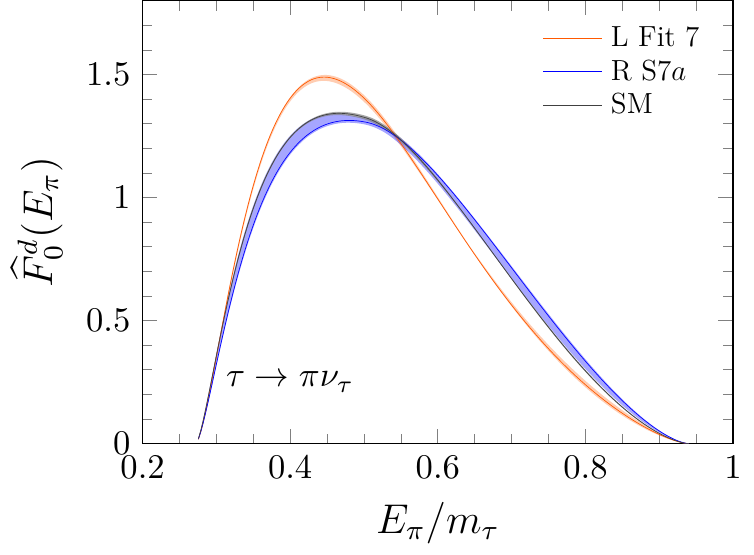}\hspace{.5cm}
\includegraphics[scale=1]{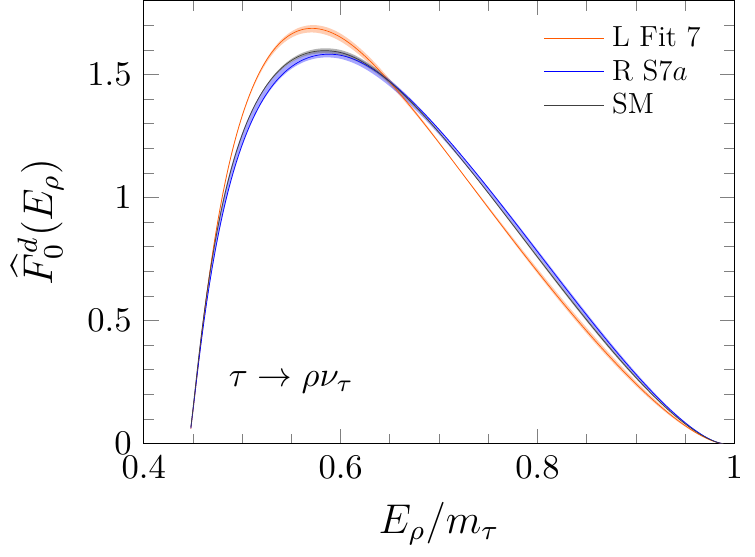}
\caption{Predictions for the $\widehat F^d_0(E_d)$ energy distribution [Eq.~\eqref{eq:defF0}]  for the 
$\Lambda_b\to\Lambda_c\tau(\mu\bar\nu_\mu\nu_\tau,\,\pi\nu_\tau,\,\rho\nu_\tau)\bar\nu_\tau$ 
decays, keeping $y=m_d/m_\tau$ finite, obtained within the SM and the 
NP scenarios corresponding to Fit 7 of Ref~\cite{Murgui:2019czp} and Fit 7a 
of Ref~\cite{Mandal:2020htr}.}
\label{fig:energia}
\end{figure}

The results for $\widehat F^d_0(E_d)$ in  the 
$\Lambda_b\to\Lambda_c\tau(\mu\bar\nu_\mu\nu_\tau,\,\pi\nu_\tau,\,\rho\nu_\tau)\bar\nu_\tau$ decays are 
presented in Fig.~\ref{fig:energia}. We observe small changes between the predictions obtained from the 
SM  and any of the NP 
models considered in this work, pointing out to a little  influence of  the $\langle P^{\rm CM}_L\rangle$ 
contribution in this distribution. 
Nevertheless, for the hadron modes, we again see  that Fit 7 of
 Ref~\cite{Murgui:2019czp} gives,  
in some regions, significantly different results 
from those obtained in the SM and Fit 7a, while the latter agrees with the SM 
within uncertainty bands.

\section{Summary and conclusions}
Using the analytical results derived in  \cite{Penalva:2021wye}, we have studied the  
$d^2 \Gamma_d /(d\omega d\cos\theta_d) $, $d\Gamma_d /d\cos\theta_d$ and $ d\Gamma_d /dE_d $ 
distributions, which are defined  in terms of the visible energy and polar angle of the charged 
particle from the $\tau$-decay  in  $b\to c \tau\, (\mu \bar \nu_\mu \nu_\tau,\pi\nu_\tau,
\rho\nu_\tau) \bar\nu_\tau$ reactions and that one expects to be measured at some point in the near
future. 
The first
two contain information on the CM transverse tau-spin ($\langle P^{\rm CM}_{T}\rangle(\omega)$), 
tau-angular  ($A_{FB,Q}(\omega)$) and  tau-angular-spin ($Z_{L,Q,\perp}(\omega)$) asymmetries 
of the $H_b\to H_c\tau\bar\nu_\tau$ parent decay. Hence, from the dynamical point of view, 
these observables are richer than the commonly used one, $d^2 \Gamma_d /(d\omega dE_d) $, 
 since the latter gives  access only to 
the  CM tau longitudinal polarization $\langle P^{\rm CM}_{L}\rangle(\omega)$. We have paid 
attention to the deviations with respect to the predictions of the SM for these new observables, 
considering NP operators constructed using both left- and right-handed neutrino fields, within 
an effective theory approach. We have presented results for  these  distributions 
in  $\Lambda_b\to\Lambda_c\tau\, (\mu \bar \nu_\mu \nu_\tau,\pi\nu_\tau,
\rho\nu_\tau)  \bar\nu_\tau$ (main text)
 and $\bar B \to D^{(*)}\tau\, (\mu \bar \nu_\mu \nu_\tau,\pi\nu_\tau,
 \rho\nu_\tau) 
 \bar\nu_\tau$ (Appendix \ref{app:bddstar}) sequential decays,  within different 
 beyond the SM scenarios, and we have discussed 
 their use to disentangle between different NP models. In this respect, we have seen that 
 $d\Gamma_d /d\cos\theta_d$, if measured with sufficiently good statistics, 
 becomes quite useful, especially in the 
 $\tau\to \pi \nu_\tau$ decay mode.  

The study carried out in this work acquires a special relevance due to the  recent  
LHCb measurement of the LFU ratio ${\cal R}_{\Lambda_c}$ in agreement, within 
errors, with the SM prediction. The experiment identified the $\tau$  using 
the three-prong hadronic $\tau^- \to\pi^-\pi^+\pi^-(\pi^0)\,\nu_\tau$ decay, 
and this result for ${\cal R}_{\Lambda_c}$, which is in conflict with the 
phenomenology from the $b$-meson sector, needs to be confirmed employing 
other reconstruction channels. 

We are aware of the difficulties in measuring the accumulated
  distributions proposed in this work for the  $\Lambda_b\to\Lambda_c\tau(\mu \bar \nu_\mu \nu_\tau,\pi\nu_\tau,\rho\nu_\tau) \bar\nu_\tau$ decay
 at LHC~\cite{Cerri:2018ypt}. As mentioned in the Introduction, the LHCb collaboration is conducting a study on this reaction 
 using the $\tau\to\mu\bar\nu_\mu\nu_\tau$ reconstruction channel. We expect that
  this would imply the measurement of some of the muon variables  and thus the determination, in the not too 
  distant future and with a certain accuracy,  of some or all, of the differential decays widths analyzed in this work. If the presence of NP is 
 confirmed, going beyond the pure measurement of 
 $R(\Lambda_c)$ (and other ratios) is essential to disentangle among different SM extensions.   Furthermore, we have also predicted accumulated distributions for the $\bar B
\to D^{(*)}$ semileptonic reactions, for which,  within the context
of the plan to increase luminosity at the LHC, the prospects look more favorable~\cite{Cerri:2018ypt}.

\section*{Acknowledgements}
This research has been supported  by the Spanish Ministerio de Ciencia e Innovaci\'on (MICINN)
and the European Regional Development Fund (ERDF) under contract PID2020-112777GB-I00 and 
PID2019-105439GB-C22, 
the EU STRONG-2020 project under the program H2020-INFRAIA-2018-1, 
grant agreement no. 824093 and by  Generalitat Valenciana under contract PROMETEO/2020/023. 
\appendix

\section{Coefficients $C^{\pi,\,\rho}_{A_{FB,Q}}(\omega)$, $C^{\pi,\,\rho}_{P_T}(\omega)$ and $C^{\pi,\,\rho}_{Z_{L,Q,\perp}}(\omega)$ for the
 $\pi\nu_\tau$ and $\rho\nu_\tau$ $\tau$-decay modes}
 \label{app:pirho}

In this appendix, we give the coefficients which define the $\widetilde F_{12}^{\pi,\,\rho}(\omega)$ distributions in terms of the tau-asymmetries, through Eqs.~\eqref{eq:F1} and ~\eqref{eq:F2},  for the $\pi\nu_\tau$  and $\rho\nu_\tau$ tau-decay modes keeping finite $y=m_d/m_\tau$. We use the analytical expressions derived in Ref.~\cite{Penalva:2021wye} for  the two dimensional $C^{\pi,\,\rho}_{A_{FB,Q}}(\omega,\xi_d)$, $C^{\pi,\,\rho}_{P_T}(\omega,\xi_d)$ and $C^{\pi,\,\rho}_{Z_{L,Q,\perp}}(\omega, \xi_d)$ functions and integrate over the variable $\xi_d$. We first discuss the coefficient of the forward-backward asymmetry,
\begin{equation}
C^{\pi,\rho}_{A_{FB}}(\omega)=\left\{
\begin{array}{c}
\frac{1+y^2}{1-y^2 }\frac{1}{\beta ^2} \left[\beta -\frac{\artanh\beta}{\gamma^2}\right],\quad y^2\le \frac{1-\beta}{1+\beta} = \frac{m^2_\tau}{q^2}\\  \\
\frac{1+y^2}{1-y^2 }\frac{1}{\beta ^2} \left[\frac{1-y^2}{1+y^2}+ \frac{\log y}{\gamma^2}\right], \quad y^2\ge \frac{1-\beta}{1+\beta}= \frac{m^2_\tau}{q^2}
\end{array}
\right.
\end{equation}
For the reactions studied in this work, we always have $y^2 \le m^2_\tau/q^2$ for all available $q^2$ values and thus, the first of the above possibilities  should be taken. The situation is repeated for the rest of the coefficients. For brevity, we only give below the expressions for the $y^2 < m^2_\tau/q^2$ case, 
\bea
C^{\pi,\rho}_{Z_L}(\omega)&=&-a_{\pi,\rho}\left[\frac{1+4 y^2+y^4}{\left(1-y^2\right)^2}\frac{1}{\gamma^2\beta^3} \left(\beta-\artanh\beta 
\right)+ \frac{2y^2}{\left(1-y^2\right)^2}\right]  \\
C^{\pi,\rho}_{P_T}(\omega)&=&-a_{\pi,\rho}\frac{2}{\pi\gamma\beta}\left[ \frac{1+4 y^2+y^4}{1-y^4}C^{\pi,\rho}_{A_{FB}}(\omega) 
-\frac{4y^2}{\left(1-y^2\right)^2}\artanh\beta\right]\\
C^{\pi,\rho}_{A_{Q}}(\omega)&=&\frac{1}{2 \beta ^2}\left[3-\beta^2+3\frac{1+\left(2+4\gamma^2\right)y^2+y^4}{2\gamma^3 \beta y  \left(1-y^2\right)}\artanh\Big(\frac{2\gamma\beta y}{1-y^2} \Big)\right]\nonumber \\
&& -\frac{3}{\gamma^2\beta ^3 } \frac{1+y^2}{1-y^2}\artanh\Big(\frac{1+y^2}{1-y^2}\beta\Big) \\
C^{\pi,\rho}_{Z_Q}(\omega)&=&-3a_{\pi,\rho}\left[\frac{ 1+y^2}{\beta ^3 \gamma^2\left(1-y^2\right)}\left(1+\frac{1}{4\gamma \beta y} \frac{1+\left(10+4\gamma^2\right)y^2+y^4}{ 1-y^2}\artanh\Big(\frac{2\gamma\beta y}{1-y^2}\Big) \right)\right.\nonumber \\
&&\left.-\frac{ 2\gamma^2\left(1+4y^2+y^4\right) +\left(1+y^2\right)^2}{2 \gamma^4 \beta ^4 \left(1-y^2\right)^2}\artanh\Big(\frac{1+y^2}{1-y^2}\beta\Big)
\right] \\
C^{\pi,\rho}_{Z_\perp}(\omega)&=&\frac{3a_{\pi,\rho}}{2\beta^4\gamma}\frac{1}{\left(1-y^2\right)^2}\left[\left(2\beta-\beta^3\right)\left(1-y^4\right)- \frac{3\left(1+y^2\right)^2+4\gamma^2y^2}{\gamma^2}\artanh\Big(\frac{1+y^2}{1-y^2}\beta\Big)\right. \nonumber\\
&&\left.   + \frac{1+y^2}{2y}\,\frac{ \left(1+y^2\right)^2+12\gamma^2y^2 }{ \gamma^3 }\artanh\Big(\frac{2\gamma\beta y}{1-y^2}\Big) \right] 
\eea
with $a_\pi=1$ and $a_\rho=(m^2_\tau-2m^2_\rho)/(m^2_\tau+2m^2_\rho)$.
Note that all 
the arguments of the $\artanh$-functions are smaller than one, since the 
above expressions are only valid for $y^2 < (1-\beta)/(1+\beta)$.
%
%
\section{Results for the $ \bar B \to D^{(*)} \tau(\mu\bar\nu_\mu\nu_\tau,\pi\nu_\tau, 
\rho\nu_\tau)\bar\nu_\tau$ sequential decays}
\label{app:bddstar}
\begin{table}
\begin{center}
\begin{tabular}{c|c|ccccc}
                        &&  SM  & L Fit 7 \cite{Murgui:2019czp}& R S7a 
			\cite{Mandal:2020htr}&\ HFLAV \cite{HFLAV:2019otj}
                       \\\hline\tstrut
 &$\Gamma_{e(\mu)}$ & ~$0.87\pm 0.03$ \\ \tstrut 
$\bar B\to D$ &$\Gamma_\tau$ & ~$0.262\pm 0.005$ 
 &~  $0.34\pm 0.04$&~ $0.292^{+0.060}_{-0.014}$  \\ \tstrut
& ${\cal R}_{D}$ & ~$0.300^{+0.005}_{-0.004} $   & ~ $0.388^{+0.044}_{-0.045}$ 
 & ~ $0.334^{+0.070}_{-0.015}$ &~~$0.340\pm0.027\pm0.013$\\   \hline\tstrut
 &$\Gamma_{e(\mu)}$ & ~$2.01^{+0.07}_{-0.08}$ \\ \tstrut 
$\bar B\to D^*$ &$\Gamma_\tau$ & ~$0.512^{+0.013}_{-0.014}$ 
 &~  $0.61\pm 0.03$&~ $0.59\pm0.03$  \\ \tstrut
& ${\cal R}_{D^*}$ & ~$0.255\pm0.003 $   & ~ $0.306\pm0.013$ 
 & ~ $0.292^{+0.014}_{-0.015}$ &~ $0.295\pm0.011\pm0.008$\\ \hline
\end{tabular}
\end{center}
\caption{Semileptonic decay widths $\Gamma_\tau=\Gamma\left(\bar B\to D^{(*)}\tau\bar\nu_\tau\right)$ 
and  $\Gamma_{e(\mu)}=\Gamma\left(\bar B\to D^{(*)}\, e(\mu)\bar\nu_{e(\mu)}\right)$ [units of  $\left(10\times |V_{cb}|^2 {\rm ps}^{-1}\right)$]
and  ratios ${\cal R}_{D^{(*)}} =\Gamma\left(\bar B\to D^{(*)}\tau\bar\nu_\tau\right)
/\Gamma\left(\bar B\to D^{(*)}\, e(\mu)\bar\nu_{e(\mu)}\right)$ 
obtained in the SM,  the NP model Fit 7 (7a) of Ref~\cite{Murgui:2019czp} (\cite{Mandal:2020htr}), which only includes left- (right-)handed 
neutrino NP operators. Errors induced by the uncertainties in
the form-factors and  Wilson Coefficients are added in quadrature. The 
${\cal R}_{D^{(*)}}$ experimental averages compiled by the 
HFLAV~\cite{HFLAV:2019otj} collaboration are also given.}
\label{tab:ratiosD}
\end{table}
\begin{table}[t]
\begin{center}
\begin{tabular}{c|c|cc}
&& $\widehat F_{1}^{\mu\bar\nu_\mu}$& $\widehat F_{2}^{\mu\bar\nu_\mu}$\\ \hline \tstrut
&SM &$-0.06029^{+0.00021}_{-0.00018}$&$-0.03539^{+0.00015}_{-0.00012}$\\ \tstrut
$\bar B\to D$&L Fit 7&$-0.0306^{+0.0015}_{-0.0012}$&
$-0.0777046^{+0.0000051}_{-0.0000004}$ \\ \tstrut
&R S7a&$-0.031^{+0.021}_{-0.042}$&$-0.027^{+0.002}_{-0.003} $\\  \hline \tstrut
&SM &$-0.1267^{+0.0012}_{-0.0014} $&$-0.0063\pm0.0003$\\ \tstrut
$\bar B\to D^*$&L Fit 7&$-0.1695^{+0.0016}_{-0.0017}$&$-0.0020\pm0.0004$\\ \tstrut
&R S7a&$-0.098^{+0.004}_{-0.016}$&$-0.0053^{+0.0007}_{-0.0019} $\\  \hline
\end{tabular}
\caption{ Predictions for the angular moments $\widehat F^{\mu\bar\nu_\mu}_{1,\,2}$ 
    for the
    $\bar B\to D^{(*)}\tau(\mu\bar\nu_\mu\nu_\tau)
    \bar\nu_\tau$ decay evaluated in the SM and the same NP scenarios 
    considered in Table~\ref{tab:ratiosD}.}
   \label{tab:hatfsDDstar1}
   \end{center}
\end{table}
\begin{table}[t]
\begin{center}
\begin{tabular}{c|c|cc|cc}
&& $\widehat F_{1}^\pi$&$\widehat F_{2}^\pi$
&$\widehat F_{1}^\rho$&$\widehat F_{2}^\rho$\\ \hline \tstrut
&SM &
$0.5427^{+0.0004}_{-0.0005}$&$ 0.0779\pm0.0003$&$0.32388^{+0.00016}_{-0.00022}$&
$0.04189^{+0.00017}_{-0.00019}$\\ \tstrut
$\bar B\to D$&L Fit 7&
$0.160^{+0.013}_{-0.016}$&$0.1699^{+0.0007}_{-0.0006} $&
$0.089^{+0.008}_{-0.010} $&$0.0910^{+0.0006}_{-0.0005}$\\ \tstrut
&R S7a&$0.45^{+0.05}_{-0.09}$
&$0.053^{+0.011}_{-0.006}$
&$0.285^{+0.015}_{-0.055}$&$0.026^{+0.008}_{-0.004}$\\  \hline \tstrut
&SM &
$0.2732^{+0.0020}_{-0.0016}$&$ 0.0146\pm0.0007$&
$0.0889^{+0.0017}_{-0.0019}$&$0.0082\pm0.0004$\\ \tstrut
$\bar B\to D^*$&L Fit 7&
$0.3200^{+0.0005}_{-0.0004}$&$0.0053\pm0.0008$&
$0.0876^{+0.0010}_{-0.0011}$&$0.0032\pm0.0004$\\ \tstrut
&R S7a&
$0.184^{+0.069}_{-0.012}$&$0.0115^{+0.0048}_{-0.0019}$
&$0.050^{+0.035}_{-0.005}$&$0.0062^{+0.0028}_{-0.0011}$\\  \hline
\end{tabular}
    \caption{ Predictions for the angular moments $\widehat F^d_{1,\,2}$ 
    for the
    $\bar B\to D^{(*)}\tau(\pi\nu_\tau,\,\rho\nu_\tau)
    \bar\nu_\tau$ decays evaluated in the SM and the same NP scenarios 
    considered in Table~\ref{tab:ratiosD}.}
   \label{tab:hatfsDDstar2}
   \end{center}
\end{table}

\begin{figure}
\includegraphics[scale=0.77]{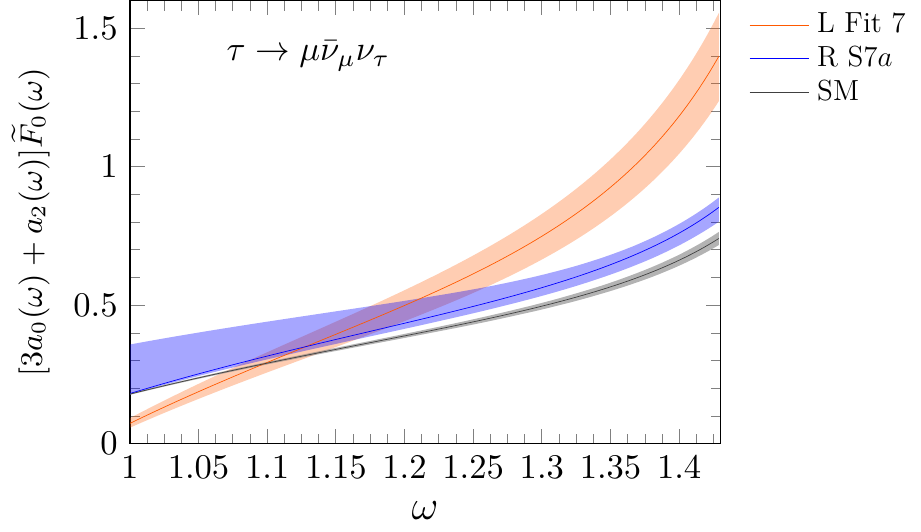}\\
\includegraphics[scale=0.77]{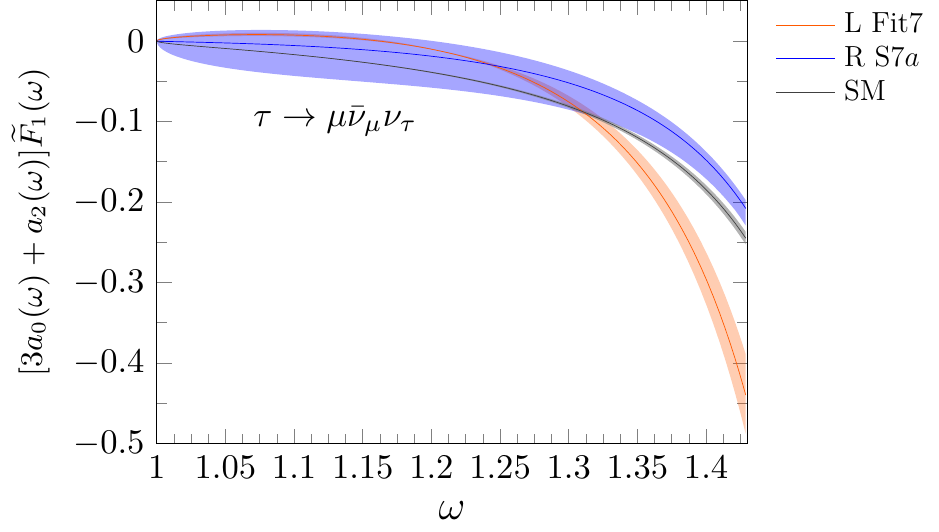}\hspace{.15cm}
\includegraphics[scale=0.77]{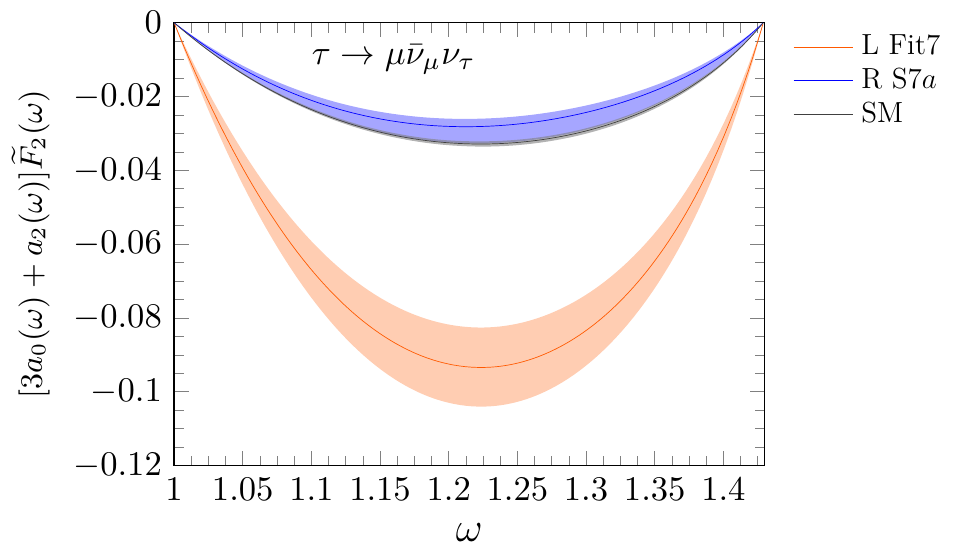}\\
\includegraphics[scale=0.77]{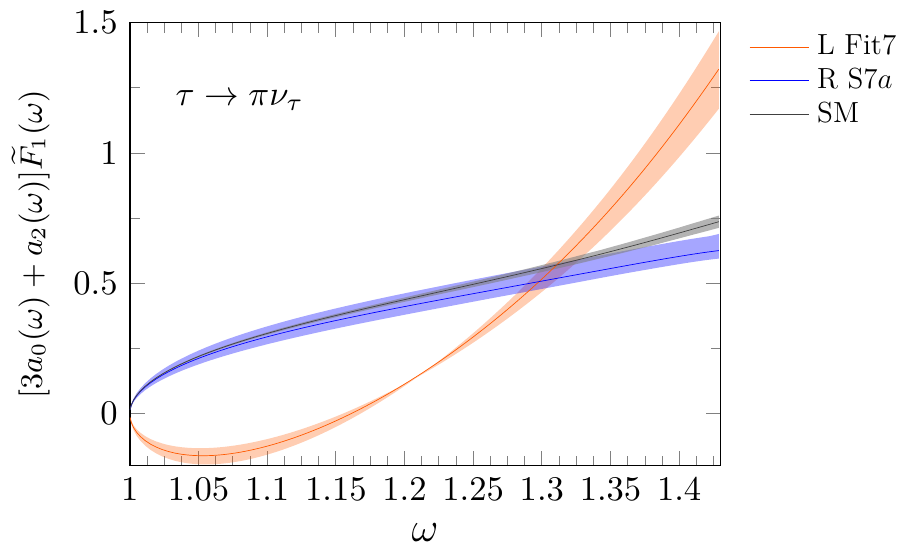}\hspace{.15cm}
\includegraphics[scale=0.77]{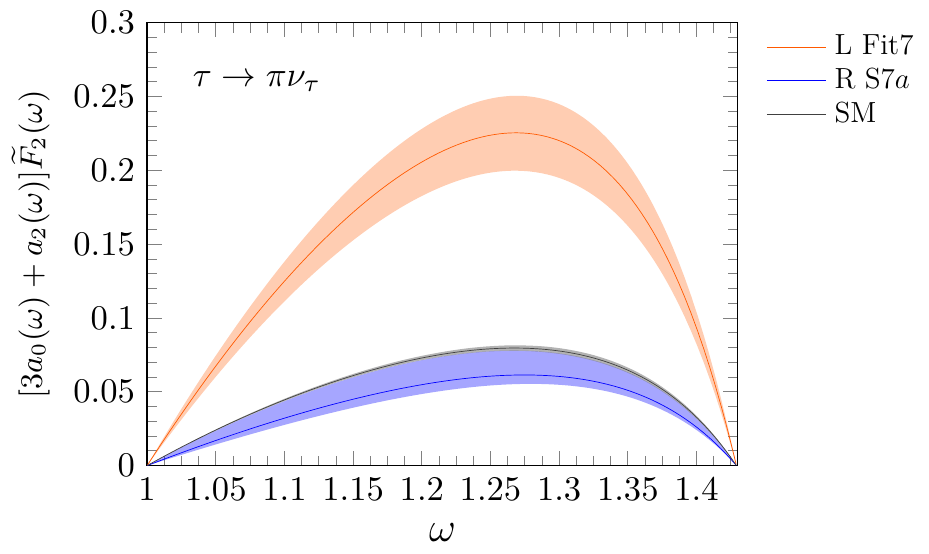}\\
\includegraphics[scale=0.77]{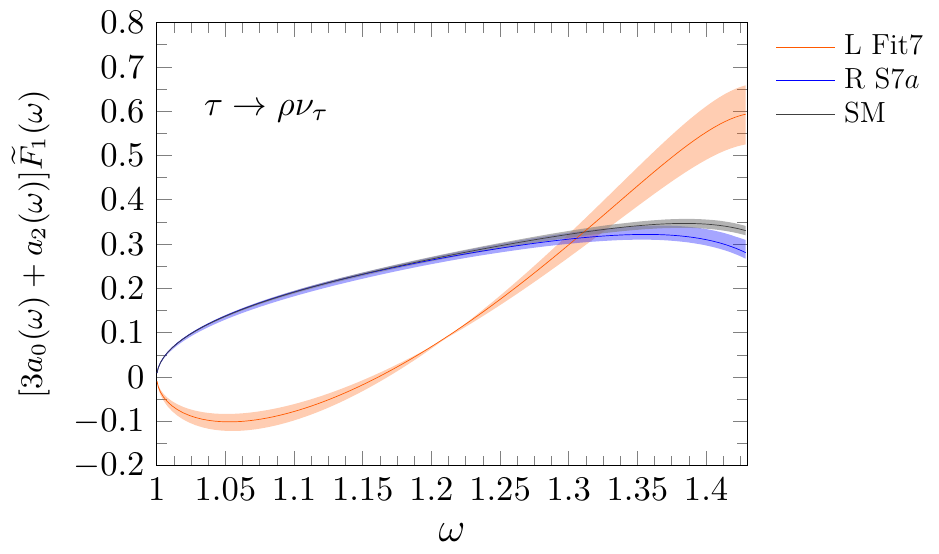}\hspace{.15cm}
\includegraphics[scale=0.77]{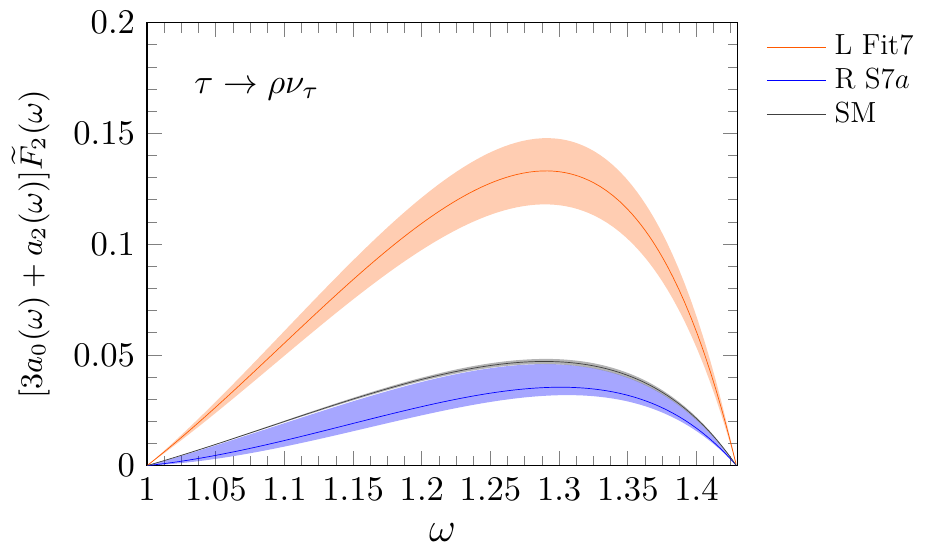}
\caption{ Results for the functions $n_0(\omega)
\widetilde F^{d}_{0,1,2}(\omega)$ evaluated for the 
$\bar B\to D\tau(\mu\bar\nu_\mu\nu_\tau,\pi\nu_\tau,\rho\nu_\tau)\bar\nu_\tau$ 
decays, 
keeping $y=m_{\mu,\pi,\rho}/m_\tau$ finite, and obtained within the SM and 
the beyond the SM scenarios of Fits 7 and 7a of Refs.~\cite{Murgui:2019czp} and \cite{Mandal:2020htr}, which only includes left- (right-)handed neutrino NP operators, respectively. Error bands account for uncertainties induced by both form-factors
and fitted Wilson coefficients (added in quadrature).  }
\label{fig:ftildemuD}
\end{figure}
\begin{figure}
\includegraphics[scale=0.77]{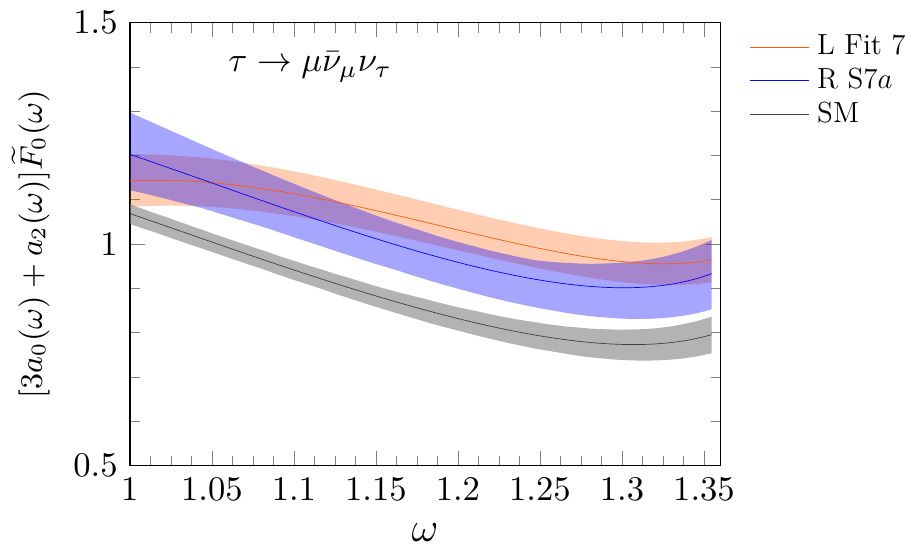}\\
\includegraphics[scale=0.77]{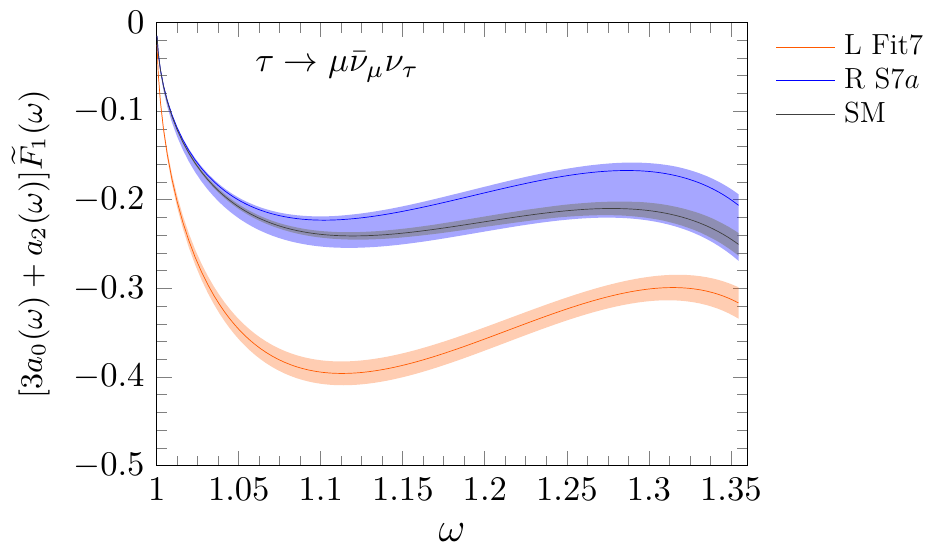}\hspace{.15cm}
\includegraphics[scale=0.77]{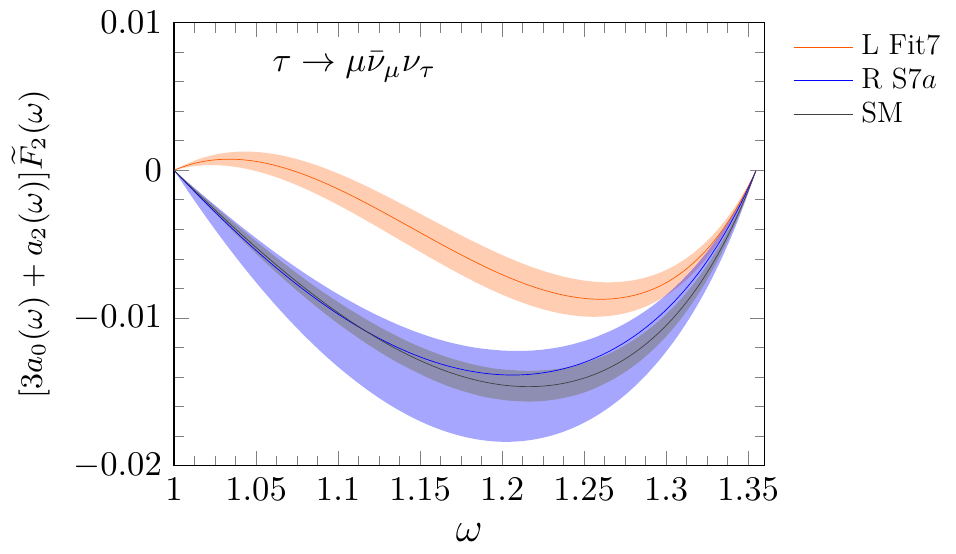}\\
\includegraphics[scale=0.77]{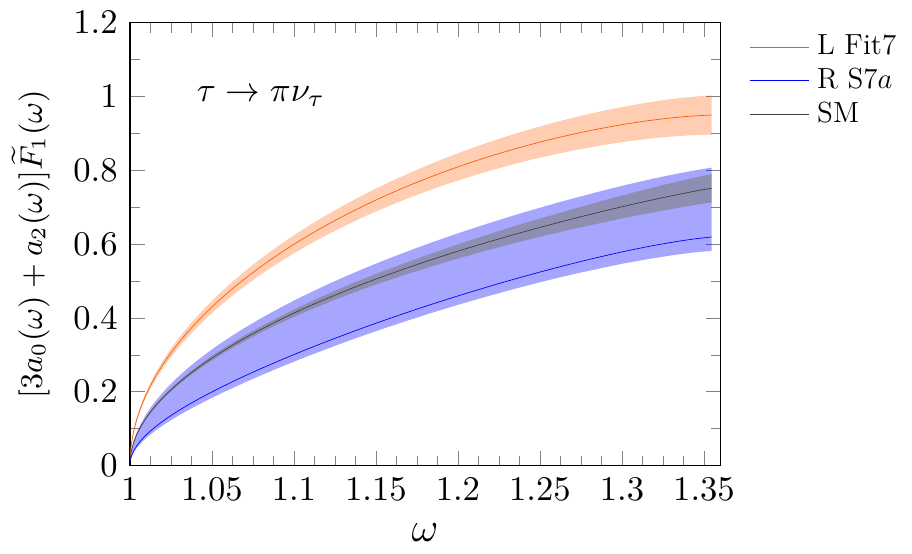}\hspace{.15cm}
\includegraphics[scale=0.77]{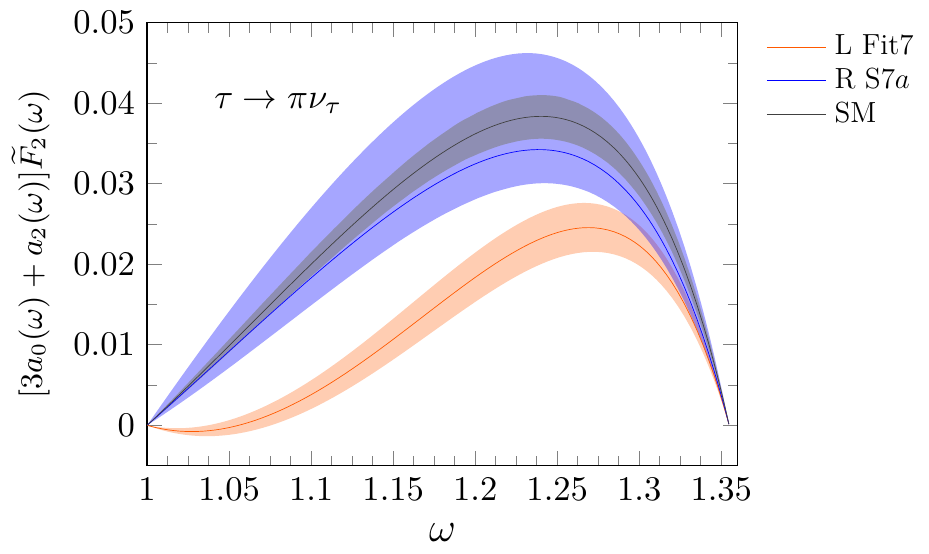}\\
\includegraphics[scale=0.77]{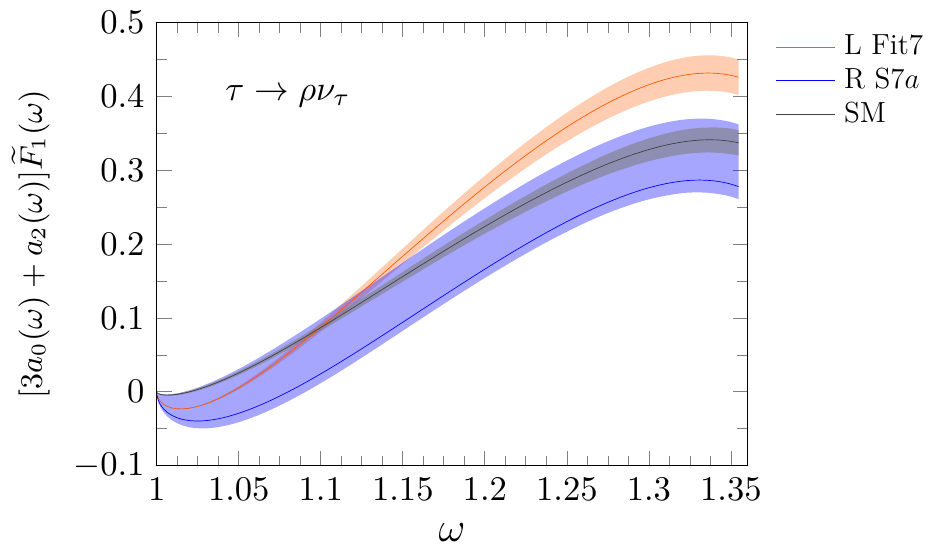}\hspace{.15cm}
\includegraphics[scale=0.77]{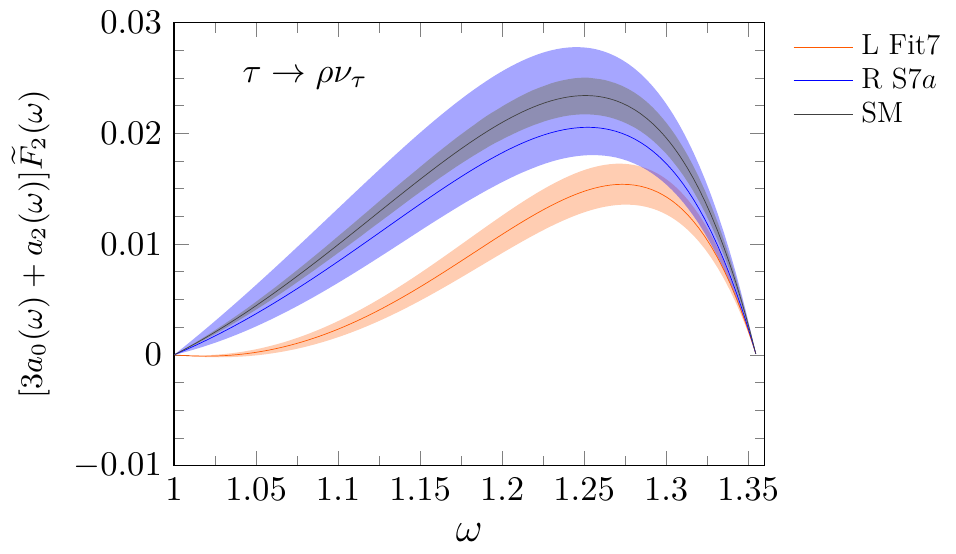}
\caption{ Same as in Fig.~\ref{fig:ftildemuD}, but for the $\bar B\to D^*\tau(\mu\bar\nu_\mu\nu_\tau,\pi\nu_\tau,\rho\nu_\tau)\bar\nu_\tau$ 
decays. }
\label{fig:ftildemuDstar}
\end{figure}

In this appendix we collect some results for the $ \bar B \to D^{(*)} \tau(\mu\bar\nu_\mu\nu_\tau\pi\nu_\tau, 
\rho\nu_\tau)\bar\nu_\tau$ sequential decays. We start by showing, in 
Figs.~\ref{fig:ftildemuD} and \ref{fig:ftildemuDstar}, the $\widetilde
F^{d}_{0,1,2}(\omega)$ functions evaluated within the SM  and the  NP models 
corresponding to Fit 7 of Ref~\cite{Murgui:2019czp} and Fit 7a
 of Ref~\cite{Mandal:2020htr}, which only includes left- (right-)handed neutrino 
 NP operators, respectively. Similarly to the $\Lambda_b\to\Lambda_c$ decay, the results for Fit 7 are very different from those
 obtained with  Fit 7a and the  SM, the latter two agreeing
 within uncertainties.

\begin{figure}
\begin{center}
\includegraphics[scale=1.0]{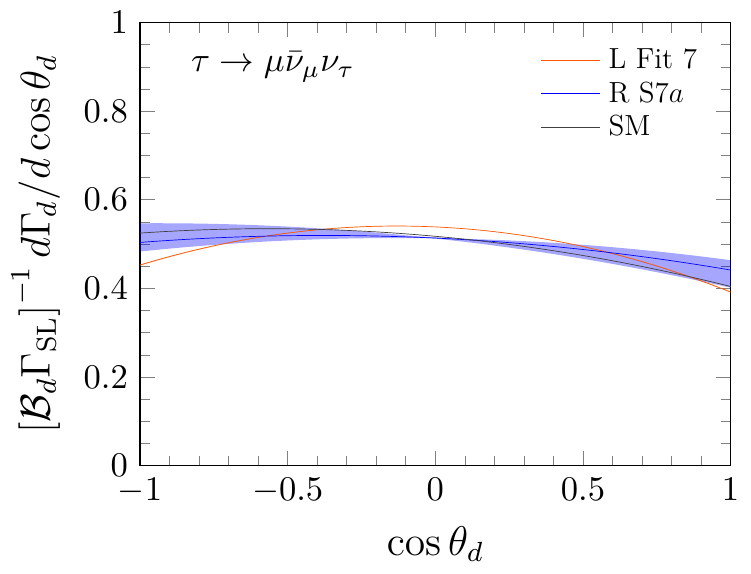}\\
\includegraphics[scale=1.0]{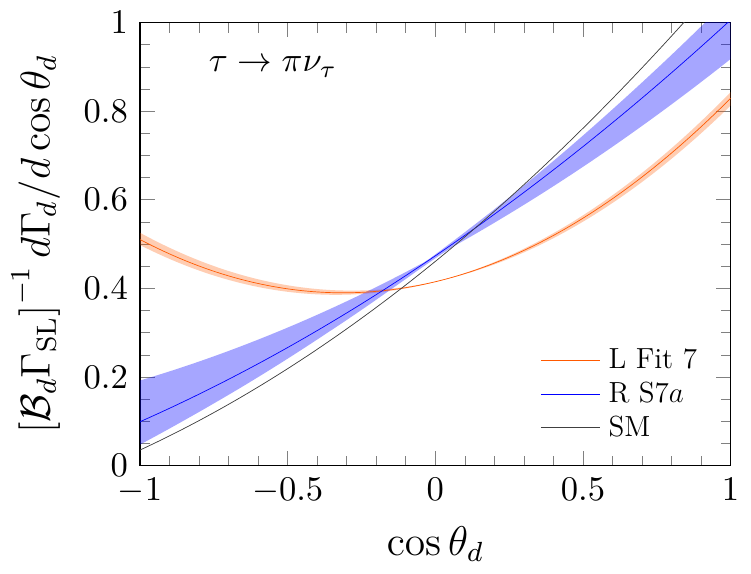}\hspace{.5cm}
\includegraphics[scale=1.0]{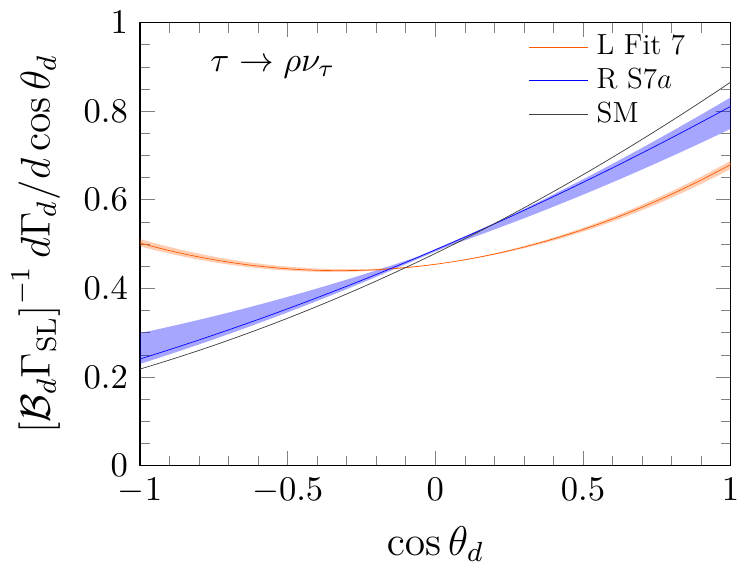}
\caption{Angular $d\Gamma/d\cos\theta_d$ distribution for the 
$\bar B\to D\tau(\mu\bar\nu_\mu\nu_\tau,\,\pi\nu_\tau,\,\rho\nu_\tau)\bar\nu_\tau$ decays, 
keeping $y=m_d/m_\tau$ to its finite value, and obtained within the SM and 
the beyond the SM scenarios of Fits 7 and 7a of Refs.~\cite{Murgui:2019czp} and \cite{Mandal:2020htr}, which only includes left- (right-)handed neutrino NP operators, respectively. Details as in Fig.~\ref{fig:angular}.}\label{fig:angularD}
\end{center}
\end{figure}

\begin{figure}
\begin{center}
\includegraphics[scale=1.0]{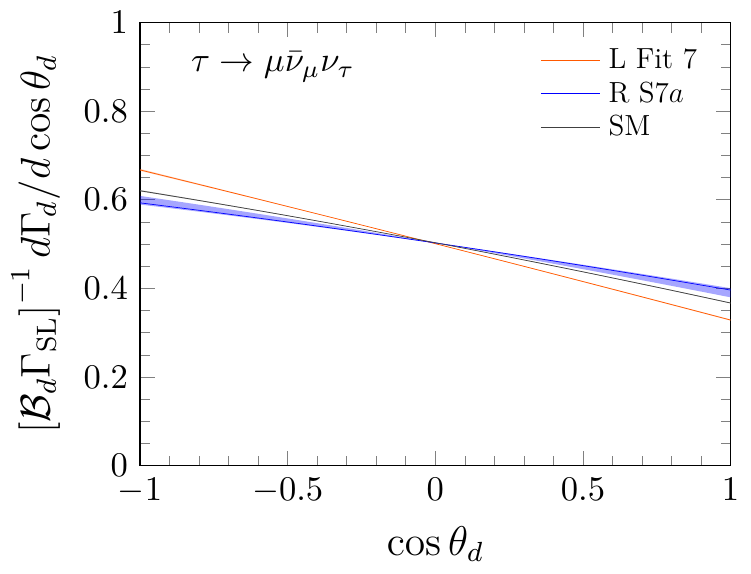}\\
\includegraphics[scale=1.0]{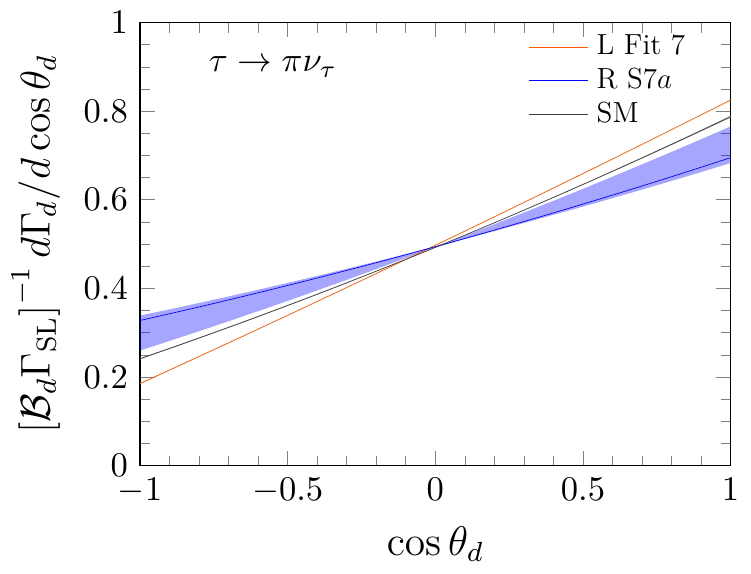}\hspace{.005cm}
\includegraphics[scale=1.0]{ang_Dstar_pion}
\caption{Same as in Fig.~\ref{fig:angularD}, but for the $\bar B\to D^*\tau(\mu\bar\nu_\mu\nu_\tau,\pi\nu_\tau,\rho\nu_\tau)\bar\nu_\tau$ 
decays.}\label{fig:angularDstar}
\end{center}
\end{figure}

In Figs.~\ref{fig:angularD} and \ref{fig:angularDstar}, we present now the $d\Gamma/d\cos\theta_d$
 distributions predicted within the SM and 
the beyond the SM scenarios of Fits 7 and 7a of Refs.~\cite{Murgui:2019czp} 
and \cite{Mandal:2020htr}, respectively. The best discriminating power  is reached for
forward and backward emission in the $\tau$-hadron decay modes for 
the $\bar B\to D$ decay.

Finally, in Table~\ref{tab:ratiosD}, we collect the values for the integrated
$\Gamma_\tau=\Gamma\left(\bar B\to D^{(*)}\tau\bar\nu_\tau\right)$ 
and  $\Gamma_{e(\mu)}=\Gamma\left[\bar B\to D^{(*)}\, e(\mu)
\bar\nu_{e(\mu)}\right]$ decay widths, as well as the ${\cal R}_{D^{(*)}}=
\Gamma\left(\bar B\to D^{(*)}\tau\bar\nu_\tau\right)
/\Gamma\left[\bar B\to D^{(*)}\, e(\mu)\bar\nu_{e(\mu)}\right]$ ratios,
 obtained in each of the
 physics scenarios considered in the figures. The corresponding results for the 
$\widehat F^d_{1,\,2}$
 angular moments are given  in
Tables~\ref{tab:hatfsDDstar1} and \ref{tab:hatfsDDstar2}, for the lepton and 
hadron modes, respectively. 

\bibliography{B2Dbib}

\begin{thebibliography}{71}%
\makeatletter
\providecommand \@ifxundefined [1]{%
 \@ifx{#1\undefined}
}%
\providecommand \@ifnum [1]{%
 \ifnum #1\expandafter \@firstoftwo
 \else \expandafter \@secondoftwo
 \fi
}%
\providecommand \@ifx [1]{%
 \ifx #1\expandafter \@firstoftwo
 \else \expandafter \@secondoftwo
 \fi
}%
\providecommand \natexlab [1]{#1}%
\providecommand \enquote  [1]{``#1''}%
\providecommand \bibnamefont  [1]{#1}%
\providecommand \bibfnamefont [1]{#1}%
\providecommand \citenamefont [1]{#1}%
\providecommand \href@noop [0]{\@secondoftwo}%
\providecommand \href [0]{\begingroup \@sanitize@url \@href}%
\providecommand \@href[1]{\@@startlink{#1}\@@href}%
\providecommand \@@href[1]{\endgroup#1\@@endlink}%
\providecommand \@sanitize@url [0]{\catcode `\\12\catcode `\$12\catcode
  `\&12\catcode `\#12\catcode `\^12\catcode `\_12\catcode `\%12\relax}%
\providecommand \@@startlink[1]{}%
\providecommand \@@endlink[0]{}%
\providecommand \url  [0]{\begingroup\@sanitize@url \@url }%
\providecommand \@url [1]{\endgroup\@href {#1}{\urlprefix }}%
\providecommand \urlprefix  [0]{URL }%
\providecommand \Eprint [0]{\href }%
\providecommand \doibase [0]{http://dx.doi.org/}%
\providecommand \selectlanguage [0]{\@gobble}%
\providecommand \bibinfo  [0]{\@secondoftwo}%
\providecommand \bibfield  [0]{\@secondoftwo}%
\providecommand \translation [1]{[#1]}%
\providecommand \BibitemOpen [0]{}%
\providecommand \bibitemStop [0]{}%
\providecommand \bibitemNoStop [0]{.\EOS\space}%
\providecommand \EOS [0]{\spacefactor3000\relax}%
\providecommand \BibitemShut  [1]{\csname bibitem#1\endcsname}%
\let\auto@bib@innerbib\@empty
\bibitem [{\citenamefont {Lees}\ \emph {et~al.}(2012)\citenamefont {Lees} \emph
  {et~al.}}]{BaBar:2012obs}%
  \BibitemOpen
  \bibfield  {author} {\bibinfo {author} {\bibfnamefont {J.~P.}\ \bibnamefont
  {Lees}} \emph {et~al.} (\bibinfo {collaboration} {BaBar}),\ }\href {\doibase
  10.1103/PhysRevLett.109.101802} {\bibfield  {journal} {\bibinfo  {journal}
  {Phys. Rev. Lett.}\ }\textbf {\bibinfo {volume} {109}},\ \bibinfo {pages}
  {101802} (\bibinfo {year} {2012})},\ \Eprint {http://arxiv.org/abs/1205.5442}
  {arXiv:1205.5442 [hep-ex]} \BibitemShut {NoStop}%
\bibitem [{\citenamefont {Lees}\ \emph {et~al.}(2013)\citenamefont {Lees} \emph
  {et~al.}}]{BaBar:2013mob}%
  \BibitemOpen
  \bibfield  {author} {\bibinfo {author} {\bibfnamefont {J.~P.}\ \bibnamefont
  {Lees}} \emph {et~al.} (\bibinfo {collaboration} {BaBar}),\ }\href {\doibase
  10.1103/PhysRevD.88.072012} {\bibfield  {journal} {\bibinfo  {journal} {Phys.
  Rev. D}\ }\textbf {\bibinfo {volume} {88}},\ \bibinfo {pages} {072012}
  (\bibinfo {year} {2013})},\ \Eprint {http://arxiv.org/abs/1303.0571}
  {arXiv:1303.0571 [hep-ex]} \BibitemShut {NoStop}%
\bibitem [{\citenamefont {Huschle}\ \emph {et~al.}(2015)\citenamefont {Huschle}
  \emph {et~al.}}]{Belle:2015qfa}%
  \BibitemOpen
  \bibfield  {author} {\bibinfo {author} {\bibfnamefont {M.}~\bibnamefont
  {Huschle}} \emph {et~al.} (\bibinfo {collaboration} {Belle}),\ }\href
  {\doibase 10.1103/PhysRevD.92.072014} {\bibfield  {journal} {\bibinfo
  {journal} {Phys. Rev. D}\ }\textbf {\bibinfo {volume} {92}},\ \bibinfo
  {pages} {072014} (\bibinfo {year} {2015})},\ \Eprint
  {http://arxiv.org/abs/1507.03233} {arXiv:1507.03233 [hep-ex]} \BibitemShut
  {NoStop}%
\bibitem [{\citenamefont {Sato}\ \emph {et~al.}(2016)\citenamefont {Sato} \emph
  {et~al.}}]{Belle:2016ure}%
  \BibitemOpen
  \bibfield  {author} {\bibinfo {author} {\bibfnamefont {Y.}~\bibnamefont
  {Sato}} \emph {et~al.} (\bibinfo {collaboration} {Belle}),\ }\href {\doibase
  10.1103/PhysRevD.94.072007} {\bibfield  {journal} {\bibinfo  {journal} {Phys.
  Rev. D}\ }\textbf {\bibinfo {volume} {94}},\ \bibinfo {pages} {072007}
  (\bibinfo {year} {2016})},\ \Eprint {http://arxiv.org/abs/1607.07923}
  {arXiv:1607.07923 [hep-ex]} \BibitemShut {NoStop}%
\bibitem [{\citenamefont {Hirose}\ \emph {et~al.}(2017)\citenamefont {Hirose}
  \emph {et~al.}}]{Belle:2016dyj}%
  \BibitemOpen
  \bibfield  {author} {\bibinfo {author} {\bibfnamefont {S.}~\bibnamefont
  {Hirose}} \emph {et~al.} (\bibinfo {collaboration} {Belle}),\ }\href
  {\doibase 10.1103/PhysRevLett.118.211801} {\bibfield  {journal} {\bibinfo
  {journal} {Phys. Rev. Lett.}\ }\textbf {\bibinfo {volume} {118}},\ \bibinfo
  {pages} {211801} (\bibinfo {year} {2017})},\ \Eprint
  {http://arxiv.org/abs/1612.00529} {arXiv:1612.00529 [hep-ex]} \BibitemShut
  {NoStop}%
\bibitem [{\citenamefont {Caria}\ \emph {et~al.}(2020)\citenamefont {Caria}
  \emph {et~al.}}]{Belle:2019rba}%
  \BibitemOpen
  \bibfield  {author} {\bibinfo {author} {\bibfnamefont {G.}~\bibnamefont
  {Caria}} \emph {et~al.} (\bibinfo {collaboration} {Belle}),\ }\href {\doibase
  10.1103/PhysRevLett.124.161803} {\bibfield  {journal} {\bibinfo  {journal}
  {Phys. Rev. Lett.}\ }\textbf {\bibinfo {volume} {124}},\ \bibinfo {pages}
  {161803} (\bibinfo {year} {2020})},\ \Eprint
  {http://arxiv.org/abs/1910.05864} {arXiv:1910.05864 [hep-ex]} \BibitemShut
  {NoStop}%
\bibitem [{\citenamefont {Aaij}\ \emph {et~al.}(2015)\citenamefont {Aaij} \emph
  {et~al.}}]{LHCb:2015gmp}%
  \BibitemOpen
  \bibfield  {author} {\bibinfo {author} {\bibfnamefont {R.}~\bibnamefont
  {Aaij}} \emph {et~al.} (\bibinfo {collaboration} {LHCb}),\ }\href {\doibase
  10.1103/PhysRevLett.115.111803} {\bibfield  {journal} {\bibinfo  {journal}
  {Phys. Rev. Lett.}\ }\textbf {\bibinfo {volume} {115}},\ \bibinfo {pages}
  {111803} (\bibinfo {year} {2015})},\ \bibinfo {note} {[Erratum:
  Phys.Rev.Lett. 115, 159901 (2015)]},\ \Eprint
  {http://arxiv.org/abs/1506.08614} {arXiv:1506.08614 [hep-ex]} \BibitemShut
  {NoStop}%
\bibitem [{\citenamefont {Aaij}\ \emph
  {et~al.}(2018{\natexlab{a}})\citenamefont {Aaij} \emph
  {et~al.}}]{LHCb:2017smo}%
  \BibitemOpen
  \bibfield  {author} {\bibinfo {author} {\bibfnamefont {R.}~\bibnamefont
  {Aaij}} \emph {et~al.} (\bibinfo {collaboration} {LHCb}),\ }\href {\doibase
  10.1103/PhysRevLett.120.171802} {\bibfield  {journal} {\bibinfo  {journal}
  {Phys. Rev. Lett.}\ }\textbf {\bibinfo {volume} {120}},\ \bibinfo {pages}
  {171802} (\bibinfo {year} {2018}{\natexlab{a}})},\ \Eprint
  {http://arxiv.org/abs/1708.08856} {arXiv:1708.08856 [hep-ex]} \BibitemShut
  {NoStop}%
\bibitem [{\citenamefont {Aaij}\ \emph
  {et~al.}(2018{\natexlab{b}})\citenamefont {Aaij} \emph
  {et~al.}}]{LHCb:2017rln}%
  \BibitemOpen
  \bibfield  {author} {\bibinfo {author} {\bibfnamefont {R.}~\bibnamefont
  {Aaij}} \emph {et~al.} (\bibinfo {collaboration} {LHCb}),\ }\href {\doibase
  10.1103/PhysRevD.97.072013} {\bibfield  {journal} {\bibinfo  {journal} {Phys.
  Rev. D}\ }\textbf {\bibinfo {volume} {97}},\ \bibinfo {pages} {072013}
  (\bibinfo {year} {2018}{\natexlab{b}})},\ \Eprint
  {http://arxiv.org/abs/1711.02505} {arXiv:1711.02505 [hep-ex]} \BibitemShut
  {NoStop}%
\bibitem [{\citenamefont {Amhis}\ \emph {et~al.}(2021)\citenamefont {Amhis}
  \emph {et~al.}}]{HFLAV:2019otj}%
  \BibitemOpen
  \bibfield  {author} {\bibinfo {author} {\bibfnamefont {Y.~S.}\ \bibnamefont
  {Amhis}} \emph {et~al.} (\bibinfo {collaboration} {HFLAV}),\ }\href {\doibase
  10.1140/epjc/s10052-020-8156-7} {\bibfield  {journal} {\bibinfo  {journal}
  {Eur. Phys. J. C}\ }\textbf {\bibinfo {volume} {81}},\ \bibinfo {pages} {226}
  (\bibinfo {year} {2021})},\ \Eprint {http://arxiv.org/abs/1909.12524}
  {arXiv:1909.12524 [hep-ex]} \BibitemShut {NoStop}%
\bibitem [{\citenamefont {Aaij}\ \emph
  {et~al.}(2018{\natexlab{c}})\citenamefont {Aaij} \emph
  {et~al.}}]{LHCb:2017vlu}%
  \BibitemOpen
  \bibfield  {author} {\bibinfo {author} {\bibfnamefont {R.}~\bibnamefont
  {Aaij}} \emph {et~al.} (\bibinfo {collaboration} {LHCb}),\ }\href {\doibase
  10.1103/PhysRevLett.120.121801} {\bibfield  {journal} {\bibinfo  {journal}
  {Phys. Rev. Lett.}\ }\textbf {\bibinfo {volume} {120}},\ \bibinfo {pages}
  {121801} (\bibinfo {year} {2018}{\natexlab{c}})},\ \Eprint
  {http://arxiv.org/abs/1711.05623} {arXiv:1711.05623 [hep-ex]} \BibitemShut
  {NoStop}%
\bibitem [{\citenamefont {Anisimov}\ \emph {et~al.}(1999)\citenamefont
  {Anisimov}, \citenamefont {Narodetsky}, \citenamefont {Semay},\ and\
  \citenamefont {Silvestre-Brac}}]{Anisimov:1998uk}%
  \BibitemOpen
  \bibfield  {author} {\bibinfo {author} {\bibfnamefont {A.~Y.}\ \bibnamefont
  {Anisimov}}, \bibinfo {author} {\bibfnamefont {I.~M.}\ \bibnamefont
  {Narodetsky}}, \bibinfo {author} {\bibfnamefont {C.}~\bibnamefont {Semay}}, \
  and\ \bibinfo {author} {\bibfnamefont {B.}~\bibnamefont {Silvestre-Brac}},\
  }\href {\doibase 10.1016/S0370-2693(99)00273-7} {\bibfield  {journal}
  {\bibinfo  {journal} {Phys. Lett.}\ }\textbf {\bibinfo {volume} {B452}},\
  \bibinfo {pages} {129} (\bibinfo {year} {1999})},\ \Eprint
  {http://arxiv.org/abs/hep-ph/9812514} {arXiv:hep-ph/9812514 [hep-ph]}
  \BibitemShut {NoStop}%
\bibitem [{\citenamefont {Ivanov}\ \emph {et~al.}(2006)\citenamefont {Ivanov},
  \citenamefont {Korner},\ and\ \citenamefont {Santorelli}}]{Ivanov:2006ni}%
  \BibitemOpen
  \bibfield  {author} {\bibinfo {author} {\bibfnamefont {M.~A.}\ \bibnamefont
  {Ivanov}}, \bibinfo {author} {\bibfnamefont {J.~G.}\ \bibnamefont {Korner}},
  \ and\ \bibinfo {author} {\bibfnamefont {P.}~\bibnamefont {Santorelli}},\
  }\href {\doibase 10.1103/PhysRevD.73.054024} {\bibfield  {journal} {\bibinfo
  {journal} {Phys. Rev.}\ }\textbf {\bibinfo {volume} {D73}},\ \bibinfo {pages}
  {054024} (\bibinfo {year} {2006})},\ \Eprint
  {http://arxiv.org/abs/hep-ph/0602050} {arXiv:hep-ph/0602050 [hep-ph]}
  \BibitemShut {NoStop}%
\bibitem [{\citenamefont {Hern\'andez}\ \emph {et~al.}(2006)\citenamefont
  {Hern\'andez}, \citenamefont {Nieves},\ and\ \citenamefont
  {Verde-Velasco}}]{Hernandez:2006gt}%
  \BibitemOpen
  \bibfield  {author} {\bibinfo {author} {\bibfnamefont {E.}~\bibnamefont
  {Hern\'andez}}, \bibinfo {author} {\bibfnamefont {J.}~\bibnamefont {Nieves}},
  \ and\ \bibinfo {author} {\bibfnamefont {J.}~\bibnamefont {Verde-Velasco}},\
  }\href {\doibase 10.1103/PhysRevD.74.074008} {\bibfield  {journal} {\bibinfo
  {journal} {Phys. Rev. D}\ }\textbf {\bibinfo {volume} {74}},\ \bibinfo
  {pages} {074008} (\bibinfo {year} {2006})},\ \Eprint
  {http://arxiv.org/abs/hep-ph/0607150} {arXiv:hep-ph/0607150} \BibitemShut
  {NoStop}%
\bibitem [{\citenamefont {Huang}\ and\ \citenamefont
  {Zuo}(2007)}]{Huang:2007kb}%
  \BibitemOpen
  \bibfield  {author} {\bibinfo {author} {\bibfnamefont {T.}~\bibnamefont
  {Huang}}\ and\ \bibinfo {author} {\bibfnamefont {F.}~\bibnamefont {Zuo}},\
  }\href {\doibase 10.1140/epjc/s10052-007-0333-4} {\bibfield  {journal}
  {\bibinfo  {journal} {Eur. Phys. J.}\ }\textbf {\bibinfo {volume} {C51}},\
  \bibinfo {pages} {833} (\bibinfo {year} {2007})},\ \Eprint
  {http://arxiv.org/abs/hep-ph/0702147} {arXiv:hep-ph/0702147 [HEP-PH]}
  \BibitemShut {NoStop}%
\bibitem [{\citenamefont {Wang}\ \emph {et~al.}(2009)\citenamefont {Wang},
  \citenamefont {Shen},\ and\ \citenamefont {Lu}}]{Wang:2008xt}%
  \BibitemOpen
  \bibfield  {author} {\bibinfo {author} {\bibfnamefont {W.}~\bibnamefont
  {Wang}}, \bibinfo {author} {\bibfnamefont {Y.-L.}\ \bibnamefont {Shen}}, \
  and\ \bibinfo {author} {\bibfnamefont {C.-D.}\ \bibnamefont {Lu}},\ }\href
  {\doibase 10.1103/PhysRevD.79.054012} {\bibfield  {journal} {\bibinfo
  {journal} {Phys. Rev.}\ }\textbf {\bibinfo {volume} {D79}},\ \bibinfo {pages}
  {054012} (\bibinfo {year} {2009})},\ \Eprint {http://arxiv.org/abs/0811.3748}
  {arXiv:0811.3748 [hep-ph]} \BibitemShut {NoStop}%
\bibitem [{\citenamefont {Wang}\ \emph {et~al.}(2013)\citenamefont {Wang},
  \citenamefont {Fan},\ and\ \citenamefont {Xiao}}]{Wen-Fei:2013uea}%
  \BibitemOpen
  \bibfield  {author} {\bibinfo {author} {\bibfnamefont {W.-F.}\ \bibnamefont
  {Wang}}, \bibinfo {author} {\bibfnamefont {Y.-Y.}\ \bibnamefont {Fan}}, \
  and\ \bibinfo {author} {\bibfnamefont {Z.-J.}\ \bibnamefont {Xiao}},\ }\href
  {\doibase 10.1088/1674-1137/37/9/093102} {\bibfield  {journal} {\bibinfo
  {journal} {Chin. Phys.}\ }\textbf {\bibinfo {volume} {C37}},\ \bibinfo
  {pages} {093102} (\bibinfo {year} {2013})},\ \Eprint
  {http://arxiv.org/abs/1212.5903} {arXiv:1212.5903 [hep-ph]} \BibitemShut
  {NoStop}%
\bibitem [{\citenamefont {Watanabe}(2018)}]{Watanabe:2017mip}%
  \BibitemOpen
  \bibfield  {author} {\bibinfo {author} {\bibfnamefont {R.}~\bibnamefont
  {Watanabe}},\ }\href {\doibase 10.1016/j.physletb.2017.11.016} {\bibfield
  {journal} {\bibinfo  {journal} {Phys. Lett. B}\ }\textbf {\bibinfo {volume}
  {776}},\ \bibinfo {pages} {5} (\bibinfo {year} {2018})},\ \Eprint
  {http://arxiv.org/abs/1709.08644} {arXiv:1709.08644 [hep-ph]} \BibitemShut
  {NoStop}%
\bibitem [{\citenamefont {Issadykov}\ and\ \citenamefont
  {Ivanov}(2018)}]{Issadykov:2018myx}%
  \BibitemOpen
  \bibfield  {author} {\bibinfo {author} {\bibfnamefont {A.}~\bibnamefont
  {Issadykov}}\ and\ \bibinfo {author} {\bibfnamefont {M.~A.}\ \bibnamefont
  {Ivanov}},\ }\href {\doibase 10.1016/j.physletb.2018.06.056} {\bibfield
  {journal} {\bibinfo  {journal} {Phys. Lett.}\ }\textbf {\bibinfo {volume}
  {B783}},\ \bibinfo {pages} {178} (\bibinfo {year} {2018})},\ \Eprint
  {http://arxiv.org/abs/1804.00472} {arXiv:1804.00472 [hep-ph]} \BibitemShut
  {NoStop}%
\bibitem [{\citenamefont {Tran}\ \emph {et~al.}(2018)\citenamefont {Tran},
  \citenamefont {Ivanov}, \citenamefont {K{\"o}rner},\ and\ \citenamefont
  {Santorelli}}]{Tran:2018kuv}%
  \BibitemOpen
  \bibfield  {author} {\bibinfo {author} {\bibfnamefont {C.-T.}\ \bibnamefont
  {Tran}}, \bibinfo {author} {\bibfnamefont {M.~A.}\ \bibnamefont {Ivanov}},
  \bibinfo {author} {\bibfnamefont {J.~G.}\ \bibnamefont {K{\"o}rner}}, \ and\
  \bibinfo {author} {\bibfnamefont {P.}~\bibnamefont {Santorelli}},\ }\href
  {\doibase 10.1103/PhysRevD.97.054014} {\bibfield  {journal} {\bibinfo
  {journal} {Phys. Rev.}\ }\textbf {\bibinfo {volume} {D97}},\ \bibinfo {pages}
  {054014} (\bibinfo {year} {2018})},\ \Eprint
  {http://arxiv.org/abs/1801.06927} {arXiv:1801.06927 [hep-ph]} \BibitemShut
  {NoStop}%
\bibitem [{\citenamefont {Hu}\ \emph {et~al.}(2020)\citenamefont {Hu},
  \citenamefont {Jin},\ and\ \citenamefont {Xiao}}]{Hu:2019qcn}%
  \BibitemOpen
  \bibfield  {author} {\bibinfo {author} {\bibfnamefont {X.-Q.}\ \bibnamefont
  {Hu}}, \bibinfo {author} {\bibfnamefont {S.-P.}\ \bibnamefont {Jin}}, \ and\
  \bibinfo {author} {\bibfnamefont {Z.-J.}\ \bibnamefont {Xiao}},\ }\href
  {\doibase 10.1088/1674-1137/44/2/023104} {\bibfield  {journal} {\bibinfo
  {journal} {Chin. Phys.}\ }\textbf {\bibinfo {volume} {C44}},\ \bibinfo
  {pages} {023104} (\bibinfo {year} {2020})},\ \Eprint
  {http://arxiv.org/abs/1904.07530} {arXiv:1904.07530 [hep-ph]} \BibitemShut
  {NoStop}%
\bibitem [{\citenamefont {Leljak}\ \emph {et~al.}(2019)\citenamefont {Leljak},
  \citenamefont {Melic},\ and\ \citenamefont {Patra}}]{Leljak:2019eyw}%
  \BibitemOpen
  \bibfield  {author} {\bibinfo {author} {\bibfnamefont {D.}~\bibnamefont
  {Leljak}}, \bibinfo {author} {\bibfnamefont {B.}~\bibnamefont {Melic}}, \
  and\ \bibinfo {author} {\bibfnamefont {M.}~\bibnamefont {Patra}},\ }\href
  {\doibase 10.1007/JHEP05(2019)094} {\bibfield  {journal} {\bibinfo  {journal}
  {JHEP}\ }\textbf {\bibinfo {volume} {05}},\ \bibinfo {pages} {094} (\bibinfo
  {year} {2019})},\ \Eprint {http://arxiv.org/abs/1901.08368} {arXiv:1901.08368
  [hep-ph]} \BibitemShut {NoStop}%
\bibitem [{\citenamefont {Azizi}\ \emph {et~al.}(2019)\citenamefont {Azizi},
  \citenamefont {Sarac},\ and\ \citenamefont {Sundu}}]{Azizi:2019aaf}%
  \BibitemOpen
  \bibfield  {author} {\bibinfo {author} {\bibfnamefont {K.}~\bibnamefont
  {Azizi}}, \bibinfo {author} {\bibfnamefont {Y.}~\bibnamefont {Sarac}}, \ and\
  \bibinfo {author} {\bibfnamefont {H.}~\bibnamefont {Sundu}},\ }\href
  {\doibase 10.1103/PhysRevD.99.113004} {\bibfield  {journal} {\bibinfo
  {journal} {Phys. Rev.}\ }\textbf {\bibinfo {volume} {D99}},\ \bibinfo {pages}
  {113004} (\bibinfo {year} {2019})},\ \Eprint
  {http://arxiv.org/abs/1904.08267} {arXiv:1904.08267 [hep-ph]} \BibitemShut
  {NoStop}%
\bibitem [{\citenamefont {Wang}\ and\ \citenamefont
  {Zhu}(2019)}]{Wang:2018duy}%
  \BibitemOpen
  \bibfield  {author} {\bibinfo {author} {\bibfnamefont {W.}~\bibnamefont
  {Wang}}\ and\ \bibinfo {author} {\bibfnamefont {R.}~\bibnamefont {Zhu}},\
  }\href {\doibase 10.1142/S0217751X19501951} {\bibfield  {journal} {\bibinfo
  {journal} {Int. J. Mod. Phys. A}\ }\textbf {\bibinfo {volume} {34}},\
  \bibinfo {pages} {1950195} (\bibinfo {year} {2019})},\ \Eprint
  {http://arxiv.org/abs/1808.10830} {arXiv:1808.10830 [hep-ph]} \BibitemShut
  {NoStop}%
\bibitem [{\citenamefont {Abdesselam}\ \emph {et~al.}(2019)\citenamefont
  {Abdesselam} \emph {et~al.}}]{Belle:2019ewo}%
  \BibitemOpen
  \bibfield  {author} {\bibinfo {author} {\bibfnamefont {A.}~\bibnamefont
  {Abdesselam}} \emph {et~al.} (\bibinfo {collaboration} {Belle}),\ }in\
  \href@noop {} {\emph {\bibinfo {booktitle} {{10th International Workshop on
  the CKM Unitarity Triangle}}}}\ (\bibinfo {year} {2019})\ \Eprint
  {http://arxiv.org/abs/1903.03102} {arXiv:1903.03102 [hep-ex]} \BibitemShut
  {NoStop}%
\bibitem [{\citenamefont {Alonso}\ \emph
  {et~al.}(2017{\natexlab{a}})\citenamefont {Alonso}, \citenamefont
  {Grinstein},\ and\ \citenamefont {Martin~Camalich}}]{Alonso:2016oyd}%
  \BibitemOpen
  \bibfield  {author} {\bibinfo {author} {\bibfnamefont {R.}~\bibnamefont
  {Alonso}}, \bibinfo {author} {\bibfnamefont {B.}~\bibnamefont {Grinstein}}, \
  and\ \bibinfo {author} {\bibfnamefont {J.}~\bibnamefont {Martin~Camalich}},\
  }\href {\doibase 10.1103/PhysRevLett.118.081802} {\bibfield  {journal}
  {\bibinfo  {journal} {Phys. Rev. Lett.}\ }\textbf {\bibinfo {volume} {118}},\
  \bibinfo {pages} {081802} (\bibinfo {year} {2017}{\natexlab{a}})},\ \Eprint
  {http://arxiv.org/abs/1611.06676} {arXiv:1611.06676 [hep-ph]} \BibitemShut
  {NoStop}%
\bibitem [{\citenamefont {Aaij}\ \emph {et~al.}(2022)\citenamefont {Aaij} \emph
  {et~al.}}]{LHCb:2022piu}%
  \BibitemOpen
  \bibfield  {author} {\bibinfo {author} {\bibfnamefont {R.}~\bibnamefont
  {Aaij}} \emph {et~al.} (\bibinfo {collaboration} {LHCb}),\ }\href@noop {} {\
  (\bibinfo {year} {2022})},\ \Eprint {http://arxiv.org/abs/2201.03497}
  {arXiv:2201.03497 [hep-ex]} \BibitemShut {NoStop}%
\bibitem [{\citenamefont {Detmold}\ \emph {et~al.}(2015)\citenamefont
  {Detmold}, \citenamefont {Lehner},\ and\ \citenamefont
  {Meinel}}]{Detmold:2015aaa}%
  \BibitemOpen
  \bibfield  {author} {\bibinfo {author} {\bibfnamefont {W.}~\bibnamefont
  {Detmold}}, \bibinfo {author} {\bibfnamefont {C.}~\bibnamefont {Lehner}}, \
  and\ \bibinfo {author} {\bibfnamefont {S.}~\bibnamefont {Meinel}},\ }\href
  {\doibase 10.1103/PhysRevD.92.034503} {\bibfield  {journal} {\bibinfo
  {journal} {Phys. Rev.}\ }\textbf {\bibinfo {volume} {D92}},\ \bibinfo {pages}
  {034503} (\bibinfo {year} {2015})},\ \Eprint
  {http://arxiv.org/abs/1503.01421} {arXiv:1503.01421 [hep-lat]} \BibitemShut
  {NoStop}%
\bibitem [{Mar()}]{Marco}%
  \BibitemOpen
  \href@noop {} {\bibinfo  {journal} {Marco Pappagallo (LHCB deputy physics
  coordinator) private communication}\ }\BibitemShut {NoStop}%
\bibitem [{\citenamefont {Nierste}\ \emph {et~al.}(2008)\citenamefont
  {Nierste}, \citenamefont {Trine},\ and\ \citenamefont
  {Westhoff}}]{Nierste:2008qe}%
  \BibitemOpen
\bibfield  {journal} {  }\bibfield  {author} {\bibinfo {author} {\bibfnamefont
  {U.}~\bibnamefont {Nierste}}, \bibinfo {author} {\bibfnamefont
  {S.}~\bibnamefont {Trine}}, \ and\ \bibinfo {author} {\bibfnamefont
  {S.}~\bibnamefont {Westhoff}},\ }\href {\doibase 10.1103/PhysRevD.78.015006}
  {\bibfield  {journal} {\bibinfo  {journal} {Phys. Rev. D}\ }\textbf {\bibinfo
  {volume} {78}},\ \bibinfo {pages} {015006} (\bibinfo {year} {2008})},\
  \Eprint {http://arxiv.org/abs/0801.4938} {arXiv:0801.4938 [hep-ph]}
  \BibitemShut {NoStop}%
\bibitem [{\citenamefont {Tanaka}\ and\ \citenamefont
  {Watanabe}(2013)}]{Tanaka:2012nw}%
  \BibitemOpen
  \bibfield  {author} {\bibinfo {author} {\bibfnamefont {M.}~\bibnamefont
  {Tanaka}}\ and\ \bibinfo {author} {\bibfnamefont {R.}~\bibnamefont
  {Watanabe}},\ }\href {\doibase 10.1103/PhysRevD.87.034028} {\bibfield
  {journal} {\bibinfo  {journal} {Phys. Rev. D}\ }\textbf {\bibinfo {volume}
  {87}},\ \bibinfo {pages} {034028} (\bibinfo {year} {2013})},\ \Eprint
  {http://arxiv.org/abs/1212.1878} {arXiv:1212.1878 [hep-ph]} \BibitemShut
  {NoStop}%
\bibitem [{\citenamefont {Fajfer}\ \emph {et~al.}(2012)\citenamefont {Fajfer},
  \citenamefont {Kamenik},\ and\ \citenamefont {Nisandzic}}]{Fajfer:2012vx}%
  \BibitemOpen
  \bibfield  {author} {\bibinfo {author} {\bibfnamefont {S.}~\bibnamefont
  {Fajfer}}, \bibinfo {author} {\bibfnamefont {J.~F.}\ \bibnamefont {Kamenik}},
  \ and\ \bibinfo {author} {\bibfnamefont {I.}~\bibnamefont {Nisandzic}},\
  }\href {\doibase 10.1103/PhysRevD.85.094025} {\bibfield  {journal} {\bibinfo
  {journal} {Phys. Rev.}\ }\textbf {\bibinfo {volume} {D85}},\ \bibinfo {pages}
  {094025} (\bibinfo {year} {2012})},\ \Eprint {http://arxiv.org/abs/1203.2654}
  {arXiv:1203.2654 [hep-ph]} \BibitemShut {NoStop}%
\bibitem [{\citenamefont {Duraisamy}\ and\ \citenamefont
  {Datta}(2013)}]{Duraisamy:2013pia}%
  \BibitemOpen
  \bibfield  {author} {\bibinfo {author} {\bibfnamefont {M.}~\bibnamefont
  {Duraisamy}}\ and\ \bibinfo {author} {\bibfnamefont {A.}~\bibnamefont
  {Datta}},\ }\href {\doibase 10.1007/JHEP09(2013)059} {\bibfield  {journal}
  {\bibinfo  {journal} {JHEP}\ }\textbf {\bibinfo {volume} {09}},\ \bibinfo
  {pages} {059} (\bibinfo {year} {2013})},\ \Eprint
  {http://arxiv.org/abs/1302.7031} {arXiv:1302.7031 [hep-ph]} \BibitemShut
  {NoStop}%
\bibitem [{\citenamefont {Duraisamy}\ \emph {et~al.}(2014)\citenamefont
  {Duraisamy}, \citenamefont {Sharma},\ and\ \citenamefont
  {Datta}}]{Duraisamy:2014sna}%
  \BibitemOpen
  \bibfield  {author} {\bibinfo {author} {\bibfnamefont {M.}~\bibnamefont
  {Duraisamy}}, \bibinfo {author} {\bibfnamefont {P.}~\bibnamefont {Sharma}}, \
  and\ \bibinfo {author} {\bibfnamefont {A.}~\bibnamefont {Datta}},\ }\href
  {\doibase 10.1103/PhysRevD.90.074013} {\bibfield  {journal} {\bibinfo
  {journal} {Phys. Rev. D}\ }\textbf {\bibinfo {volume} {90}},\ \bibinfo
  {pages} {074013} (\bibinfo {year} {2014})},\ \Eprint
  {http://arxiv.org/abs/1405.3719} {arXiv:1405.3719 [hep-ph]} \BibitemShut
  {NoStop}%
\bibitem [{\citenamefont {Becirevic}\ \emph {et~al.}(2019)\citenamefont
  {Becirevic}, \citenamefont {Fajfer}, \citenamefont {Nisandzic},\ and\
  \citenamefont {Tayduganov}}]{Becirevic:2016hea}%
  \BibitemOpen
  \bibfield  {author} {\bibinfo {author} {\bibfnamefont {D.}~\bibnamefont
  {Becirevic}}, \bibinfo {author} {\bibfnamefont {S.}~\bibnamefont {Fajfer}},
  \bibinfo {author} {\bibfnamefont {I.}~\bibnamefont {Nisandzic}}, \ and\
  \bibinfo {author} {\bibfnamefont {A.}~\bibnamefont {Tayduganov}},\ }\href
  {\doibase 10.1016/j.nuclphysb.2019.114707} {\bibfield  {journal} {\bibinfo
  {journal} {Nucl. Phys. B}\ }\textbf {\bibinfo {volume} {946}},\ \bibinfo
  {pages} {114707} (\bibinfo {year} {2019})},\ \Eprint
  {http://arxiv.org/abs/1602.03030} {arXiv:1602.03030 [hep-ph]} \BibitemShut
  {NoStop}%
\bibitem [{\citenamefont {Ligeti}\ \emph {et~al.}(2017)\citenamefont {Ligeti},
  \citenamefont {Papucci},\ and\ \citenamefont {Robinson}}]{Ligeti:2016npd}%
  \BibitemOpen
  \bibfield  {author} {\bibinfo {author} {\bibfnamefont {Z.}~\bibnamefont
  {Ligeti}}, \bibinfo {author} {\bibfnamefont {M.}~\bibnamefont {Papucci}}, \
  and\ \bibinfo {author} {\bibfnamefont {D.~J.}\ \bibnamefont {Robinson}},\
  }\href {\doibase 10.1007/JHEP01(2017)083} {\bibfield  {journal} {\bibinfo
  {journal} {JHEP}\ }\textbf {\bibinfo {volume} {01}},\ \bibinfo {pages} {083}
  (\bibinfo {year} {2017})},\ \Eprint {http://arxiv.org/abs/1610.02045}
  {arXiv:1610.02045 [hep-ph]} \BibitemShut {NoStop}%
\bibitem [{\citenamefont {Ivanov}\ \emph {et~al.}(2017)\citenamefont {Ivanov},
  \citenamefont {K\"orner},\ and\ \citenamefont {Tran}}]{Ivanov:2017mrj}%
  \BibitemOpen
  \bibfield  {author} {\bibinfo {author} {\bibfnamefont {M.~A.}\ \bibnamefont
  {Ivanov}}, \bibinfo {author} {\bibfnamefont {J.~G.}\ \bibnamefont
  {K\"orner}}, \ and\ \bibinfo {author} {\bibfnamefont {C.-T.}\ \bibnamefont
  {Tran}},\ }\href {\doibase 10.1103/PhysRevD.95.036021} {\bibfield  {journal}
  {\bibinfo  {journal} {Phys. Rev. D}\ }\textbf {\bibinfo {volume} {95}},\
  \bibinfo {pages} {036021} (\bibinfo {year} {2017})},\ \Eprint
  {http://arxiv.org/abs/1701.02937} {arXiv:1701.02937 [hep-ph]} \BibitemShut
  {NoStop}%
\bibitem [{\citenamefont {Bernlochner}\ \emph {et~al.}(2017)\citenamefont
  {Bernlochner}, \citenamefont {Ligeti}, \citenamefont {Papucci},\ and\
  \citenamefont {Robinson}}]{Bernlochner:2017jka}%
  \BibitemOpen
  \bibfield  {author} {\bibinfo {author} {\bibfnamefont {F.~U.}\ \bibnamefont
  {Bernlochner}}, \bibinfo {author} {\bibfnamefont {Z.}~\bibnamefont {Ligeti}},
  \bibinfo {author} {\bibfnamefont {M.}~\bibnamefont {Papucci}}, \ and\
  \bibinfo {author} {\bibfnamefont {D.~J.}\ \bibnamefont {Robinson}},\ }\href
  {\doibase 10.1103/PhysRevD.95.115008, 10.1103/PhysRevD.97.059902} {\bibfield
  {journal} {\bibinfo  {journal} {Phys. Rev.}\ }\textbf {\bibinfo {volume}
  {D95}},\ \bibinfo {pages} {115008} (\bibinfo {year} {2017})},\ \bibinfo
  {note} {[erratum: Phys. Rev.D97,no.5,059902(2018)]},\ \Eprint
  {http://arxiv.org/abs/1703.05330} {arXiv:1703.05330 [hep-ph]} \BibitemShut
  {NoStop}%
\bibitem [{\citenamefont {Blanke}\ \emph
  {et~al.}(2019{\natexlab{a}})\citenamefont {Blanke}, \citenamefont
  {Crivellin}, \citenamefont {de~Boer}, \citenamefont {Kitahara}, \citenamefont
  {Moscati}, \citenamefont {Nierste},\ and\ \citenamefont
  {Ni\v{s}and\v{z}i\'c}}]{Blanke:2018yud}%
  \BibitemOpen
  \bibfield  {author} {\bibinfo {author} {\bibfnamefont {M.}~\bibnamefont
  {Blanke}}, \bibinfo {author} {\bibfnamefont {A.}~\bibnamefont {Crivellin}},
  \bibinfo {author} {\bibfnamefont {S.}~\bibnamefont {de~Boer}}, \bibinfo
  {author} {\bibfnamefont {T.}~\bibnamefont {Kitahara}}, \bibinfo {author}
  {\bibfnamefont {M.}~\bibnamefont {Moscati}}, \bibinfo {author} {\bibfnamefont
  {U.}~\bibnamefont {Nierste}}, \ and\ \bibinfo {author} {\bibfnamefont
  {I.}~\bibnamefont {Ni\v{s}and\v{z}i\'c}},\ }\href {\doibase
  10.1103/PhysRevD.99.075006} {\bibfield  {journal} {\bibinfo  {journal} {Phys.
  Rev.}\ }\textbf {\bibinfo {volume} {D99}},\ \bibinfo {pages} {075006}
  (\bibinfo {year} {2019}{\natexlab{a}})},\ \Eprint
  {http://arxiv.org/abs/1811.09603} {arXiv:1811.09603 [hep-ph]} \BibitemShut
  {NoStop}%
\bibitem [{\citenamefont {Bhattacharya}\ \emph {et~al.}(2019)\citenamefont
  {Bhattacharya}, \citenamefont {Nandi},\ and\ \citenamefont
  {Kumar~Patra}}]{Bhattacharya:2018kig}%
  \BibitemOpen
  \bibfield  {author} {\bibinfo {author} {\bibfnamefont {S.}~\bibnamefont
  {Bhattacharya}}, \bibinfo {author} {\bibfnamefont {S.}~\bibnamefont {Nandi}},
  \ and\ \bibinfo {author} {\bibfnamefont {S.}~\bibnamefont {Kumar~Patra}},\
  }\href {\doibase 10.1140/epjc/s10052-019-6767-7} {\bibfield  {journal}
  {\bibinfo  {journal} {Eur. Phys. J. C}\ }\textbf {\bibinfo {volume} {79}},\
  \bibinfo {pages} {268} (\bibinfo {year} {2019})},\ \Eprint
  {http://arxiv.org/abs/1805.08222} {arXiv:1805.08222 [hep-ph]} \BibitemShut
  {NoStop}%
\bibitem [{\citenamefont {Colangelo}\ and\ \citenamefont
  {De~Fazio}(2018)}]{Colangelo:2018cnj}%
  \BibitemOpen
  \bibfield  {author} {\bibinfo {author} {\bibfnamefont {P.}~\bibnamefont
  {Colangelo}}\ and\ \bibinfo {author} {\bibfnamefont {F.}~\bibnamefont
  {De~Fazio}},\ }\href {\doibase 10.1007/JHEP06(2018)082} {\bibfield  {journal}
  {\bibinfo  {journal} {JHEP}\ }\textbf {\bibinfo {volume} {06}},\ \bibinfo
  {pages} {082} (\bibinfo {year} {2018})},\ \Eprint
  {http://arxiv.org/abs/1801.10468} {arXiv:1801.10468 [hep-ph]} \BibitemShut
  {NoStop}%
\bibitem [{\citenamefont {Murgui}\ \emph {et~al.}(2019)\citenamefont {Murgui},
  \citenamefont {Pen\~uelas}, \citenamefont {Jung},\ and\ \citenamefont
  {Pich}}]{Murgui:2019czp}%
  \BibitemOpen
  \bibfield  {author} {\bibinfo {author} {\bibfnamefont {C.}~\bibnamefont
  {Murgui}}, \bibinfo {author} {\bibfnamefont {A.}~\bibnamefont {Pen\~uelas}},
  \bibinfo {author} {\bibfnamefont {M.}~\bibnamefont {Jung}}, \ and\ \bibinfo
  {author} {\bibfnamefont {A.}~\bibnamefont {Pich}},\ }\href {\doibase
  10.1007/JHEP09(2019)103} {\bibfield  {journal} {\bibinfo  {journal} {JHEP}\
  }\textbf {\bibinfo {volume} {09}},\ \bibinfo {pages} {103} (\bibinfo {year}
  {2019})},\ \Eprint {http://arxiv.org/abs/1904.09311} {arXiv:1904.09311
  [hep-ph]} \BibitemShut {NoStop}%
\bibitem [{\citenamefont {Shi}\ \emph {et~al.}(2019)\citenamefont {Shi},
  \citenamefont {Geng}, \citenamefont {Grinstein}, \citenamefont {J{\"a}ger},\
  and\ \citenamefont {Martin~Camalich}}]{Shi:2019gxi}%
  \BibitemOpen
  \bibfield  {author} {\bibinfo {author} {\bibfnamefont {R.-X.}\ \bibnamefont
  {Shi}}, \bibinfo {author} {\bibfnamefont {L.-S.}\ \bibnamefont {Geng}},
  \bibinfo {author} {\bibfnamefont {B.}~\bibnamefont {Grinstein}}, \bibinfo
  {author} {\bibfnamefont {S.}~\bibnamefont {J{\"a}ger}}, \ and\ \bibinfo
  {author} {\bibfnamefont {J.}~\bibnamefont {Martin~Camalich}},\ }\href
  {\doibase 10.1007/JHEP12(2019)065} {\bibfield  {journal} {\bibinfo  {journal}
  {JHEP}\ }\textbf {\bibinfo {volume} {12}},\ \bibinfo {pages} {065} (\bibinfo
  {year} {2019})},\ \Eprint {http://arxiv.org/abs/1905.08498} {arXiv:1905.08498
  [hep-ph]} \BibitemShut {NoStop}%
\bibitem [{\citenamefont {Alok}\ \emph {et~al.}(2020)\citenamefont {Alok},
  \citenamefont {Kumar}, \citenamefont {Kumbhakar},\ and\ \citenamefont
  {Uma~Sankar}}]{Alok:2019uqc}%
  \BibitemOpen
  \bibfield  {author} {\bibinfo {author} {\bibfnamefont {A.~K.}\ \bibnamefont
  {Alok}}, \bibinfo {author} {\bibfnamefont {D.}~\bibnamefont {Kumar}},
  \bibinfo {author} {\bibfnamefont {S.}~\bibnamefont {Kumbhakar}}, \ and\
  \bibinfo {author} {\bibfnamefont {S.}~\bibnamefont {Uma~Sankar}},\ }\href
  {\doibase 10.1016/j.nuclphysb.2020.114957} {\bibfield  {journal} {\bibinfo
  {journal} {Nucl. Phys. B}\ }\textbf {\bibinfo {volume} {953}},\ \bibinfo
  {pages} {114957} (\bibinfo {year} {2020})},\ \Eprint
  {http://arxiv.org/abs/1903.10486} {arXiv:1903.10486 [hep-ph]} \BibitemShut
  {NoStop}%
\bibitem [{\citenamefont {Mandal}\ \emph {et~al.}(2020)\citenamefont {Mandal},
  \citenamefont {Murgui}, \citenamefont {Pe\~nuelas},\ and\ \citenamefont
  {Pich}}]{Mandal:2020htr}%
  \BibitemOpen
  \bibfield  {author} {\bibinfo {author} {\bibfnamefont {R.}~\bibnamefont
  {Mandal}}, \bibinfo {author} {\bibfnamefont {C.}~\bibnamefont {Murgui}},
  \bibinfo {author} {\bibfnamefont {A.}~\bibnamefont {Pe\~nuelas}}, \ and\
  \bibinfo {author} {\bibfnamefont {A.}~\bibnamefont {Pich}},\ }\href {\doibase
  10.1007/JHEP08(2020)022} {\bibfield  {journal} {\bibinfo  {journal} {JHEP}\
  }\textbf {\bibinfo {volume} {08}},\ \bibinfo {pages} {022} (\bibinfo {year}
  {2020})},\ \Eprint {http://arxiv.org/abs/2004.06726} {arXiv:2004.06726
  [hep-ph]} \BibitemShut {NoStop}%
\bibitem [{\citenamefont {Kumbhakar}(2021)}]{Kumbhakar:2020jdz}%
  \BibitemOpen
  \bibfield  {author} {\bibinfo {author} {\bibfnamefont {S.}~\bibnamefont
  {Kumbhakar}},\ }\href {\doibase 10.1016/j.nuclphysb.2020.115297} {\bibfield
  {journal} {\bibinfo  {journal} {Nucl. Phys. B}\ }\textbf {\bibinfo {volume}
  {963}},\ \bibinfo {pages} {115297} (\bibinfo {year} {2021})},\ \Eprint
  {http://arxiv.org/abs/2007.08132} {arXiv:2007.08132 [hep-ph]} \BibitemShut
  {NoStop}%
\bibitem [{\citenamefont {Iguro}\ and\ \citenamefont
  {Watanabe}(2020)}]{Iguro:2020cpg}%
  \BibitemOpen
  \bibfield  {author} {\bibinfo {author} {\bibfnamefont {S.}~\bibnamefont
  {Iguro}}\ and\ \bibinfo {author} {\bibfnamefont {R.}~\bibnamefont
  {Watanabe}},\ }\href {\doibase 10.1007/JHEP08(2020)006} {\bibfield  {journal}
  {\bibinfo  {journal} {JHEP}\ }\textbf {\bibinfo {volume} {08}},\ \bibinfo
  {pages} {006} (\bibinfo {year} {2020})},\ \Eprint
  {http://arxiv.org/abs/2004.10208} {arXiv:2004.10208 [hep-ph]} \BibitemShut
  {NoStop}%
\bibitem [{\citenamefont {Bhattacharya}\ \emph {et~al.}(2020)\citenamefont
  {Bhattacharya}, \citenamefont {Datta}, \citenamefont {Kamali},\ and\
  \citenamefont {London}}]{Bhattacharya:2020lfm}%
  \BibitemOpen
  \bibfield  {author} {\bibinfo {author} {\bibfnamefont {B.}~\bibnamefont
  {Bhattacharya}}, \bibinfo {author} {\bibfnamefont {A.}~\bibnamefont {Datta}},
  \bibinfo {author} {\bibfnamefont {S.}~\bibnamefont {Kamali}}, \ and\ \bibinfo
  {author} {\bibfnamefont {D.}~\bibnamefont {London}},\ }\href {\doibase
  10.1007/JHEP07(2020)194} {\bibfield  {journal} {\bibinfo  {journal} {JHEP}\
  }\textbf {\bibinfo {volume} {07}},\ \bibinfo {pages} {194} (\bibinfo {year}
  {2020})},\ \Eprint {http://arxiv.org/abs/2005.03032} {arXiv:2005.03032
  [hep-ph]} \BibitemShut {NoStop}%
\bibitem [{\citenamefont {Penalva}\ \emph
  {et~al.}(2021{\natexlab{a}})\citenamefont {Penalva}, \citenamefont
  {Hern\'andez},\ and\ \citenamefont {Nieves}}]{Penalva:2021gef}%
  \BibitemOpen
  \bibfield  {author} {\bibinfo {author} {\bibfnamefont {N.}~\bibnamefont
  {Penalva}}, \bibinfo {author} {\bibfnamefont {E.}~\bibnamefont
  {Hern\'andez}}, \ and\ \bibinfo {author} {\bibfnamefont {J.}~\bibnamefont
  {Nieves}},\ }\href {\doibase 10.1007/JHEP06(2021)118} {\bibfield  {journal}
  {\bibinfo  {journal} {JHEP}\ }\textbf {\bibinfo {volume} {06}},\ \bibinfo
  {pages} {118} (\bibinfo {year} {2021}{\natexlab{a}})},\ \Eprint
  {http://arxiv.org/abs/2103.01857} {arXiv:2103.01857 [hep-ph]} \BibitemShut
  {NoStop}%
\bibitem [{\citenamefont {Penalva}\ \emph
  {et~al.}(2020{\natexlab{a}})\citenamefont {Penalva}, \citenamefont
  {Hern\'andez},\ and\ \citenamefont {Nieves}}]{Penalva:2020ftd}%
  \BibitemOpen
  \bibfield  {author} {\bibinfo {author} {\bibfnamefont {N.}~\bibnamefont
  {Penalva}}, \bibinfo {author} {\bibfnamefont {E.}~\bibnamefont
  {Hern\'andez}}, \ and\ \bibinfo {author} {\bibfnamefont {J.}~\bibnamefont
  {Nieves}},\ }\href {\doibase 10.1103/PhysRevD.102.096016} {\bibfield
  {journal} {\bibinfo  {journal} {Phys. Rev. D}\ }\textbf {\bibinfo {volume}
  {102}},\ \bibinfo {pages} {096016} (\bibinfo {year} {2020}{\natexlab{a}})},\
  \Eprint {http://arxiv.org/abs/2007.12590} {arXiv:2007.12590 [hep-ph]}
  \BibitemShut {NoStop}%
\bibitem [{\citenamefont {Dutta}\ and\ \citenamefont
  {Bhol}(2017)}]{Dutta:2017xmj}%
  \BibitemOpen
  \bibfield  {author} {\bibinfo {author} {\bibfnamefont {R.}~\bibnamefont
  {Dutta}}\ and\ \bibinfo {author} {\bibfnamefont {A.}~\bibnamefont {Bhol}},\
  }\href {\doibase 10.1103/PhysRevD.96.076001} {\bibfield  {journal} {\bibinfo
  {journal} {Phys. Rev. D}\ }\textbf {\bibinfo {volume} {96}},\ \bibinfo
  {pages} {076001} (\bibinfo {year} {2017})},\ \Eprint
  {http://arxiv.org/abs/1701.08598} {arXiv:1701.08598 [hep-ph]} \BibitemShut
  {NoStop}%
\bibitem [{\citenamefont {Harrison}\ \emph {et~al.}(2020)\citenamefont
  {Harrison}, \citenamefont {Davies},\ and\ \citenamefont
  {Lytle}}]{Harrison:2020nrv}%
  \BibitemOpen
  \bibfield  {author} {\bibinfo {author} {\bibfnamefont {J.}~\bibnamefont
  {Harrison}}, \bibinfo {author} {\bibfnamefont {C.~T.}\ \bibnamefont
  {Davies}}, \ and\ \bibinfo {author} {\bibfnamefont {A.}~\bibnamefont {Lytle}}
  (\bibinfo {collaboration} {LATTICE-HPQCD}),\ }\href {\doibase
  10.1103/PhysRevLett.125.222003} {\bibfield  {journal} {\bibinfo  {journal}
  {Phys. Rev. Lett.}\ }\textbf {\bibinfo {volume} {125}},\ \bibinfo {pages}
  {222003} (\bibinfo {year} {2020})},\ \Eprint
  {http://arxiv.org/abs/2007.06956} {arXiv:2007.06956 [hep-lat]} \BibitemShut
  {NoStop}%
\bibitem [{\citenamefont {Dutta}(2016)}]{Dutta:2015ueb}%
  \BibitemOpen
  \bibfield  {author} {\bibinfo {author} {\bibfnamefont {R.}~\bibnamefont
  {Dutta}},\ }\href {\doibase 10.1103/PhysRevD.93.054003} {\bibfield  {journal}
  {\bibinfo  {journal} {Phys. Rev. D}\ }\textbf {\bibinfo {volume} {93}},\
  \bibinfo {pages} {054003} (\bibinfo {year} {2016})},\ \Eprint
  {http://arxiv.org/abs/1512.04034} {arXiv:1512.04034 [hep-ph]} \BibitemShut
  {NoStop}%
\bibitem [{\citenamefont {Shivashankara}\ \emph {et~al.}(2015)\citenamefont
  {Shivashankara}, \citenamefont {Wu},\ and\ \citenamefont
  {Datta}}]{Shivashankara:2015cta}%
  \BibitemOpen
  \bibfield  {author} {\bibinfo {author} {\bibfnamefont {S.}~\bibnamefont
  {Shivashankara}}, \bibinfo {author} {\bibfnamefont {W.}~\bibnamefont {Wu}}, \
  and\ \bibinfo {author} {\bibfnamefont {A.}~\bibnamefont {Datta}},\ }\href
  {\doibase 10.1103/PhysRevD.91.115003} {\bibfield  {journal} {\bibinfo
  {journal} {Phys. Rev. D}\ }\textbf {\bibinfo {volume} {91}},\ \bibinfo
  {pages} {115003} (\bibinfo {year} {2015})},\ \Eprint
  {http://arxiv.org/abs/1502.07230} {arXiv:1502.07230 [hep-ph]} \BibitemShut
  {NoStop}%
\bibitem [{\citenamefont {Li}\ \emph {et~al.}(2017)\citenamefont {Li},
  \citenamefont {Yang},\ and\ \citenamefont {Zhang}}]{Li:2016pdv}%
  \BibitemOpen
  \bibfield  {author} {\bibinfo {author} {\bibfnamefont {X.-Q.}\ \bibnamefont
  {Li}}, \bibinfo {author} {\bibfnamefont {Y.-D.}\ \bibnamefont {Yang}}, \ and\
  \bibinfo {author} {\bibfnamefont {X.}~\bibnamefont {Zhang}},\ }\href
  {\doibase 10.1007/JHEP02(2017)068} {\bibfield  {journal} {\bibinfo  {journal}
  {JHEP}\ }\textbf {\bibinfo {volume} {02}},\ \bibinfo {pages} {068} (\bibinfo
  {year} {2017})},\ \Eprint {http://arxiv.org/abs/1611.01635} {arXiv:1611.01635
  [hep-ph]} \BibitemShut {NoStop}%
\bibitem [{\citenamefont {Datta}\ \emph {et~al.}(2017)\citenamefont {Datta},
  \citenamefont {Kamali}, \citenamefont {Meinel},\ and\ \citenamefont
  {Rashed}}]{Datta:2017aue}%
  \BibitemOpen
  \bibfield  {author} {\bibinfo {author} {\bibfnamefont {A.}~\bibnamefont
  {Datta}}, \bibinfo {author} {\bibfnamefont {S.}~\bibnamefont {Kamali}},
  \bibinfo {author} {\bibfnamefont {S.}~\bibnamefont {Meinel}}, \ and\ \bibinfo
  {author} {\bibfnamefont {A.}~\bibnamefont {Rashed}},\ }\href {\doibase
  10.1007/JHEP08(2017)131} {\bibfield  {journal} {\bibinfo  {journal} {JHEP}\
  }\textbf {\bibinfo {volume} {08}},\ \bibinfo {pages} {131} (\bibinfo {year}
  {2017})},\ \Eprint {http://arxiv.org/abs/1702.02243} {arXiv:1702.02243
  [hep-ph]} \BibitemShut {NoStop}%
\bibitem [{\citenamefont {Ray}\ \emph {et~al.}(2019)\citenamefont {Ray},
  \citenamefont {Sahoo},\ and\ \citenamefont {Mohanta}}]{Ray:2018hrx}%
  \BibitemOpen
  \bibfield  {author} {\bibinfo {author} {\bibfnamefont {A.}~\bibnamefont
  {Ray}}, \bibinfo {author} {\bibfnamefont {S.}~\bibnamefont {Sahoo}}, \ and\
  \bibinfo {author} {\bibfnamefont {R.}~\bibnamefont {Mohanta}},\ }\href
  {\doibase 10.1103/PhysRevD.99.015015} {\bibfield  {journal} {\bibinfo
  {journal} {Phys. Rev.}\ }\textbf {\bibinfo {volume} {D99}},\ \bibinfo {pages}
  {015015} (\bibinfo {year} {2019})},\ \Eprint
  {http://arxiv.org/abs/1812.08314} {arXiv:1812.08314 [hep-ph]} \BibitemShut
  {NoStop}%
\bibitem [{\citenamefont {Bernlochner}\ \emph {et~al.}(2019)\citenamefont
  {Bernlochner}, \citenamefont {Ligeti}, \citenamefont {Robinson},\ and\
  \citenamefont {Sutcliffe}}]{Bernlochner:2018bfn}%
  \BibitemOpen
  \bibfield  {author} {\bibinfo {author} {\bibfnamefont {F.~U.}\ \bibnamefont
  {Bernlochner}}, \bibinfo {author} {\bibfnamefont {Z.}~\bibnamefont {Ligeti}},
  \bibinfo {author} {\bibfnamefont {D.~J.}\ \bibnamefont {Robinson}}, \ and\
  \bibinfo {author} {\bibfnamefont {W.~L.}\ \bibnamefont {Sutcliffe}},\ }\href
  {\doibase 10.1103/PhysRevD.99.055008} {\bibfield  {journal} {\bibinfo
  {journal} {Phys. Rev.}\ }\textbf {\bibinfo {volume} {D99}},\ \bibinfo {pages}
  {055008} (\bibinfo {year} {2019})},\ \Eprint
  {http://arxiv.org/abs/1812.07593} {arXiv:1812.07593 [hep-ph]} \BibitemShut
  {NoStop}%
\bibitem [{\citenamefont {Di~Salvo}\ \emph {et~al.}(2018)\citenamefont
  {Di~Salvo}, \citenamefont {Fontanelli},\ and\ \citenamefont
  {Ajaltouni}}]{DiSalvo:2018ngq}%
  \BibitemOpen
  \bibfield  {author} {\bibinfo {author} {\bibfnamefont {E.}~\bibnamefont
  {Di~Salvo}}, \bibinfo {author} {\bibfnamefont {F.}~\bibnamefont
  {Fontanelli}}, \ and\ \bibinfo {author} {\bibfnamefont {Z.~J.}\ \bibnamefont
  {Ajaltouni}},\ }\href {\doibase 10.1142/S0217751X18501695} {\bibfield
  {journal} {\bibinfo  {journal} {Int. J. Mod. Phys.}\ }\textbf {\bibinfo
  {volume} {A33}},\ \bibinfo {pages} {1850169} (\bibinfo {year} {2018})},\
  \Eprint {http://arxiv.org/abs/1804.05592} {arXiv:1804.05592 [hep-ph]}
  \BibitemShut {NoStop}%
\bibitem [{\citenamefont {Blanke}\ \emph
  {et~al.}(2019{\natexlab{b}})\citenamefont {Blanke}, \citenamefont
  {Crivellin}, \citenamefont {Kitahara}, \citenamefont {Moscati}, \citenamefont
  {Nierste},\ and\ \citenamefont {Ni\v{s}and\v{z}i\'c}}]{Blanke:2019qrx}%
  \BibitemOpen
  \bibfield  {author} {\bibinfo {author} {\bibfnamefont {M.}~\bibnamefont
  {Blanke}}, \bibinfo {author} {\bibfnamefont {A.}~\bibnamefont {Crivellin}},
  \bibinfo {author} {\bibfnamefont {T.}~\bibnamefont {Kitahara}}, \bibinfo
  {author} {\bibfnamefont {M.}~\bibnamefont {Moscati}}, \bibinfo {author}
  {\bibfnamefont {U.}~\bibnamefont {Nierste}}, \ and\ \bibinfo {author}
  {\bibfnamefont {I.}~\bibnamefont {Ni\v{s}and\v{z}i\'c}},\ }\href {\doibase
  10.1103/PhysRevD.100.035035} {\bibfield  {journal} {\bibinfo  {journal}
  {Phys. Rev.}\ }\textbf {\bibinfo {volume} {D100}},\ \bibinfo {pages} {035035}
  (\bibinfo {year} {2019}{\natexlab{b}})},\ \Eprint
  {http://arxiv.org/abs/1905.08253} {arXiv:1905.08253 [hep-ph]} \BibitemShut
  {NoStop}%
\bibitem [{\citenamefont {B{\"o}er}\ \emph {et~al.}(2019)\citenamefont
  {B{\"o}er}, \citenamefont {Kokulu}, \citenamefont {Toelstede},\ and\
  \citenamefont {van Dyk}}]{Boer:2019zmp}%
  \BibitemOpen
  \bibfield  {author} {\bibinfo {author} {\bibfnamefont {P.}~\bibnamefont
  {B{\"o}er}}, \bibinfo {author} {\bibfnamefont {A.}~\bibnamefont {Kokulu}},
  \bibinfo {author} {\bibfnamefont {J.-N.}\ \bibnamefont {Toelstede}}, \ and\
  \bibinfo {author} {\bibfnamefont {D.}~\bibnamefont {van Dyk}},\ }\href@noop
  {} {\  (\bibinfo {year} {2019})},\ \Eprint {http://arxiv.org/abs/1907.12554}
  {arXiv:1907.12554 [hep-ph]} \BibitemShut {NoStop}%
\bibitem [{\citenamefont {Mu}\ \emph {et~al.}(2019)\citenamefont {Mu},
  \citenamefont {Li}, \citenamefont {Zou},\ and\ \citenamefont
  {Zhu}}]{Mu:2019bin}%
  \BibitemOpen
  \bibfield  {author} {\bibinfo {author} {\bibfnamefont {X.-L.}\ \bibnamefont
  {Mu}}, \bibinfo {author} {\bibfnamefont {Y.}~\bibnamefont {Li}}, \bibinfo
  {author} {\bibfnamefont {Z.-T.}\ \bibnamefont {Zou}}, \ and\ \bibinfo
  {author} {\bibfnamefont {B.}~\bibnamefont {Zhu}},\ }\href {\doibase
  10.1103/PhysRevD.100.113004} {\bibfield  {journal} {\bibinfo  {journal}
  {Phys. Rev. D}\ }\textbf {\bibinfo {volume} {100}},\ \bibinfo {pages}
  {113004} (\bibinfo {year} {2019})},\ \Eprint
  {http://arxiv.org/abs/1909.10769} {arXiv:1909.10769 [hep-ph]} \BibitemShut
  {NoStop}%
\bibitem [{\citenamefont {Hu}\ \emph {et~al.}(2021)\citenamefont {Hu},
  \citenamefont {Li}, \citenamefont {Yang},\ and\ \citenamefont
  {Zheng}}]{Hu:2020axt}%
  \BibitemOpen
  \bibfield  {author} {\bibinfo {author} {\bibfnamefont {Q.-Y.}\ \bibnamefont
  {Hu}}, \bibinfo {author} {\bibfnamefont {X.-Q.}\ \bibnamefont {Li}}, \bibinfo
  {author} {\bibfnamefont {Y.-D.}\ \bibnamefont {Yang}}, \ and\ \bibinfo
  {author} {\bibfnamefont {D.-H.}\ \bibnamefont {Zheng}},\ }\href {\doibase
  10.1007/JHEP02(2021)183} {\bibfield  {journal} {\bibinfo  {journal} {JHEP}\
  }\textbf {\bibinfo {volume} {02}},\ \bibinfo {pages} {183} (\bibinfo {year}
  {2021})},\ \Eprint {http://arxiv.org/abs/2011.05912} {arXiv:2011.05912
  [hep-ph]} \BibitemShut {NoStop}%
\bibitem [{\citenamefont {Penalva}\ \emph {et~al.}(2019)\citenamefont
  {Penalva}, \citenamefont {Hern\'andez},\ and\ \citenamefont
  {Nieves}}]{Penalva:2019rgt}%
  \BibitemOpen
  \bibfield  {author} {\bibinfo {author} {\bibfnamefont {N.}~\bibnamefont
  {Penalva}}, \bibinfo {author} {\bibfnamefont {E.}~\bibnamefont
  {Hern\'andez}}, \ and\ \bibinfo {author} {\bibfnamefont {J.}~\bibnamefont
  {Nieves}},\ }\href {\doibase 10.1103/PhysRevD.100.113007} {\bibfield
  {journal} {\bibinfo  {journal} {Phys. Rev.}\ }\textbf {\bibinfo {volume}
  {D100}},\ \bibinfo {pages} {113007} (\bibinfo {year} {2019})},\ \Eprint
  {http://arxiv.org/abs/1908.02328} {arXiv:1908.02328 [hep-ph]} \BibitemShut
  {NoStop}%
\bibitem [{\citenamefont {Penalva}\ \emph
  {et~al.}(2020{\natexlab{b}})\citenamefont {Penalva}, \citenamefont
  {Hern\'andez},\ and\ \citenamefont {Nieves}}]{Penalva:2020xup}%
  \BibitemOpen
  \bibfield  {author} {\bibinfo {author} {\bibfnamefont {N.}~\bibnamefont
  {Penalva}}, \bibinfo {author} {\bibfnamefont {E.}~\bibnamefont
  {Hern\'andez}}, \ and\ \bibinfo {author} {\bibfnamefont {J.}~\bibnamefont
  {Nieves}},\ }\href {\doibase 10.1103/PhysRevD.101.113004} {\bibfield
  {journal} {\bibinfo  {journal} {Phys. Rev. D}\ }\textbf {\bibinfo {volume}
  {101}},\ \bibinfo {pages} {113004} (\bibinfo {year} {2020}{\natexlab{b}})},\
  \Eprint {http://arxiv.org/abs/2004.08253} {arXiv:2004.08253 [hep-ph]}
  \BibitemShut {NoStop}%
\bibitem [{\citenamefont {Alonso}\ \emph {et~al.}(2016)\citenamefont {Alonso},
  \citenamefont {Kobach},\ and\ \citenamefont
  {Martin~Camalich}}]{Alonso:2016gym}%
  \BibitemOpen
  \bibfield  {author} {\bibinfo {author} {\bibfnamefont {R.}~\bibnamefont
  {Alonso}}, \bibinfo {author} {\bibfnamefont {A.}~\bibnamefont {Kobach}}, \
  and\ \bibinfo {author} {\bibfnamefont {J.}~\bibnamefont {Martin~Camalich}},\
  }\href {\doibase 10.1103/PhysRevD.94.094021} {\bibfield  {journal} {\bibinfo
  {journal} {Phys. Rev. D}\ }\textbf {\bibinfo {volume} {94}},\ \bibinfo
  {pages} {094021} (\bibinfo {year} {2016})},\ \Eprint
  {http://arxiv.org/abs/1602.07671} {arXiv:1602.07671 [hep-ph]} \BibitemShut
  {NoStop}%
\bibitem [{\citenamefont {Alonso}\ \emph
  {et~al.}(2017{\natexlab{b}})\citenamefont {Alonso}, \citenamefont
  {Martin~Camalich},\ and\ \citenamefont {Westhoff}}]{Alonso:2017ktd}%
  \BibitemOpen
  \bibfield  {author} {\bibinfo {author} {\bibfnamefont {R.}~\bibnamefont
  {Alonso}}, \bibinfo {author} {\bibfnamefont {J.}~\bibnamefont
  {Martin~Camalich}}, \ and\ \bibinfo {author} {\bibfnamefont {S.}~\bibnamefont
  {Westhoff}},\ }\href {\doibase 10.1103/PhysRevD.95.093006} {\bibfield
  {journal} {\bibinfo  {journal} {Phys. Rev. D}\ }\textbf {\bibinfo {volume}
  {95}},\ \bibinfo {pages} {093006} (\bibinfo {year} {2017}{\natexlab{b}})},\
  \Eprint {http://arxiv.org/abs/1702.02773} {arXiv:1702.02773 [hep-ph]}
  \BibitemShut {NoStop}%
\bibitem [{\citenamefont {Asadi}\ \emph {et~al.}(2020)\citenamefont {Asadi},
  \citenamefont {Hallin}, \citenamefont {Martin~Camalich}, \citenamefont
  {Shih},\ and\ \citenamefont {Westhoff}}]{Asadi:2020fdo}%
  \BibitemOpen
  \bibfield  {author} {\bibinfo {author} {\bibfnamefont {P.}~\bibnamefont
  {Asadi}}, \bibinfo {author} {\bibfnamefont {A.}~\bibnamefont {Hallin}},
  \bibinfo {author} {\bibfnamefont {J.}~\bibnamefont {Martin~Camalich}},
  \bibinfo {author} {\bibfnamefont {D.}~\bibnamefont {Shih}}, \ and\ \bibinfo
  {author} {\bibfnamefont {S.}~\bibnamefont {Westhoff}},\ }\href {\doibase
  10.1103/PhysRevD.102.095028} {\bibfield  {journal} {\bibinfo  {journal}
  {Phys. Rev. D}\ }\textbf {\bibinfo {volume} {102}},\ \bibinfo {pages}
  {095028} (\bibinfo {year} {2020})},\ \Eprint
  {http://arxiv.org/abs/2006.16416} {arXiv:2006.16416 [hep-ph]} \BibitemShut
  {NoStop}%
\bibitem [{\citenamefont {Penalva}\ \emph
  {et~al.}(2021{\natexlab{b}})\citenamefont {Penalva}, \citenamefont
  {Hern\'andez},\ and\ \citenamefont {Nieves}}]{Penalva:2021wye}%
  \BibitemOpen
  \bibfield  {author} {\bibinfo {author} {\bibfnamefont {N.}~\bibnamefont
  {Penalva}}, \bibinfo {author} {\bibfnamefont {E.}~\bibnamefont
  {Hern\'andez}}, \ and\ \bibinfo {author} {\bibfnamefont {J.}~\bibnamefont
  {Nieves}},\ }\href {\doibase 10.1007/JHEP10(2021)122} {\bibfield  {journal}
  {\bibinfo  {journal} {JHEP}\ }\textbf {\bibinfo {volume} {10}},\ \bibinfo
  {pages} {122} (\bibinfo {year} {2021}{\natexlab{b}})},\ \Eprint
  {http://arxiv.org/abs/2107.13406} {arXiv:2107.13406 [hep-ph]} \BibitemShut
  {NoStop}%
\bibitem [{\citenamefont {Tanaka}\ and\ \citenamefont
  {Watanabe}(2010)}]{Tanaka:2010se}%
  \BibitemOpen
  \bibfield  {author} {\bibinfo {author} {\bibfnamefont {M.}~\bibnamefont
  {Tanaka}}\ and\ \bibinfo {author} {\bibfnamefont {R.}~\bibnamefont
  {Watanabe}},\ }\href {\doibase 10.1103/PhysRevD.82.034027} {\bibfield
  {journal} {\bibinfo  {journal} {Phys. Rev. D}\ }\textbf {\bibinfo {volume}
  {82}},\ \bibinfo {pages} {034027} (\bibinfo {year} {2010})},\ \Eprint
  {http://arxiv.org/abs/1005.4306} {arXiv:1005.4306 [hep-ph]} \BibitemShut
  {NoStop}%
\bibitem [{\citenamefont {Cerri}\ \emph {et~al.}(2019)\citenamefont {Cerri}
  \emph {et~al.}}]{Cerri:2018ypt}%
  \BibitemOpen
  \bibfield  {author} {\bibinfo {author} {\bibfnamefont {A.}~\bibnamefont
  {Cerri}} \emph {et~al.},\ }\enquote {\bibinfo {title} {{Report from Working
  Group 4}: {Opportunities in Flavour Physics at the HL-LHC and HE-LHC}},}\ in\
  \href {\doibase 10.23731/CYRM-2019-007.867} {\emph {\bibinfo {booktitle}
  {{Report on the Physics at the HL-LHC,and Perspectives for the HE-LHC}}}},\
  Vol.~\bibinfo {volume} {7},\ \bibinfo {editor} {edited by\ \bibinfo {editor}
  {\bibfnamefont {A.}~\bibnamefont {Dainese}}, \bibinfo {editor} {\bibfnamefont
  {M.}~\bibnamefont {Mangano}}, \bibinfo {editor} {\bibfnamefont {A.~B.}\
  \bibnamefont {Meyer}}, \bibinfo {editor} {\bibfnamefont {A.}~\bibnamefont
  {Nisati}}, \bibinfo {editor} {\bibfnamefont {G.}~\bibnamefont {Salam}}, \
  and\ \bibinfo {editor} {\bibfnamefont {M.~A.}\ \bibnamefont {Vesterinen}}}\
  (\bibinfo {year} {2019})\ pp.\ \bibinfo {pages} {867--1158},\ \Eprint
  {http://arxiv.org/abs/1812.07638} {arXiv:1812.07638 [hep-ph]} \BibitemShut
  {NoStop}%
\end{thebibliography}%

\end{document}